\documentclass[aps,pra,twocolumn,footinbib,amsmath,amssymb,amsfonts]{revtex4-2} 

\usepackage{color,framed,graphicx}
\usepackage{amsmath, amssymb, amsfonts,mathrsfs, amsthm, dsfont}
\usepackage{physics}
\usepackage{lipsum}
\usepackage[hypertexnames=false]{hyperref}
\usepackage{enumerate}
\usepackage{array}
\usepackage{verbatim}

\usepackage{tikz-cd}
\usepackage{tikz}
\usetikzlibrary{arrows, decorations.pathmorphing, decorations.markings, decorations.pathreplacing, calligraphy}
\usepackage{float}
\usepackage{relsize}

\newtheoremstyle{naturethm}%
  {6pt}   
  {6pt}   
  {\normalfont} 
  {}      
  {}      
  {:}     
  {0.5em} 
  {\itshape\thmname{#1}\ \thmnumber{#2}}

\theoremstyle{naturethm}

\newtheorem{theorem}{Theorem}

\newtheorem{definition}{Definition}

\begin{document}

\title{On the Fundamental Resource for Exponential Advantage in Quantum Channel Learning}

\author{Minsoo Kim}
\affiliation{Department of Physics, Korea Advanced Institute of Science and Technology, Daejeon 34141, Korea}

\author{Changhun Oh}\email[]{changhun0218@gmail.com}
\affiliation{Department of Physics, Korea Advanced Institute of Science and Technology, Daejeon 34141, Korea}

\date{\today}

\begin{abstract}

Quantum resources enable us to achieve an exponential advantage in learning the properties of unknown physical systems by employing quantum memory.
While entanglement with quantum memory is recognized as a necessary qualitative resource, its quantitative role remains less understood. 
In this work, we distinguish between two fundamental resources provided by quantum memory---entanglement and ancilla qubits---and analyze their separate contributions to the sampling complexity of quantum learning. 
Focusing on the task of Pauli channel learning, a prototypical example of quantum channel learning, remarkably, we prove that vanishingly small entanglement in the input state already suffices to accomplish the learning task with only a polynomial number of channel queries in the number of qubits.
In contrast, we show that without a sufficient number of ancilla qubits, even learning partial information about the channel demands an exponentially large sample complexity.
Thus, our findings reveal that while a large amount of entanglement is not necessary, the dimension of the quantum memory is a crucial resource.
Hence, by identifying how the two resources contribute differently, our work offers deeper insight into the nature of the quantum learning advantage.

\end{abstract}

\maketitle

\section{Introduction}

Quantum advantage arises from the utilization of quantum effects, manifesting in diverse tasks such as accelerating computation by quantum computing
\cite{shor1994algorithms, lloyd1996universal, harrow2017quantum, arute2019quantum, zhong2020quantum, wu2021strong, madsen2022quantum, morvan2024phase, zhong2021phaseprogrammable, deng2023gaussian, decross2024computational},
and improving sensitivity by quantum metrology
\cite{giovannetti2006quantum, giovannetti2011advances, polino2020photonic, degen2017quantum}.
In addition to the above, a particularly promising approach to realize a quantum advantage, recently attracting much attention, is quantum learning, which leverages quantum effects to achieve high efficiency in learning unknown physical systems that classical approaches cannot achieve from an information perspective
\cite{huang2021informationtheoretic, huang2022quantum}.
Various forms of quantum learning advantage have been investigated, including expectation value estimation~\cite{huang2021informationtheoretic, huang2022quantum}, learning quantum channels~\cite{flammia2020efficient, flammia2021pauli, chen2021exponential, chen2022quantum, chen2024tight, chen2025efficient}, and extension of learning techniques to continuous-variable systems for the characterization of quantum states~\cite{coroi2025exponential} and channels~\cite{oh2024entanglementenabled}.
Especially, quantum channel learning has attracted increasing attention due to its utility in learning errors of quantum devices
\cite{wallman2016noise, hashim2021randomized}
and mitigating noise
\cite{vandenberg2023probabilistic, ferracin2024efficiently, kim2023evidence}
for quantum computing.

\begin{figure}[t]
    \centerline{\includegraphics[scale=1.0]{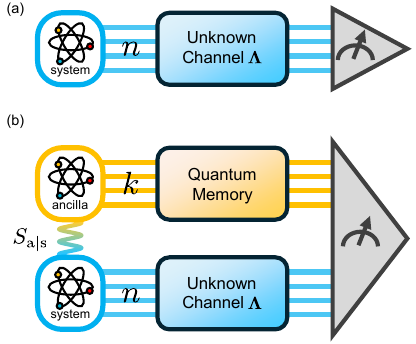}}
    \caption{
        \textbf{Schematic illustration of learning a quantum channel}.
        (a) Learning a quantum channel acting on an $n$-qubit system without quantum memory.
        (b) Quantum channel learning with the assistance of quantum memory.
        The availability of quantum memory allows the use of $k$ ancilla qubits as a resource.
        Another resource, the entanglement entropy between the ancilla and the system, is denoted by $S_{\text{a}|\text{s}}$.
        The channel acts only on the system, while the ancilla is stored in the quantum memory.
        Joint measurements, such as Bell measurements, are also permitted.
    }
    \label{Fig1}
\end{figure}

The quantum advantage in channel learning is often defined by the accessibility to quantum memory~\cite{huang2021informationtheoretic,huang2022quantum}; thus, quantum memory is regarded as a quantum resource in this context.
Hence, two different families of learning schemes are considered and compared with each other~\cite{chen2021exponential, chen2024tight, oh2024entanglementenabled, coroi2025exponential},
which are illustrated in Fig.~\ref{Fig1}.
As depicted in Fig.~\ref{Fig1}(a), learning an unknown $n$-qubit quantum channel $\mathbf{\Lambda}$ involves preparing input states, applying $\mathbf{\Lambda}$, and estimating its parameters from measurement outcomes~\cite{huang2021informationtheoretic}.
Without quantum memory, the required sample complexity for learning---the number of channel applications---often scales exponentially with~$n$
\cite{flammia2020efficient, chen2021exponential, chen2022quantum, chen2024tight, chen2025efficient}.
In contrast, by using quantum memory [Fig.~\ref{Fig1}(b)], the sample complexity can be significantly reduced.
As an example, Pauli channel learning, which is a crucial task for various applications
\cite{wallman2016noise, hashim2021randomized, vandenberg2023probabilistic, ferracin2024efficiently, kim2023evidence}, can be accomplished with only $O(n)$  samples by employing a $2n$-qubit Bell pair as input, while any scheme in Fig.~\ref{Fig1}(a) requires $\Omega(2^{n})$ samples~\cite{chen2022quantum, chen2025efficient}, establishing exponential quantum advantage through quantum memory.
In addition, this advantage has recently been demonstrated experimentally~\cite{liu2025quantum, seif2024entanglementenhanced}.

Accordingly, quantum memory has been identified as a key resource in quantum learning.
In fact, in the literature, entanglement-enabled learning and ancilla-assisted learning are used interchangeably in this context
\cite{chen2021exponential, chen2024tight, oh2024entanglementenabled, coroi2025exponential},
which implicitly treats the two resources as equivalent.
From an information-theoretic perspective, however, two distinct types of resources emerge when quantum memory is provided: the \textit{ancilla qubits} in the memory and the \textit{entanglement} between these ancilla qubits and the system.
In many cases, these two resources can be regarded as effectively identical, such as when Bell pairs are used as input probes.
However, this equivalence does not necessarily hold in general, and understanding its breakdown is the main focus of this work.

In this work, we consider Pauli channel learning and show that the two resources, ancilla qubits and entanglement between the ancilla and the system, contribute in fundamentally distinct ways to the exponential quantum advantage.
In particular, we first prove that while the sample complexity must be exponential when the number of ancilla qubits is limited~\cite{chen2021exponential, chen2022quantum, chen2025efficient}, even with a vanishingly small amount of entanglement, a polynomial number of samples is sufficient to learn a Pauli channel.
This highlights that a large amount of entanglement is not necessary for the exponential quantum advantage as long as the number of ancilla qubits is sufficient.
To understand the necessary number of ancilla qubits in practice, we consider learning a subset of the channel parameters---specifically, the Pauli eigenvalues associated with low-weight Pauli strings (see Definition~\ref{def: Pauli channel learning}). 
We then prove that even in this easier setting, if the number of ancilla qubits is limited, the sample complexity must be exponential.
This result emphasizes that the number of ancilla qubits plays an even more crucial role than previously recognized.
Therefore, by revealing the distinct contributions of entanglement and ancilla qubit number to the exponential quantum learning advantage, our work provides a comprehensive connection between quantum resources and the sample complexity of channel learning.

\section{Results}

\subsection{Pauli Channel Learning Setup}

We begin by introducing the definitions of Pauli strings, Pauli channels, and Pauli channel learning, which are the main subject of this work.
Each Pauli operator $P_{a} \in \{I, X, Y, Z\}$ is labeled by a 2-bit string $a:=a_{x}a_{z} \in \mathds{Z}_{2}^{2}$, and expressed as $P_{a} = i^{a_{x}a_{z}} X^{a_{x}}Z^{a_{z}}$.
This extends to an $n$-qubit system via the Pauli string $P_{\mathbf{a}} = \bigotimes_{j=1}^{n}P_{a_{j}}$ determined by a $2n$-bit string $\mathbf{a} := a_{1}a_{2} \cdots a_{n} = a_{1,x}a_{1,z}a_{2,x}a_{2,z} \cdots a_{n,x}a_{n,z} \in \mathds{Z}_{2}^{2n}$.
Any $P_{\mathbf{a}}$ and $P_{\mathbf{b}}$ satisfy $P_{\mathbf{a}} P_{\mathbf{b}} = (-1)^{\langle \mathbf{a}, \mathbf{b}\rangle} P_{\mathbf{b}}P_{\mathbf{a}}$, where $\langle \mathbf{a}, \mathbf{b}\rangle := \sum_{j=1}^{n}(a_{j,x}b_{j,z} + a_{j,z}b_{j,x})\mod 2$.
The Pauli weight $|\mathbf{a}|$ is defined as the number of non-identity operators in the Pauli string $P_{\mathbf{a}}$.
A Pauli channel $\mathbf{\Lambda}(\cdot)$ is defined as
\begin{equation}
    \label{def of Pauli channel}
    \mathbf{\Lambda}(\cdot)
    := \sum_{\mathbf{a} \in \mathds{Z}_{2}^{2n}} p({\mathbf{a}}) P_{\mathbf{a}}(\cdot)P_{\mathbf{a}}
    = \frac{1}{2^{n}}\sum_{\mathbf{b} \in \mathds{Z}_{2}^{2n}}
    \lambda(\mathbf{b}) \Tr[P_{\mathbf{b}}(\cdot)]P_{\mathbf{b}},
\end{equation}
where $p(\mathbf{a})$ is a Pauli error rate, and $\lambda(\mathbf{b})$ is a Pauli eigenvalue.
They are related via the Walsh-Hadamard transform, given by $p({\mathbf{a}}) = \frac{1}{4^{n}} \sum_{\mathbf{b} \in \mathds{Z}_{2}^{2n}} \lambda(\mathbf{b}) (-1)^{\langle \mathbf{a}, \mathbf{b}\rangle}$~\cite{flammia2020efficient}.

As shown in Fig.~\ref{Fig1}(b), we treat the ancilla as an additional register consisting of $k$ qubits.
Since the channel acts only on the system, the output state corresponding to an input state $\rho_{\text{in}}$ is given by $(\mathds{1}_{\text{anc}}\otimes \mathbf{\Lambda}) (\rho_{\text{in}}) = \frac{1}{2^{n}} \sum_{\mathbf{b}} \lambda(\mathbf{b}) \Tr_{\text{sys}}[(I_{\text{anc}}\otimes P_{\mathbf{b}}) \rho_{\text{in}}] \otimes P_{\mathbf{b}}$, where $\mathds{1}_{\text{anc}}$ and $I_{\text{anc}}$ denote the identity channel and operator on the ancilla, respectively, and $\Tr_{\text{sys}}$ denotes the partial trace over the system.

Based on the definition of the Pauli channel in Eq.~\eqref{def of Pauli channel}, we define the Pauli channel learning task. 
\begin{definition}[($\varepsilon, \delta, w$)-Pauli channel learning task]
    \label{def: Pauli channel learning}
    \normalfont
    We are given access to $N$ copies of the Pauli channel $\mathbf{\Lambda}$.
    Classical data are collected by preparing an input state, applying a single copy of the channel, and measuring the output.
    In each round, both the input and measurement POVM can be chosen adaptively based on prior measurement outcomes.
    After $N$ measurements, the goal is to provide an estimate $\hat{\lambda}(\mathbf{b})$ satisfying \( |\hat{\lambda}(\mathbf{b}) - \lambda(\mathbf{b})| \leq \varepsilon \) for any \( \mathbf{b} \in \mathds{Z}_2^{2n} \) such that \( |\mathbf{b}| \leq w \) with success probability at least \( 1 - \delta \).
\end{definition}
\noindent
In this setting, the number of channel queries required to accomplish the task, denoted by $N$, is referred to as the sample complexity.
The motivation for estimating $\lambda(\mathbf{b})$ for low $|\mathbf{b}|$ stems from realistic physical error models~\cite{flammia2020efficient}.
The number of parameters to learn is determined by the maximum weight $w$; when $w=n$, the learning task requires estimating a total of $4^{n}$ parameters.
We consider only $0\leq k\leq n$, since $k=n$ ancilla qubits are sufficient to accomplish the task with $N = O(n)$ even when $w=n$~\cite{chen2022quantum}. 
In addition, as implied in the definition, this work focuses on non-concatenated applications of the channel, although concatenated applications might potentially reduce the sample complexity~\cite{chen2025efficient}, which we leave as future work.

\subsection{Learning with restricted entanglement}
We first analyze the sample complexity in Pauli channel learning depending on the entanglement between the system and the ancilla.
Here, the entanglement is defined as the entanglement entropy of the input probe state, and we denote it as $S_{\text{a}|\text{s}}$ [Fig.~\ref{Fig1}(b)].
As highlighted in several previous works~\cite{huang2021informationtheoretic,chen2022quantum,chen2021exponential, chen2024tight, chen2025efficient}, if $S_{\text{a}|\text{s}}=0$, an exponentially large $N$ is required to accomplish the ($\varepsilon, \delta, w=n$)-Pauli channel learning task.
Consequently, one might expect that even if enough $k=n$ ancilla qubits are available, the exponential $N$ may still be necessary when the input state has small $S_{\text{a}|\text{s}}$, i.e., it is close to a separable state.
Remarkably, our first main result shows that the exponential advantage can be achieved using only small $S_{\text{a}|\text{s}}$, even for estimating all $\lambda(\mathbf{b})$, i.e., $w=n$.
More precisely, our theorem below reveals that even when $S_{\text{a}|\text{s}}$ in the input state is inverse-polynomially small, the Pauli channel can be learned with polynomially many samples in $n$.
\begin{theorem}
    \normalfont
    \label{theorem 1}
    For an $n$-qubit system with $k=n$ ancilla qubits,
    there exists a scheme that accomplishes the ($\varepsilon, \delta, n$)-Pauli channel learning task with sample complexity $N = O(n \alpha^{-2} \times \varepsilon^{-2}\log \delta^{-1})$ by using input states, each with entanglement $S_{\text{a}|\text{s}} = \Theta(n\alpha)$, where $\alpha = \Theta(1/\text{poly}(n))$.
\end{theorem}
\noindent
Hence, Theorem~\ref{theorem 1} implies that the exponential advantage remains attainable even when highly entangled states are inaccessible. 
This reveals that the contributions of entanglement and ancilla qubit number to the exponential advantage are fundamentally distinct, which becomes evident by comparing the two cases:
(1) If only $k$ ancilla qubits are allowed, even if maximally entangled states are used, i.e., $S_{\text{a}|\text{s}} = k$, $N=\Omega(2^{(n-k)/3})$ is required~\cite{chen2022quantum, chen2021exponential, chen2025efficient} (the lower bound will be improved later).
(2) In contrast, when $n$ ancilla qubits are provided, by preparing input states such that each has the same amount of entanglement $S_{\text{a}|\text{s}} = k$ (i.e., setting $\alpha = k/n$), the learning task can be accomplished with a sample complexity that scales polynomially with $n$.

Note that a smaller $S_{\text{a}|\text{s}}$ leads to a larger $N$ as a trade-off.
Thus, the total entanglement resource of the $N$ copies of the input state cannot be arbitrarily reduced.
For instance, when we set $\alpha=1$, each input state has entanglement $S_{\text{a}|\text{s}} = \Theta(n)$, and $O(n)$ such states are required; consequently, the total entanglement is $O(n^{2})$ (this corresponds to the case where a Bell pair is employed in~\cite{chen2022quantum}).
However, if we take $\alpha=1/n^{c}$ ($c>0$), each input has $S_{\text{a}|\text{s}} = \Theta(n^{1-c})$ and $O(n^{1+2c})$ such states are required, so the total required entanglement is $O(n^{2+c})$.

{\it Proof Sketch.} (See Supplemental Material (SM) Sec.~S2~\cite{supple} for the full proof)
To prove Theorem~\ref{theorem 1}, we provide an explicit input state $|{\Psi_{\text{in}}(\alpha)}\rangle$ parameterized by a constant $\alpha$, with $S_{\text{a}|\text{s}} = \Theta(n\alpha)$.
The input state is a superposition of the $2n$-qubit Bell pair $|{\Psi_{\text{B}}}\rangle$ and a separable state $|{\Psi_{\text{sep}}}\rangle$, defined as
\begin{equation}
    \label{negligibly entangled input}
    |{\Psi_{\text{in}}(\alpha)}\rangle
    := \sqrt{\alpha'} |{\Psi_{\text{B}}}\rangle + \sqrt{1-\alpha} |{\Psi_{\text{sep}}}\rangle,
\end{equation}
where $\sqrt{\alpha'}:=\sqrt{\alpha + (1-\alpha)/2^{n}} - \sqrt{(1-\alpha)/2^{n}}$.
Here, the separable state $|{\Psi_{\text{sep}}}\rangle$ is defined as
\begin{equation}
\begin{aligned}
    \label{special separable state}
    |{\Psi_{\text{sep}}}\rangle \langle{\Psi_{\text{sep}}}|
    :=\underbrace{\left[ \rho_{\text{sep}}^{\text{T}} \right]^{\otimes n}}_{\text{ancilla}}
	\otimes \underbrace{\left[ \rho_{\text{sep}} \right]^{\otimes n}}_{\text{system}}, \\
    \rho_{\text{sep}} = \frac{1}{2}\left(I + \frac{1}{\sqrt{3}}(X+Y+Z)\right).
\end{aligned}
\end{equation}
Note that $\rho_{\text{sep}}$ is a pure product state, and setting $\alpha=1$ recovers $|{\Psi_{\text{in}}(\alpha=1)}\rangle = |{\Psi_{\text{B}}}\rangle$. 
As a result of the superposition of the two states with weight $\alpha$, the entanglement $S_{\text{a}|\text{s}}$ is reduced from $n$ (of the Bell pair) to $\Theta(n\alpha)$. 
For measurement, we use the Bell measurement POVM $\{E_{\mathbf{v}}\}_{\mathbf{v} \in \mathds{Z}_{2}^{2n}}$, where $E_{\mathbf{v}} = \frac{1}{4^{n}} \sum_{\mathbf{a} \in \mathds{Z}_{2}^{2n}} (-1)^{\langle \mathbf{a}, \mathbf{v} \rangle} P_{\mathbf{a}}^{\text{T}}\otimes P_{\mathbf{a}}$~\cite{chen2022quantum}.

When the input state is $|{\Psi_{\text{in}}(\alpha)}\rangle$ in Eq.~\eqref{negligibly entangled input} and the measurement is performed using $\{E_{\mathbf{v}}\}_{\mathbf{v} \in \mathds{Z}_{2}^{2n}}$, the probability $\Pr(\mathbf{v})$ of obtaining outcome $\mathbf{v}$ is related to the Pauli eigenvalue $\lambda(\mathbf{b})$ as follows:
\begin{equation}
    \lambda(\mathbf{b})
    = \sum_{\mathbf{v} \in \mathds{Z}_{2}^{2n}}
    \Pr(\mathbf{v}) \frac{(-1)^{\langle \mathbf{b}, \mathbf{v} \rangle}}{\mathcal{E}(\mathbf{b})}, \quad
    \mathcal{E}(\mathbf{b}) = \alpha + (1-\alpha)3^{-|\mathbf{b}|}.
\end{equation}
Therefore, we obtain an unbiased estimator $\hat{\lambda}(\mathbf{b})$, which is given by $\hat{\lambda}(\mathbf{b}) = \frac{1}{N}\sum_{l=1}^{N} \frac{(-1)^{\langle \mathbf{b}, \mathbf{v}^{(l)} \rangle}}{\mathcal{E}(\mathbf{b})}$, where $\{ \mathbf{v}^{(l)} \}_{l=1}^{N}$ are the outcomes of $N$ measurements.
Applying Hoeffding's inequality, the sample complexity $N(\mathbf{b})$ sufficient to estimate a single parameter $\lambda(\mathbf{b})$ within error $\varepsilon$ with success probability at least $1-\delta$ (see Definition~\ref{def: Pauli channel learning}) is
\begin{equation}
    \label{sample complexity to one parameter}
    N(\mathbf{b})
    = O\left(\frac{1}{\mathcal{E}^{2}(\mathbf{b})} \times \varepsilon^{-2}\log \delta^{-1}\right).
\end{equation}
From the union bound, to ensure this level of precision and confidence for any $\lambda(\mathbf{b})$ among the $4^{n}$ parameters, the sufficient sample complexity is $N = O(n \alpha^{-2} \times \varepsilon^{-2}\log \delta^{-1})$.

Although our main example is the specific state given in Eq.~\eqref{negligibly entangled input}, a broad class of states shares the same property: having an inverse polynomially small entanglement while enabling the learning task to be accomplished with polynomial sample complexity in $n$.
As a generalization of the input state in Eq.~\eqref{negligibly entangled input}, we find that the state $|{\Psi_{\text{in}}( \alpha,  \varphi, \varphi^{*} )}\rangle
:= \sqrt{\alpha'} |\Psi_{\text{B}}\rangle + \sqrt{1-\alpha} | \varphi \rangle \otimes (| \varphi \rangle)^{*}$, where $\alpha = \Theta(1/\text{poly}(n))$ and $| \varphi \rangle$ is an arbitrary state in the $n$-qubit Hilbert space, retains $S_{\text{a}|\text{s}} = \Theta(n\alpha)$ and $\mathcal{E}(\mathbf{b}) = \Omega(\alpha)$ (see SM Sec.~S2 D~\cite{supple}).
In addition, we show that the Werner state~\cite{werner1989quantum} also exhibits this property (see SM Sec.~S2 E~\cite{supple}).
Since it is a mixed state, we use the entanglement of formation as the entanglement measure, instead of the entanglement entropy $S_{\text{a}|\text{s}}$~\cite{bennett1996concentrating, terhal2000entanglement}.
Although our analysis focuses on the Bell-measurement POVM, employing more general POVMs may enable a broader set of input states to satisfy these properties.

\subsection{Learning with a restricted number of ancilla qubits}

In the previous section, we showed that when the system is assisted by the $k=n$ ancilla qubits, with only limited entanglement, the full set of Pauli channel parameters can be learned within polynomial $N$.
In contrast, if $k$ is insufficient, it is known that an exponential $N$ is necessary to accomplish ($\varepsilon, \delta, w=n$)-Pauli channel learning task~\cite{chen2022quantum, chen2025efficient}.
In the following theorem, we show that under the restriction on $k$, learning even a subset of the parameters (i.e., $w<n$) requires an exponentially large $N$.
\begin{theorem}
    \normalfont
    \label{theorem 2}
    To accomplish the ($\varepsilon, \delta, w$)-Pauli channel learning task by using $k$ ancilla qubits, the lower bound on the required sample complexity $N$ is
    \begin{equation}
        \label{lower bound with k and w}
        N =
        \begin{cases}
            \Omega \left( \frac{1}{\sqrt{n}} 2^{-k} 3^{w} \times \varepsilon^{-2}(1-2\delta)  \right) & w \leq n/2 \\ 
            \Omega \left(
                2^{-k} \frac{\sum_{u=0}^{w} \binom{n}{u}3^{u}}{2^{n}}
                \times \varepsilon^{-2}(1-2\delta)  
            \right) & w > n/2
        \end{cases}.
    \end{equation}
\end{theorem}
\begin{figure}[t]
    \centerline{\includegraphics[scale=1.0]{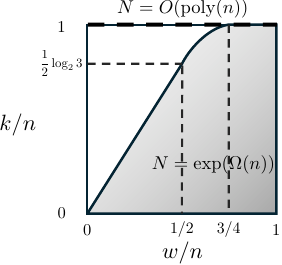}}
    \caption{
        \textbf{Illustration of the regime in which exponential sample complexity arises to accomplish the ($\varepsilon, \delta, w$)-Pauli channel learning task.}
        The regime is denoted as a function of the maximum weight $w$ and the number of ancilla qubits $k$.
        In this figure, we focus on the case $k$ and $w$ scale proportionally with $n$.
        As stated in Theorem~\ref{theorem 2}, the boundary of this regime is linear for $w \leq n/2$, and becomes concave for $w > n/2$.
        According to Eq.~\eqref{lower bound on CN(P(w))}, within our stabilizer-covering scheme, polynomial sample complexity is achievable only when $k=n$.
        This case is indicated by the black dashed line.
    }
    \label{Fig2}
\end{figure}
\noindent
In Fig.~\ref{Fig2}, we illustrate the regime where the exponential $N$ is required in the maximum weight-ancilla qubit number ($w-k$) plane, according to Eq.~\eqref{lower bound with k and w}.
As shown in Fig.~\ref{Fig2}, Theorem~\ref{theorem 2} highlights that a sufficient $k$ is crucial for achieving the exponential advantage.
When $k$ is limited, even for relatively simple tasks with $w<n$, an exponentially large $N$ is required regardless of the entanglement of the input state.
This presents a complementary case to that in Theorem~\ref{theorem 1}, which considers the situation where $k$ is large enough but $S_{\text{a}|\text{s}}$ is limited.
Therefore, we concretely establish that the two scenarios---limited $k$ and limited $S_{\text{a}|\text{s}}$---are essentially different in terms of their impact on the exponential learning advantage.

Furthermore, for the case $w=n$, we improve the lower bound from the previously suggested $N=\Omega(2^{(n-k)/3})$ in~\cite{chen2022quantum, chen2025efficient} to $N=\Omega(2^{n-k})$.
It follows from the fact that when $w=n$, the summation $\sum_{u=0}^{w}\binom{n}{u}3^{u}$ in Eq.~\eqref{lower bound with k and w} becomes $4^{n}$.
In this case, since the upper bound $N = O(n2^{n-k})$ is given in~\cite{chen2022quantum}, our lower bound $N = \Omega(2^{n-k})$ is tight up to the linear factor $n$.

{\it Proof Sketch.} (The detailed proof is provided in SM Sec.~S3~\cite{supple}.)
To establish the lower bound on $N$, we introduce a hypothesis-testing game, as discussed in
\cite{oh2024entanglementenabled,coroi2025exponential,liu2025quantum, chen2024tight,chen2025efficient}.
We consider two types of channels: the completely depolarizing channel $\mathbf{\Lambda}_{\text{dep}}$ and a channel with a single non-trivial eigenvalue, denoted by $\mathbf{\Lambda}_{(\mathbf{e},s)}$~\cite{chen2022quantum}.
These channels are defined as
\begin{equation}
    \mathbf{\Lambda}_{\text{dep}} : \lambda(\mathbf{b}) = \delta_{\mathbf{b}, \mathbf{0}},
    ~~
    \mathbf{\Lambda}_{(\mathbf{e}, s)} : \lambda(\mathbf{b})
    = \delta_{\mathbf{b}, \mathbf{0}} + 2s \varepsilon \delta_{\mathbf{b}, \mathbf{e}},
\end{equation}
where $\mathbf{e} \in \mathds{Z}_{2}^{2n}$ and $s \in \{-1, 1\}$ is a sign.
For the hypothesis-testing game formulation of the particular ($\varepsilon, \delta, w$)-Pauli channel learning task, we introduce the probability distribution
\begin{equation}
    \label{def of Pr(e)}
    \Pr(\mathbf{e}) = \frac{1}{(1+3x)^{n}}x^{|\mathbf{e}|},
\end{equation}
where $0< x \leq 1$ is a tunable parameter.
The hypothesis-testing game is set up as follows: 
(1) According to $\Pr(\mathbf{e})$, a referee samples $\mathbf{e}$ and chooses the sign $s$ uniformly at random.
(2) The referee selects $\mathbf{\Lambda}_{\text{dep}}$ or $\mathbf{\Lambda}_{(\mathbf{e}, s)}$ with equal probability and sends $N$ copies of the chosen channel to the player.
(3) The player collects $N$ measurement outcomes by performing one measurement on each copy.
(4) Finally, the referee reveals the sampled $\mathbf{e}$ and asks the player to determine whether the channel is $\mathbf{\Lambda}_{\text{dep}}$ or not.

If an ($\varepsilon, \delta, w$)-learning scheme exists, then the player can win this game with probability at least $1-\delta$ whenever $|\mathbf{e}| \leq w$ is sampled, by checking whether $|\lambda(\mathbf{b})| < \varepsilon$ for any $\mathbf{b}$ such that $|\mathbf{b}| \leq w$.
Therefore, the player's winning probability $\Pr(\text{win})$ satisfies $\Pr(\text{win}) \geq \Pr( |\mathbf{e}| \leq w)\times(1-\delta) + (1-\Pr( |\mathbf{e}| \leq w))\times\frac{1}{2}$, where the first term covers the case $|\mathbf{e}| \leq w$, and the second term corresponds to the complementary case.
According to Le Cam's two-point method~\cite{lecam1973convergence}, the total variation difference (TVD) provides an upper bound on $\Pr(\text{win})$, where the TVD quantifies the difference between the output distributions of $\mathbf{\Lambda}_{\text{dep}}$ and $\mathbf{\Lambda}_{(\mathbf{e}, s)}$.
The bound is given by $\frac{1}{2}(1+\text{TVD}) \geq \Pr(\text{win})$, and as a result, we have
\begin{equation}
    \label{TVD and delta}
    \text{TVD} \geq \Pr( |\mathbf{e}| \leq w)(1-2\delta),
\end{equation}
where $\Pr( |\mathbf{e}| \leq w) = \frac{1}{(1+3x)^{n}}\sum_{u=0}^{w} \binom{n}{u}(3x)^{u}$.
From Eq.~\eqref{TVD and delta}, the lower bound on $N$ can be derived by finding an upper bound on the TVD as a function of $N$, $k$, $n$, $w$, and $x$. 
Our key improvement in the proof technique is introducing the probability distribution in Eq.~\eqref{def of Pr(e)}.
Since Eq.~\eqref{TVD and delta} holds for any $\Pr(\mathbf{e})$, we can choose the value of $x$ that maximizes the resulting lower bound on $N$.
Using the optimal choice of $x$, we derive Eq.~\eqref{lower bound with k and w}.

We find that the lower bound in Theorem~\ref{theorem 2} is not tight, and resolving a specific technical step would enable its improvement.
In SM Sec. S3 E~\cite{supple}, we present a detailed formulation of the lower bound we expect to be achievable.
If the expectation outlined in SM holds, then the parameter $k$ plays an even more critical role: when $k$ and $w$ scale proportionally with $n$, an exponential number of channel queries is required whenever $k< n$, irrespective of the value of $w$; in other words, the boundary shown in Fig.~\ref{Fig2} converges to $k/n=1$.

To more precisely characterize the role of the number of ancilla qubits in the ($\varepsilon, \delta, w$)-Pauli channel learning, we further investigate an upper bound on the sample complexity $N$.
In particular, through the derivation of the upper bound in Theorem~\ref{theorem 3}, we show that the lower bound in Eq.~\eqref{lower bound with k and w} is tight when $k=0$, up to a small polynomial.
\begin{theorem}
    \normalfont
    \label{theorem 3}
    When $k=0$, the upper bound on $N$ to accomplish the ($\varepsilon, \delta, w$)-Pauli channel learning task is
    \begin{equation}
        \label{upper bound for k=0}
        N =
        \begin{cases}
            O(n^{2} 3^{w} \times \varepsilon^{-2}\log\delta^{-1} ) & w \leq n/2 \\
            O\left(
                n^{5/2} \frac{ \sum_{u=0}^{w} \binom{n}{u} 3^{u} }{ 2^{n} }
                \times \varepsilon^{-2}\log\delta^{-1}
            \right) & w > n/2
        \end{cases}.
    \end{equation}
\end{theorem}
\noindent
When $k=0$, the lower bound in Eq.~\eqref{lower bound with k and w} matches the upper bound in Eq.~\eqref{upper bound for k=0} up to a polynomial factor of~$n$. 
Theorem~\ref{theorem 3} implies that when $w = \Theta(\log(n))$, although the number of parameters to be learned scales quasi-polynomially, the sample complexity grows only polynomially.
Additionally, by combining with the lower bound in Theorem~\ref{theorem 2}, we reveal that the scaling of the sample complexity exhibits a transition at the threshold $w = n/2$.
We prove the theorem by explicitly constructing the learning scheme using the concept of the stabilizer covering.

{\it Proof Sketch.} (Further details can be found in SM Sec.~S4~\cite{supple}.)
To derive the upper bound on the sample complexity, we employ the concept of the stabilizer covering~\cite{flammia2020efficient}.
Given a set of Pauli strings $\mathsf{P}$, a set $\mathsf{C} = \{\mathsf{S}_{i}\}_{i}$ of stabilizer groups $\mathsf{S}_{i}$ is called a \textit{stabilizer covering} of $\mathsf{P}$ if it satisfies $\mathsf{P} \subseteq \bigcup_{\mathsf{S}_{i} \in \mathsf{C}} \mathsf{S}_{i}$.
A stabilizer covering of $\mathsf{P}$ is not unique, and we denote by $\mathrm{CN}(\mathsf{P})$ the minimal size $|\mathsf{C}|$ among all stabilizer coverings of $\mathsf{P}$.

For the task of learning all $\lambda(\mathbf{b})$ such that $P_{\mathbf{b}} \in \mathsf{P}$, the stabilizer covering provides an upper bound on $N$ as~\cite{chen2022quantum}
\begin{equation}
    \label{N for CN}
    N = O\left( n\times \mathrm{CN}(\mathsf{P})\times \varepsilon^{-2} \log \delta^{-1} \right).
\end{equation}
The proof is as follows: given a stabilizer group $\mathsf{S}_{i}$, all $\lambda(\mathbf{b})$ such that $P_{\mathbf{b}} \in \mathsf{S}_{i}$ can be estimated by using only $N=O(n\times \varepsilon^{-2} \log\delta^{-1})$ samples.
Accordingly, for a stabilizer covering $\mathsf{C}$ of $\mathsf{P}$ satisfying $|\mathsf{C}| = \mathrm{CN}(\mathsf{P})$, repeating the above estimation for each $\mathsf{S}_{i} \in \mathsf{C}$ enables us to estimate all $\lambda(\mathbf{b})$ such that $P_{\mathbf{b}} \in \mathsf{P}$, since every element in $\mathsf{P}$ is contained in some $\mathsf{S}_{i} \in \mathsf{C}$.

To obtain the upper bound on $\mathrm{CN}(\mathsf{P})$, we develop the concept of a \textit{uniform stabilizer covering}.
We refer to a stabilizer covering $\mathsf{U} = \{\mathsf{S}_{i}\}_{i}$ of $\mathsf{P}$ as \textit{uniform} if it satisfies the following conditions:
(1) For every $\mathsf{S}_{i} \in \mathsf{U}$, $|\mathsf{S}_{i} \cap \mathsf{P}| = \Sigma$, and we call $\Sigma$ the \textit{covering power}.
(2) For all $P_{\mathbf{a}} \in \mathsf{P}$, $|\{ \mathsf{S}_{i} \in \mathsf{U} : P_{\mathbf{a}} \in \mathsf{S}_{i} \}| = R$, and according to condition~(1), the relation $|\mathsf{U}|\times \Sigma = |\mathsf{P}| \times R$ holds.
By extending the theory of covering arrays
\cite{johnson1974approximation, lovasz1975ratio, stein1974two, sarkar2017upper},
we prove that for a given $\mathsf{P}$, if a uniform stabilizer covering $\mathsf{U}$ with covering power $\Sigma$ exists, an upper bound on $\mathrm{CN}(\mathsf{P})$ is given by
\begin{equation}
    \label{CN for Sigma}
    \mathrm{CN}(\mathsf{P})
    \leq \left\lceil \frac{|\mathsf{P}| \log |\mathsf{P}|}{\Sigma} \right\rceil.
\end{equation}
Furthermore, we provide a heuristic density-based greedy algorithm~\cite{bryce2007density,bryce2009densitybased} to find a stabilizer covering of size specified in Eq.~\eqref{CN for Sigma}.

\begin{figure}[h]
\centerline{\includegraphics[scale=1]{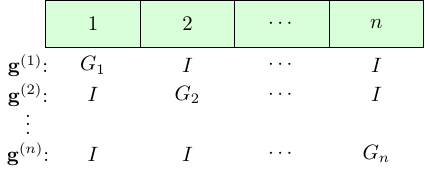}}
\caption{
\textbf{Illustration of the set $\mathsf{G}^{(n)}(\mathbb{G})$.}
Each number in the box labels a qubit. 
$G_{j}$ denotes the $j$-th element of $\mathbb{G}$, and $\mathbf{g}^{(j)}$ is an element of $\mathsf{G}^{(n)}(\mathbb{G})$ that applies $G_{j}$ to qubit $j$ and the identity elsewhere.
}
\label{Fig3}
\end{figure}

\begin{figure}[h]
\begin{center}
\centerline{\includegraphics[scale=1]{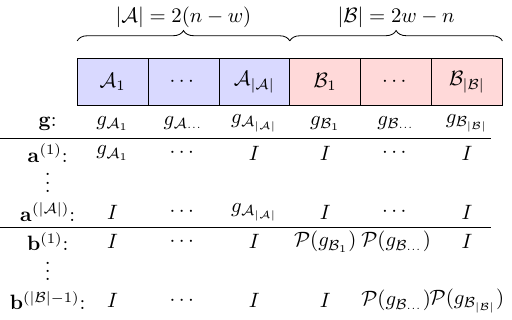}}
\caption{
\textbf{Illustration of $\mathbf{g}$, $\mathsf{G}^{ (\mathrm{A}) }(\mathbf{g}, \mathcal{A})$, and $\mathsf{G}^{ (\mathrm{B}) }(\mathbf{g}, \mathcal{B})$.}
Each boxed number indicates the corresponding qubit index.
Although we draw $\mathcal{A} = \{\mathcal{A}_{j}\}_{j=1}^{2(n-w)}$ and $\mathcal{B} = \{\mathcal{B}_{j}\}_{j=1}^{2w-n}$ as contiguous subsets of qubits for simplicity, all of two subsets are considered.
Each element of $\mathsf{G}^{ (\mathrm{A}) }(\mathbf{g}, \mathcal{A})$ is denoted by $\mathbf{a}^{ (j) }$, whose $\mathcal{A}_{j}$-th component is $g_{ \mathcal{A}_{j} }$, and all other components are $I$.
We denote each element of $\mathsf{G}^{ (\mathrm{B}) }(\mathbf{g}, \mathcal{B})$ by $\mathbf{b}^{ (j) }$, where $b_{\mathcal{B}_{j}}^{ (j) } = \mathcal{P}(g_{ \mathcal{B}_{j} })$, $b_{\mathcal{B}_{j+1}}^{ (j) } = \mathcal{P}(g_{ \mathcal{B}_{j+1} })$, and all other components are $I$.
Here, $\mathcal{P}(X) = Y$, $\mathcal{P}(Y) = Z$, and $\mathcal{P}(Z) = X$. 
}
\label{Fig4}
\end{center}
\end{figure}

For the ($\varepsilon, \delta, w$)-Pauli channel learning, we consider a set $ \mathsf{P}(w) := \{ P_{\mathbf{a}} : |P_{\mathbf{a}}| = w \} $ where $|\mathsf{P}(w)| = \binom{n}{w}3^{w}$.
For the set $\mathsf{P}(w)$, we construct two uniform stabilizer coverings, namely $\mathsf{U}^{ (w \leq n/2) }$ with covering power $\binom{n}{w}$ and $\mathsf{U}^{ (w > n/2) }$ with covering power $\Omega(\frac{1}{\sqrt{n}}2^{n})$, corresponding to the regimes $w \leq n/2$ and $w > n/2$, respectively.
We briefly outline their constructions as follows:
$\mathsf{U}^{ (w\leq n/2) }$ is defined as the collection of $\mathsf{S}^{(n)}(\mathbb{G})$ over all $\mathbb{G}$, where $\mathbb{G}$ is a tuple consisting of $n$ non-identity Pauli operators, with a total of $3^{n}$ such tuples.
Here, $\mathsf{S}^{(n)}(\mathbb{G})$ is a stabilizer group generated by $\mathsf{G}^{(n)}(\mathbb{G})$, where $\mathsf{G}^{(n)}(\mathbb{G})$ is a set that consists of $n$ weight-1 Pauli strings (see Fig.~\ref{Fig3}).
When $w > n/2$, $\mathsf{U}^{ (w > n/2) }$ is defined as the collection of $\mathsf{S}^{ (\mathrm{A,B}) }( \mathbf{g}, (\mathcal{A}, \mathcal{B}) )$ over all $\mathbf{g}$ and $(\mathcal{A}, \mathcal{B})$, where $\mathbf{g}$ is a weight-$n$ Pauli string, and $(\mathcal{A}, \mathcal{B})$ is a partition of the $n$ system qubits such that $|\mathcal{A}| = 2(n-w)$ and $|\mathcal{B}| = 2w-n$.
Here, $\mathsf{S}^{ (\mathrm{A,B}) }( \mathbf{g}, (\mathcal{A}, \mathcal{B}) )$ is a stabilizer group generated by the union of $\{ \mathbf{g} \}$, $\mathsf{G}^{ (\mathrm{A}) }(\mathbf{g}, \mathcal{A})$, and $\mathsf{G}^{ (\mathrm{B}) }(\mathbf{g}, \mathcal{B})$, where $\mathsf{G}^{ (\mathrm{A}) }(\mathbf{g}, \mathcal{A})$ is a set of $|\mathcal{A}|$ weight-1 Pauli strings and $\mathsf{G}^{ (\mathrm{B}) }(\mathbf{g}, \mathcal{B})$ is a set of $|\mathcal{B}|-1$ weight-2 Pauli strings (see Fig.~\ref{Fig4}).

Combining Eq.~\eqref{CN for Sigma} with the covering powers of $\mathsf{U}^{ (w \leq n/2) }$ and $\mathsf{U}^{ (w > n/2) }$, we derive
\begin{equation}
\begin{gathered}
    \label{bound on CN}
    \mathrm{CN}(\mathsf{P}(w))
    =
    \begin{cases}
        O( n 3^{w} ) & w \leq n/2 \\
        O\left( n^{3/2} \frac{\binom{n}{w}3^{w}}{2^{n}} \right) & w > n/2
    \end{cases},
\end{gathered}
\end{equation}
where we used $\log |\mathsf{P}(w)| = O(n)$.
Finally, by using Eqs.~\eqref{N for CN} and~\eqref{bound on CN}, along with the inequality $\mathrm{CN}(\bigcup_{u=0}^{w} \mathsf{P}(u)) \leq  \sum_{u=0}^{w} \mathrm{CN}(\mathsf{P}(u))$, we derive the upper bound on $N$ stated in Theorem~\ref{theorem 3}.

In addition, from Eqs.~\eqref{lower bound with k and w} and~\eqref{N for CN}, we conclude that our bound on $\text{CN}(\mathsf{P}(w))$ in Eq.~\eqref{bound on CN} is tight within a polynomial factor of $n$.
Although finding the exact value of $\mathrm{CN}(\mathsf{P})$ for an arbitrary set $\mathsf{P}$ is NP-hard~\cite{verteletskyi2020measurement}, by exploiting the specific structure of weight-$w$ Pauli strings, we derive this tight bound.

Additionally, we derive an upper bound on $N$ for the case $k>0$ by generalizing the strategy developed for the case $k=0$.
In particular, by extending the concept of the uniform stabilizer covering, we show that Eqs.~\eqref{N for CN} and~\eqref{CN for Sigma} also hold when the $k$-qubit ancilla is used (see SM Sec.~S5 A~\cite{supple}).
Accordingly, we construct a uniform stabilizer covering of $\mathsf{P}(w)$ and compute the corresponding covering power.
Specifically, we find two uniform stabilizer coverings $\mathsf{U}^{ (2w \leq k+n) }$ and $\mathsf{U}^{ (2w > k+n) }$, for the regimes $2w \leq k+n$ and $2w > k+n$, respectively.
Their construction leverages the $k$ ancilla qubits by forming a $2k$-qubit Bell pair with $k$ system qubits.
Then, for the remaining $n-k$ system qubits, each stabilizer group in $\mathsf{U}^{ (2w \leq k+n) }$ is designed following the same strategy as in $\mathsf{U}^{ (w \leq n/2) }$, and those in $\mathsf{U}^{ (2w > k+n) }$ are built analogously to $\mathsf{U}^{ (w > n/2) }$.
Although the resulting upper bound does not match the lower bound in Theorem~\ref{theorem 2}, we present explicit forms of $\mathsf{U}^{ (2w \leq k+n) }$ and $\mathsf{U}^{ (2w > k+n) }$, together with their covering powers (see SM Sec.~S5 B~\cite{supple}).
According to our construction, when $k$ and $w$ scale proportionally with $n$, the ($\varepsilon, \delta, w$)-Pauli channel learning task can be accomplished with $N = O(\text{poly}(n))$ only for the case $k=n$.
In detail, we find that 
\begin{equation}
    \label{lower bound on CN(P(w))}
    \left\lceil \frac{|\mathsf{P}(w)| \log |\mathsf{P}(w)|}{\Sigma} \right\rceil
    = \Omega(n^{3/2} 3^{w\left(1-\frac{k}{n}\right)}),
\end{equation}
where the left-hand side is related with Eq.~\eqref{CN for Sigma} (see SM Sec.~S5 C~\cite{supple}).
This result is illustrated in Fig.~\ref{Fig2}.

\section{Discussion}

We establish that the two fundamental quantum resources, namely the entanglement in the input state and the number of ancilla qubits, have different contributions to the exponential learning advantage.
Specifically, we prove that the exponential advantage in Pauli channel learning can be achieved using input states with only inverse-polynomially small entanglement.
In contrast, if the number of ancilla qubits is insufficient, even for the easier task of learning a subset of channel parameters, an exponential sample complexity is required.
Our results are expected to be useful for quantum channel learning under resource constraints in the NISQ era, such as limited entanglement (e.g., noisy Bell states) or a restricted number of ancilla qubits.

We expect that our result---exponential advantage using only slightly entangled input states---can be extended to a wide range of quantum systems.
Potential extensions include learning more general quantum channels beyond the Pauli channel, such as qudit systems and continuous variable systems.
Rather than channel learning, applying a similar approach to quantum state learning presents an interesting direction.

Understanding whether concatenated applications of the channel can further reduce the input state entanglement is an intriguing topic for future investigation.
Recent studies have shown that such concatenation can reduce the number of measurements; however, the required number of channel applications remains exponential when the number of ancilla qubits is insufficient~\cite{chen2024tight, chen2025efficient}.
Despite these findings, its effect on the input state entanglement has not been investigated.

Determining a tight bound on the sample complexity in the presence of ancilla qubits remains an open problem.
Even though addressing a technical issue could lead to a tighter lower bound, as described in SM Sec.~S3 E~\cite{supple}, the resulting bound still does not reach the upper bound.
Hence, we expect that further advances in the proof techniques for both the upper and lower bounds may ultimately close the remaining gap.

We note that the exponential advantages established in Theorems~\ref{theorem 1} and~\ref{theorem 3} can be destroyed by state-preparation and measurement (SPAM) noise.
More specifically, under depolarizing noise, the exponential advantage becomes fragile even with access to $n$-ancilla qubits, as demonstrated in~\cite{chen2024tight}.
Future research may focus on developing a SPAM-robust learning algorithm and analyzing the resulting sample complexity scaling under limitations on entanglement or the number of ancilla qubits.

Another intriguing future direction is to integrate our framework with SPAM-robust randomized benchmarking (RB) techniques.
In particular, Ref.~\cite{chen2022quantum} proposes an RB method that utilizes an $n$-qubit ancilla for sample-efficient benchmarking.
This scheme encompasses the case where the input state is a noisy Bell state, and our input state in Eq.~\eqref{negligibly entangled input} also falls within this framework.
In addition, since our algorithm associated with Theorem~\ref{theorem 3} is based on stabilizer covering, the RB method introduced in Ref.~\cite{flammia2020efficient} can naturally be extended to our setting.
We leave a detailed analysis of how the efficiency of the RB scheme depends on SPAM noise for future work.

\section*{Data availability}
No data has been generated in this work.

\bibliography{On_the_Fundamental_Resource_for_Exponential_Advantage_in_Quantum_Channel_Learning}

\acknowledgments
M.K. and C.O. were supported by the National
Research Foundation of Korea Grants (No. RS-2024-00431768 and No. RS-2025-00515456) funded by the Korean government (Ministry of Science and ICT~(MSIT)) and the Institute of Information \& Communications Technology Planning \& Evaluation (IITP) Grants funded by the Korea government (MSIT) (No. IITP-2025-RS-2025-02283189 and IITP-2025-RS-2025-02263264).
This work was supported by Global Partnership Program of Leading Universities in Quantum Science and Technology (RS-2025-08542968) through the National Research Foundation of Korea(NRF) funded by the Korean government (Ministry of Science and ICT(MSIT)).

\section*{Author contributions}
M.K. carried out the detailed calculations.
C.O. conceptualized the idea and supervised the overall project.
All authors contributed to writing the manuscript.

\section*{Competing interests}
The authors declare no competing interests.

\end{document}






\title{On the Fundamental Resource for Exponential Advantage in Quantum Channel Learning: Supplemental Material}

\author{Minsoo Kim}
\affiliation{Department of Physics, Korea Advanced Institute of Science and Technology, Daejeon 34141, Korea}

\author{Changhun Oh}\email[]{changhun0218@gmail.com}
\affiliation{Department of Physics, Korea Advanced Institute of Science and Technology, Daejeon 34141, Korea}

\maketitle

\newcommand{\beginsup}{
        \setcounter{equation}{0}
        \renewcommand{\theequation}{S\arabic{equation}}
        \setcounter{table}{0}
        \renewcommand{\thetable}{S\arabic{table}}
        \setcounter{figure}{0}
        \renewcommand{\thefigure}{S\arabic{figure}}
        \setcounter{definition}{0}
        \renewcommand{\thedefinition}{S\arabic{definition}}
        \setcounter{theorem}{0}
        \renewcommand{\thetheorem}{S\arabic{theorem}}
        \setcounter{lemma}{0}
        \renewcommand{\thelemma}{S\arabic{lemma}}
     }
\beginsup

\tableofcontents

\section{Preliminary}

In this section, we provide definitions and detailed descriptions of Pauli strings and Pauli channels.
In addition, we introduce the Bell basis, which gives a convenient representation for analyzing Pauli strings and Pauli channels.

\subsection{Pauli string}
\label{Sec: S1 A}

To establish the definition of Pauli strings, we follow a notation similar to that used in~\cite{chen2022quantum}. 
A Pauli operator $P_{a} \in \{I, X, Y, Z\}$ acting on a single-qubit Hilbert space is represented by a 2-bit string $a = a_{x}a_{z} \in \mathds{Z}_{2}^{2}$, where $P_{a} = i^{a_{x}a_{z}} X^{a_{x}} Z^{a_{z}}$.
Since $XZ = -iY$, the phase factor $i^{a_{x}a_{z}}$ is included to ensure hermiticity.
This definition can be extended to an $n$-qubit Hilbert space.
A Pauli string $P_{\mathbf{a}}$ corresponding to a $2n$-bit string is defined as
\begin{equation}
    \label{def of Pauli string}
    P_{\mathbf{a}} := \bigotimes_{j=1}^{n}i^{a_{j,x}a_{j,z}}X^{a_{j,x}}Z^{a_{j,z}}, \quad
	\mathbf{a} := a_{1} a_{2} \cdots a_{n} = a_{1,x}a_{1,z}a_{2,x}a_{2,z}\cdots a_{n,x}a_{n,z} \in \mathds{Z}_{2}^{2n}.
\end{equation}
Consequently, all Pauli strings are Hermitian and unitary $2^{n} \times 2^{n}$ matrices, and satisfy the relation $P_{\mathbf{a}}^{2} = P_{\mathbf{a}}P_{\mathbf{a}}^{\dagger} = P_{\mathbf{0}} = I^{\otimes n}$.
Throughout this material, bold symbols $\mathbf{a}$ denote $2n$-bit strings, whereas regular symbols $a$ denote 2-bit strings.
Also, we denote a Pauli string $P_{\mathbf{a}}$ simply by its corresponding $2n$-bit string $\mathbf{a}$, as defined in Eq.~\eqref{def of Pauli string}, and use $\mathbf{a}$ and $P_{\mathbf{a}}$ interchangeably whenever there is no ambiguity.
The Pauli weight $|\mathbf{a}|$ of a Pauli string $P_{\mathbf{a}}$ is defined as the number of non-identity Pauli matrices ($X$, $Y$, and $Z$) present in the string.

The trace of the product of two Pauli strings is $\Tr(P_{\mathbf{a}}P_{\mathbf{b}}) = 2^{n} \delta_{\mathbf{a}, \mathbf{b}}$.
Since the Pauli strings form a complete and orthonormal basis in the space of the Hermitian operators on an $n$-qubit Hilbert space, any Hermitian operator $O$ can be expressed as
\begin{equation}
    \label{Pauli string representation of Hermitian operator}
    O = \sum_{\mathbf{a}} o_{\mathbf{a}} P_{\mathbf{a}}, \quad
    o_{\mathbf{a}} = \frac{1}{2^{n}} \Tr(O P_{\mathbf{a}}).
\end{equation}

Any two Pauli strings either commute or anticommute. 
The commutation relation between two Pauli strings $P_{\mathbf{a}}$ and $P_{\mathbf{b}}$ is determined by their symplectic inner product $\langle \mathbf{a}, \mathbf{b}\rangle$, defined as~\cite{nielsen2010quantum}
\begin{equation}
    \label{commuation relation}
    \langle \mathbf{a}, \mathbf{b}\rangle
    := \sum_{j=1}^{n}(a_{j,x}b_{j,z} + a_{j,z}b_{j,x}) \mod 2, ~~
    P_{\mathbf{a}}P_{\mathbf{b}}
    = (-1)^{\langle \mathbf{a}, \mathbf{b}\rangle} P_{\mathbf{b}}P_{\mathbf{a}}.
\end{equation}
Using the symplectic inner product, we define the Walsh-Hadamard transform $\mathcal{F}(\mathbf{b})$ of a function $f(\mathbf{a})$, along with its inverse transform:
\begin{equation}
    \label{Walsh-Hadamard transform of bit string}
    \mathcal{F}(\mathbf{b}) := \sum_{\mathbf{a} \in \mathds{Z}_{2}^{2n}} (-1)^{\langle \mathbf{a}, \mathbf{b}\rangle} f(\mathbf{a})
    ,\quad 
    f(\mathbf{a})
    = \frac{1}{4^{n}} \sum_{\mathbf{b} \in \mathds{Z}_{2}^{2n}} (-1)^{\langle \mathbf{a}, \mathbf{b}\rangle} \mathcal{F}(\mathbf{b}).
\end{equation}
The inverse transformation can be proven by using the identity $\sum_{\mathbf{b} \in \mathds{Z}_{2}^{2n}} (-1)^{\langle \mathbf{a}, \mathbf{b}\rangle} = 4^{n}\delta_{\mathbf{a}, \mathbf{0}}$.

\subsection{Pauli channel}

A Pauli channel $\mathbf{\Lambda}(\cdot)$ acting on an $n$-qubit system is defined by its Pauli error rates $\{ p({\mathbf{a}}) \}_{\mathbf{a} \in \mathds{Z}_{2}^{2n}}$ as
\begin{equation}
    \mathbf{\Lambda}(\cdot)
    := \sum_{\mathbf{a} \in \mathds{Z}_{2}^{2n}} p({\mathbf{a}}) P_{\mathbf{a}}(\cdot)P_{\mathbf{a}}.
\end{equation}
This channel is completely positive and trace-preserving when the conditions $p_{\mathbf{a}} \geq 0$ for all $\mathbf{a}$ and $\sum_{\mathbf{a}} p_{\mathbf{a}}=1$ are satisfied.
Since the input state $\rho_{\text{in}}$ to the Pauli channel is Hermitian, the resulting output state $\mathbf{\Lambda}(\rho_{\text{in}})$ is also Hermitian.
Thus, using Eq.~\eqref{Pauli string representation of Hermitian operator}, the Pauli channel can be represented as
\begin{equation}
\begin{aligned}
    \mathbf{\Lambda}(\rho_{\text{in}})
    &= \sum_{\mathbf{a}} p({\mathbf{a}})
    P_{\mathbf{a}}\rho_{\text{in}}P_{\mathbf{a}} \\
    &= \frac{1}{2^{n}} \sum_{\mathbf{b}} \sum_{\mathbf{a}} p({\mathbf{a}})
    \Tr( P_{\mathbf{a}} P_{\mathbf{b}} P_{\mathbf{a}} \rho_{\text{in}} ) P_{\mathbf{b}}
    \quad (\because \text{by Eq.}~\eqref{Pauli string representation of Hermitian operator})
    \\
    &= \frac{1}{2^{n}} \sum_{\mathbf{b}}
    \underbrace{
    \sum_{\mathbf{a}} p({\mathbf{a}})
    (-1)^{\langle \mathbf{a}, \mathbf{b}\rangle} 
    }_{\lambda(\mathbf{b})}
    \Tr( P_{\mathbf{b}} \rho_{\text{in}} ) P_{\mathbf{b}}
    \quad (\because \text{by Eq.}~\eqref{commuation relation}),
\end{aligned}
\end{equation}
where the set $\{ \lambda(\mathbf{b}) \}_{\mathbf{b} \in \mathds{Z}_{2}^{2n}}$ is referred to as Pauli eigenvalues~\cite{flammia2020efficient, flammia2021pauli}.
By using the Pauli eigenvalues, an alternative definition of the Pauli channel is
\begin{equation}
    \mathbf{\Lambda}(\cdot)
    = \frac{1}{2^{n}}\sum_{\mathbf{b} \in \mathds{Z}_{2}^{2n}}
    \lambda(\mathbf{b}) \Tr[P_{\mathbf{b}}(\cdot)]P_{\mathbf{b}},
\end{equation}
where $\lambda(\mathbf{b})$ is related to $p(\mathbf{a})$ as
\begin{equation}
    \lambda(\mathbf{b})
    = \sum_{\mathbf{a} \in \mathds{Z}_{2}^{2n}} p({\mathbf{a}})
    (-1)^{\langle \mathbf{a}, \mathbf{b}\rangle}, \quad
    p({\mathbf{a}})
    = \frac{1}{4^{n}} \sum_{\mathbf{b} \in \mathds{Z}_{2}^{2n}}
    \lambda(\mathbf{b}) (-1)^{\langle \mathbf{a}, \mathbf{b}\rangle}.
\end{equation}
These are precisely the Walsh-Hadamard transform and its inverse, as given in Eq.~\eqref{Walsh-Hadamard transform of bit string}.

When quantum memory is provided through ancilla qubits and a quantum channel $\mathbf{\Lambda}$ is applied exclusively to the system, the channel operation on the joint Hilbert space is described by
\begin{equation}
    \label{definition of channel with ancilla}
    (\mathds{1}_{\text{anc}}\otimes \mathbf{\Lambda}) (\rho_{\text{in}})
    = \frac{1}{2^{n}} \sum_{\mathbf{b}} \lambda(\mathbf{b})
    \Tr_{\text{sys}}((I_{\text{anc}}\otimes P_{\mathbf{b}}) \rho_{\text{in}})
    \otimes P_{\mathbf{b}},
\end{equation}
where $\mathds{1}_{\text{anc}}$ denotes the identity channel acting on the ancilla, $\Tr_{\text{sys}}$ indicates the partial trace over the system, and $I_{\text{anc}}$ is the identity operator on the ancilla.
If the ancilla consists of $k$ qubits, then $I_{\text{anc}} = I^{\otimes k}$.
Similarly, we define the identity operator on the $n$-qubit system, $I_{\text{sys}} = I^{\otimes n}$.



\subsection{Bell basis}

We introduce the Bell basis~\cite{nielsen2010quantum}, which consists of four mutually orthogonal states spanning the 2-qubit Hilbert space:
\begin{equation}
    \label{single qubit Bell basis}
\begin{aligned}
    |{\Psi_{I}}\rangle
    &= \frac{1}{\sqrt{2}} ( |{0}\rangle_{\text{anc}}\otimes |{0}\rangle_{\text{sys}}
    + |{1}\rangle_{\text{anc}}\otimes |{1}\rangle_{\text{sys}} ) \\
	|{\Psi_{X}}\rangle
    &= \frac{1}{\sqrt{2}} ( |{0}\rangle_{\text{anc}}\otimes |{1}\rangle_{\text{sys}}
    + |{1}\rangle_{\text{anc}}\otimes |{0}\rangle_{\text{sys}} ) \\
	|{\Psi_{Y}}\rangle
    &= \frac{i}{\sqrt{2}} ( |{0}\rangle_{\text{anc}}\otimes |{1}\rangle_{\text{sys}}
    - |{1}\rangle_{\text{anc}}\otimes |{0}\rangle_{\text{sys}} ) \\
	|{\Psi_{Z}}\rangle
    &= \frac{1}{\sqrt{2}} ( |{0}\rangle_{\text{anc}}\otimes |{0}\rangle_{\text{sys}}
    - |{1}\rangle_{\text{anc}}\otimes |{1}\rangle_{\text{sys}} ),
\end{aligned}
\end{equation}
where $\{|0\rangle _{\text{anc}},|1\rangle _{\text{anc}}\}$ and $\{|0\rangle _{\text{sys}},|1\rangle _{\text{sys}}\}$ denote the computational basis of the ancilla and system, respectively.
These states correspond to the normalized vectorization of Pauli matrices, which are described as
\begin{equation}
    |{\Psi_{a}}\rangle = \frac{1}{\sqrt{2}} (I \otimes P_{a}) |{\Omega}\rangle,\
    ~\text{where}~ |{\Omega}\rangle := |{00}\rangle + |{11}\rangle.
\end{equation}
It can be generalized to the $n$-qubit case as
\begin{equation}
    \label{n qubit Bell basis}
    |{\Psi_{\mathbf{a}}}\rangle
    = \bigotimes_{j=1}^{n}|{\Psi_{a_{j}}}\rangle,
\end{equation}
where $|{\Psi_{a_{j}}} \rangle$ denotes the Bell basis state between the $j$-th ancilla qubit and the $j$-th system qubit.
The set $\{|{\Psi_{\mathbf{a}}}\rangle\}_{\mathbf{a}\in \mathds{Z}_{2}^{2n}}$ forms a complete orthonormal basis for the combined Hilbert space of an $n$-qubit ancilla and an $n$-qubit system.
In particular, the choice $\mathbf{a} = \mathbf{0}$ corresponds to the $2n$-qubit Bell pair, defined as $|{\Psi_{\text{B}}}\rangle := |{\Psi_{\mathbf{0}}}\rangle$.
The state $|{\Psi_{\text{B}}}\rangle$ was employed for Pauli channel learning in~\cite{chen2022quantum}.

From this definition, each state $|{\Psi_{\mathbf{a}}}\rangle$ is an eigenstate of every operator $P_{\mathbf{b}}^{\text{T}}\otimes P_{\mathbf{b}}$, satisfying
\begin{equation}
    P_{\mathbf{b}}^{\text{T}}\otimes P_{\mathbf{b}} |{\Psi_{\mathbf{a}}}\rangle
    = (-1)^{\langle \mathbf{a}, \mathbf{b}\rangle} |{\Psi_{\mathbf{a}}}\rangle.
\end{equation}
Therefore, the spectral decomposition of the operator $P_{\mathbf{b}}^{\text{T}}\otimes P_{\mathbf{b}}$ can be succinctly expressed in the Bell basis as follows:
\begin{equation}
    \label{density matrix of Bell basis state}
    P_{\mathbf{b}}^{\text{T}}\otimes P_{\mathbf{b}}
    = \sum_{ \mathbf{a} \in \mathds{Z}_{2}^{2n} } (-1)^{\langle \mathbf{a}, \mathbf{b}\rangle} 
    |{\Psi_{\mathbf{a}}}\rangle \langle{\Psi_{\mathbf{a}}}| , \quad
    |{\Psi_{\mathbf{a}}}\rangle \langle{\Psi_{\mathbf{a}}}|
    = \frac{1}{4^{n}} \sum_{ \mathbf{b} \in \mathds{Z}_{2}^{2n} } (-1)^{\langle \mathbf{a}, \mathbf{b}\rangle} 
    P_{\mathbf{b}}^{\text{T}}\otimes P_{\mathbf{b}}
\end{equation}
where the last equality is a result of the Walsh-Hadamard transform given in Eq.~\eqref{Walsh-Hadamard transform of bit string}.

\section{Proof of Theorem 1}

In this section, we provide a detailed proof of Theorem 1 in the main text. 
We explicitly present the form of the input state $|\Psi_{\text{in}}(\alpha)\rangle$ (Eq.~(2) in the main text) and the Bell measurement POVM $\{ E_{\mathbf{v}} \}_{\mathbf{v} \in \mathds{Z}_{2}^{2n}}$.
From the reduced density matrix of the input state, we compute the entanglement entropy $S_{\text{a}|\text{s}}$ between the ancilla and the system.
Using the input state and the POVM, we construct an estimator $\hat{\lambda}(\mathbf{b})$ for the Pauli eigenvalues $\lambda(\mathbf{b})$, and by applying Hoeffding's inequality, we determine the required sample complexity $N$ to accomplish the learning task.
Finally, by combining the derived values of $S_{\text{a}|\text{s}}$ and $N$, we complete the proof of Theorem 1.

As discussed in the main text, there exists a broad class of states exhibiting the properties described in Theorem 1. 
Since we provide a constructive method to compute the entanglement and the sample complexity in this section, it is possible to find alternative states tailored to specific purposes.

\subsection{Input state with inverse-polynomially small entanglement}

Employing the Bell basis introduced in Eqs.~\eqref{single qubit Bell basis} and~\eqref{n qubit Bell basis}, we express our input state as 
\begin{equation}
    \label{input state}
    |{\Psi_{\text{in}}(\alpha) }\rangle := \sum_{ \mathbf{a} \in \mathds{Z}_{2}^{2n} }
    c_{\mathbf{a}}(\alpha) |{\Psi_{\mathbf{a}}}\rangle, \quad
    |c_{\mathbf{a}}(\alpha)|^{2}
    := \alpha \delta_{\mathbf{a}, \mathbf{0}}
    + (1-\alpha) \left(\frac{1}{2}\right)^{n-|\mathbf{a}|} \left(\frac{1}{6}\right)^{|\mathbf{a}|},
\end{equation}
where $0\leq \alpha \leq 1$ is a constant we can choose arbitrarily.
In this section, we compute the entanglement $S_{\text{a}|\text{s}}$ of the input state defined in Eq.~\eqref{input state} as a function of $\alpha$ and $n$.
As shown in Eq.~\eqref{input state}, our input state consists of two terms: one corresponding to $\alpha=1$ (the first term), and the other to $\alpha=0$ (the second term).
We first describe the physical properties of each term, providing intuitive motivation for constructing Eq.~\eqref{input state}.
We further show that Eq.~\eqref{input state} is equivalent to the simplified form given in Eq.~(2) in the main text.
Then, by combining these two terms, we calculate exactly the entanglement $S_{\text{a}|\text{s}}$.

When $\alpha = 1$, the input state in Eq.~\eqref{input state} is reduced to the $2n$-qubit Bell pair, i.e., $|{\Psi_{\text{in}}(\alpha=1)}\rangle = |{\Psi}_{\text{B}}\rangle$. 
From Eq.~\eqref{density matrix of Bell basis state}, the density matrix of the $2n$-qubit Bell pair is
\begin{equation}
    \label{rho Bell}
    |{\Psi_{\text{B}}}\rangle \langle{\Psi_{\text{B}}}|
    = \frac{1}{4^{n}} \sum_{\mathbf{a} \in \mathds{Z}_{2}^{2n}} P_{\mathbf{a}}^{\text{T}} \otimes P_{\mathbf{a}}.
\end{equation}
As previously mentioned, $|{\Psi_{\text{B}}}\rangle$ is used as an input in~\cite{chen2022quantum}, with the entanglement given by $S_{\text{a}|\text{s}} = n$ (we compute the entanglement using base-2 logarithms), corresponding to a maximally entangled state.
By using $|{\Psi_{\text{B}}}\rangle$ as input, the Pauli learning task can be completed within $O(n)$ sample complexity~\cite{chen2022quantum}.

The case $\alpha = 0$ corresponds to $|{\Psi_{\text{in}}(\alpha=0)}\rangle = |{\Psi_{\text{sep}}}\rangle$, which is defined as
\begin{equation}
    \label{psi sep}
\begin{aligned}
    |{\Psi_{\text{sep}}}\rangle
    &:= \sum_{ \mathbf{a} \in \mathds{Z}_{2}^{2n} }
    \left(\frac{1}{\sqrt{2}}\right)^{n-|\mathbf{a}|} \left(\frac{1}{\sqrt{6}}\right)^{|\mathbf{a}|}
    | {\Psi_{\mathbf{a}}} \rangle \\ 
    &= \sum_{a_{1} \in \mathds{Z}_{2}^{2}} \cdots \sum_{a_{n} \in \mathds{Z}_{2}^{2}}
    \left(\frac{1}{\sqrt{2}}\right)^{n-|\mathbf{a}|} \left(\frac{1}{\sqrt{6}}\right)^{|\mathbf{a}|}
    \bigotimes_{j=1}^{n}|{\Psi_{a_{j}}}\rangle \\ 
    &= \bigotimes_{j=1}^{n}
    \left[
        \sum_{a_{j} \in \mathds{Z}_{2}^{2}}
        \left(\frac{1}{\sqrt{2}}\right)^{1-|a_{j}|} \left(\frac{1}{\sqrt{6}}\right)^{|a_{j}|}
        |{\Psi_{a_{j}}}\rangle
    \right] \\ 
    &= \bigotimes_{j=1}^{n}
    \left[
        \frac{1}{\sqrt{2}} |{ \Psi_{ I_{j} } }\rangle
        + \frac{1}{\sqrt{6}} \left( |{ \Psi_{ X_{j} } }\rangle + |{ \Psi_{ Y_{j} } }\rangle + |{ \Psi_{ Z_{j} } }\rangle \right)
    \right] \\
    &= \underbrace{ \left[\frac{(1+i)|{0}\rangle + (\sqrt{3}-1)|{1}\rangle}{\sqrt{6-2\sqrt{3}}}\right]^{\otimes n} }_{\text{ancilla}}
    \otimes 
    \underbrace{ \left[\frac{(1-i)|{0}\rangle + (\sqrt{3}-1)|{1}\rangle}{\sqrt{6-2\sqrt{3}}}\right]^{\otimes n} }_{\text{system}}.
\end{aligned}
\end{equation}
Note that $|{\Psi_{\text{sep}}}\rangle$ is separable, and its density matrix representation can be written as
\begin{equation}
    \label{rho sep}
    |{\Psi_{\text{sep}}}\rangle \langle{\Psi_{\text{sep}}}|
    = \underbrace{\left[\frac{1}{2}\left(I + \frac{1}{\sqrt{3}}(X+Y+Z)\right)^{\text{T}}\right]^{\otimes n}}_{\text{ancilla}}
    \otimes \underbrace{\left[\frac{1}{2}\left(I + \frac{1}{\sqrt{3}}(X+Y+Z)\right)\right]^{\otimes n}}_{\text{system}},
\end{equation}
which corresponds to Eq.~(3) in the main text.
The expression in Eq.~\eqref{rho sep} can also be obtained directly from the definition of the Bell basis, as these basis states represent the vectorized Pauli strings.
By using the definitions of the two states $|{\Psi_{\text{B}}}\rangle$ and $|{\Psi_{\text{sep}}}\rangle$, the input state is expressed as
\begin{equation}
    \label{alpha prime in input}
    |{\Psi_{\text{in}}(\alpha) }\rangle
    = \left( \sqrt{\alpha + \frac{1-\alpha}{2^{n}}} - \sqrt{\frac{1-\alpha}{2^{n}}} \right) |{\Psi_{\text{B}}}\rangle
    + \sqrt{1-\alpha} |{\Psi_{\text{sep}}}\rangle,
\end{equation}
which is equivalent to Eq.~(2) in the main text.
The normalization condition $\langle \Psi_{\text{in}}(\alpha) | \Psi_{\text{in}}(\alpha) \rangle = 1$ follows from the fact that $\langle \Psi_{\text{B}} | \Psi_{\text{sep}} \rangle = \frac{1}{\sqrt{2^{n}}}$. 

To compute the entanglement $S_{\text{a}|\text{s}}$, we first obtain the reduced density matrix $ \Tr_{\text{sys}} ( |{\Psi_{\text{in}}(\alpha) }\rangle \langle{\Psi_{\text{in}}(\alpha)}| ) $.
In the Bell basis, the partial trace over the system can be straightforwardly evaluated as
\begin{equation}
    \Tr_{\text{sys}} ( |{\Psi_{a}}\rangle \langle{\Psi_{a'}}| ) = \frac{1}{2} P_{a}P_{a'}, \quad 
    \Tr_{\text{sys}} ( |{\Psi_{\mathbf{a}}}\rangle \langle{\Psi_{\mathbf{a}'}}| )
    = \frac{1}{2^{n}} P_{\mathbf{a}}P_{\mathbf{a}'}.
\end{equation}
Thus, for the state $|{\Psi_{\text{in}}(\alpha) }\rangle$ given in Eq.~\eqref{input state}, the reduced density matrix is
\begin{equation}
    \label{RDM}
\begin{aligned}
    \Tr_{\text{sys}} ( |{\Psi_{\text{in}}(\alpha) }\rangle \langle{\Psi_{\text{in}}(\alpha)}|
    &= \sum_{\mathbf{a}, \mathbf{a}'} c_{\mathbf{a}}(\alpha)c_{\mathbf{a}'}^{*}(\alpha)
    \Tr_{\text{sys}} ( |{\Psi_{\mathbf{a}}}\rangle \langle{\Psi_{\mathbf{a}'}}| ) \\
    &= \frac{1}{2^{n}} \sum_{\mathbf{a}, \mathbf{a}'} c_{\mathbf{a}}(\alpha)c_{\mathbf{a}'}^{*}(\alpha) P_{\mathbf{a}}P_{\mathbf{a}'} \\
    &= \frac{1}{2^{n}} \left( \sum_{\mathbf{a}} c_{\mathbf{a}}(\alpha) P_{\mathbf{a}}\right)
    \left( \sum_{\mathbf{a}'} c_{\mathbf{a}'}(\alpha) P_{\mathbf{a}'} \right)^{\dagger}.
\end{aligned}
\end{equation}
We compute the following quantity to evaluate Eq.~\eqref{RDM}.
\begin{equation}
\begin{aligned}
    \sum_{\mathbf{a}} c_{\mathbf{a}}(\alpha) P_{\mathbf{a}}
    &= c_{\mathbf{0}}(\alpha) I^{\otimes n} + \sum_{\mathbf{a}\neq \mathbf{0}} c_{\mathbf{a}}(\alpha) P_{\mathbf{a}} \\ 
    &= \sqrt{\alpha + \frac{1-\alpha}{2^{n}}} I^{\otimes n}
    + \sqrt{1-\alpha} \sum_{\mathbf{a}} \left(\frac{1}{\sqrt{2}}\right)^{n-|\mathbf{a}|}
    \left(\frac{1}{\sqrt{6}}\right)^{|\mathbf{a}|} P_{\mathbf{a}} - \sqrt{\frac{1-\alpha}{2^{n}}} I^{\otimes n} \\ 
    &= \left( \sqrt{\alpha + \frac{1-\alpha}{2^{n}}} - \sqrt{\frac{1-\alpha}{2^{n}}} \right) I^{\otimes n}
    + \sqrt{1-\alpha} \left[ \frac{1}{\sqrt{2}}I + \frac{1}{\sqrt{6}}(X+Y+Z) \right]^{\otimes n}.
\end{aligned}
\end{equation}
As shown in Eqs.~\eqref{psi sep} and~\eqref{rho sep}, $\left[ \frac{1}{\sqrt{2}}I + \frac{1}{\sqrt{6}}(X+Y+Z) \right]^{\otimes n}$ is a rank-1 matrix and it has only one non-zero eigenvalue $(\sqrt{2})^{n}$.
By using this property, the eigenvalues of $\sum_{\mathbf{a}} c_{\mathbf{a}}(\alpha) P_{\mathbf{a}}$ are as follows:
\begin{equation}
    \label{RDM eigenvalues}
    \text{eig} \left( \sum_{\mathbf{a}} c_{\mathbf{a}}(\alpha) P_{\mathbf{a}} \right)
    = 
    \left\lbrace
    \sqrt{\alpha + \frac{1-\alpha}{2^{n}}} - \sqrt{\frac{1-\alpha}{2^{n}}}~(2^{n}-1~\text{degeneracy}), \quad
    \sqrt{\alpha + \frac{1-\alpha}{2^{n}}} - \sqrt{\frac{1-\alpha}{2^{n}}} + 2^{n}\sqrt{\frac{1-\alpha}{2^{n}}}
    \right\rbrace.
\end{equation}
According to Eq.~\eqref{RDM}, the eigenvalues of the reduced density matrix are the squares of the eigenvalues of $\sum_{\mathbf{a}} c_{\mathbf{a}}(\alpha) P_{\mathbf{a}}$ in Eq.~\eqref{RDM eigenvalues}, multiplied by the constant factor $1/2^{n}$.
For large $n \gg 1$ with $\alpha = \Theta(1/\text{poly}(n))$, the eigenvalue spectrum consists of $2^{n}-1$ eigenvalues approximately equal to $\frac{1}{2^{n}}\alpha$, and a single eigenvalue equal to $1-\alpha$.
Thus, the entanglement entropy is given by
\begin{equation}
    \label{entanglement entropy}
\begin{aligned}
    S_{\text{a}|\text{s}}
    &= \Theta \left( -(2^{n}-1)\frac{\alpha}{2^{n}}\log_{2}(\frac{\alpha}{2^{n}}) - (1-\alpha)\log_{2}(1-\alpha) \right) \\
    &= \Theta \left( n\alpha -\alpha \log_{2} \alpha - (1-\alpha)\log_{2}(1-\alpha) \right) \\ 
    &= \Theta \left( n\alpha + \frac{H(\alpha)}{\log 2} \right) \\ 
    &= \Theta \left( n\alpha \right),
\end{aligned}
\end{equation}
where $H(\alpha) := -\alpha\log \alpha - (1-\alpha)\log (1-\alpha)$ is the binary Shannon entropy.

\subsection{Construction of unbiased estimator}

To obtain the measurement outcomes for the Pauli channel learning, we employ the Bell measurement POVM $\{ E_{\mathbf{v}} \}_{\mathbf{v} \in \mathds{Z}_{2}^{2n}}$, where each POVM element corresponds to the Bell basis state labeled by $\mathbf{v}$~\cite{chen2022quantum}.
Each POVM element is defined as
\begin{equation}
    \label{Bell POVM}
    E_{\mathbf{v}}
    := |\Psi_{\mathbf{v}} \rangle \langle \Psi_{\mathbf{v}}|
    = \frac{1}{4^{n}} \sum_{\mathbf{a} \in \mathds{Z}_{2}^{2n}} (-1)^{\langle \mathbf{a}, \mathbf{v} \rangle}
    P_{\mathbf{a}}^{\text{T}} \otimes P_{\mathbf{a}},
\end{equation}
where the last equality is equivalent to Eq.~\eqref{density matrix of Bell basis state}.
According to the channel operation in Eq.~\eqref{definition of channel with ancilla}, the probability $\Pr(\mathbf{v})$ of obtaining the measurement outcome $\mathbf{v}$ is given by
\begin{equation}
    \label{probability of outcomes}
\begin{aligned}
    \Pr(\mathbf{v})
    &= \Tr[
        E_{\mathbf{v}}
        (\mathds{1}_{\text{anc}}\otimes \mathbf{\Lambda}) (\rho_{\text{in}})
    ] \\
    &= \frac{1}{8^{n}} \sum_{\mathbf{a}, \mathbf{b}}
    (-1)^{\langle \mathbf{a}, \mathbf{v} \rangle} \lambda(\mathbf{b})
    \Tr[P_{\mathbf{a}}^{\text{T}} \Tr_{\text{sys}}((I_{\text{anc}}\otimes P_{\mathbf{b}}) \rho_{\text{in}})]
    \Tr(P_{\mathbf{a}} P_{\mathbf{b}}) \\ 
    &= \frac{1}{8^{n}} \sum_{\mathbf{a}, \mathbf{b}}
    (-1)^{\langle \mathbf{a}, \mathbf{v} \rangle} \lambda(\mathbf{b})
    \Tr((P_{\mathbf{a}}^{\text{T}}\otimes P_{\mathbf{b}}) \rho_{\text{in}})
    \times 2^{n} \delta_{\mathbf{a}, \mathbf{b}} \\
    &= \frac{1}{4^{n}} \sum_{\mathbf{b}}
    (-1)^{\langle \mathbf{b}, \mathbf{v} \rangle} \lambda(\mathbf{b})
    \Tr((P_{\mathbf{b}}^{\text{T}}\otimes P_{\mathbf{b}}) \rho_{\text{in}}).
\end{aligned}
\end{equation}
Using the inverse transformation given in Eq.~\eqref{Walsh-Hadamard transform of bit string}, we find the desired expression for $\lambda(\mathbf{b})$ as
\begin{equation}
    \label{def of E function}
    \lambda(\mathbf{b})
    = \sum_{\mathbf{v}} \Pr(\mathbf{v}) \frac{(-1)^{\langle \mathbf{b}, \mathbf{v} \rangle}}{\mathcal{E}(\mathbf{b})}, ~~ \text{where }
    \mathcal{E}(\mathbf{b}) := \Tr((P_{\mathbf{b}}^{\text{T}}\otimes P_{\mathbf{b}}) \rho_{\text{in}}).
\end{equation}
Therefore, we find an unbiased estimator $\hat{\lambda}(\mathbf{b})$ as
\begin{equation}
    \label{unbiased estimator}
    \hat{\lambda}(\mathbf{b})
    = \frac{1}{N}\sum_{l=1}^{N}
    \frac{(-1)^{\langle \mathbf{b}, \mathbf{v}^{(l)} \rangle}}{\mathcal{E}(\mathbf{b})},
\end{equation}
where $\{\mathbf{v}^{(l)}\}_{l=1}^{N}$ are $N$ measurement outcomes.
Here, by using the properties of the Bell basis given in Eq.~\eqref{density matrix of Bell basis state} and exploiting Eq.~\eqref{Walsh-Hadamard transform of bit string}, $\mathcal{E}(\mathbf{b})$ can be computed as
\begin{equation}
    \mathcal{E}(\mathbf{b})
    = \Tr( (P_{\mathbf{b}}^{\text{T}}\otimes P_{\mathbf{b}}) \rho_{\text{in}} )
    = \sum_{\mathbf{a}} (-1)^{\langle \mathbf{a}, \mathbf{b}\rangle}
    \langle \Psi_{\mathbf{a}}|\rho_{\text{in}}|\Psi_{\mathbf{a}} \rangle, \quad
    \langle \Psi_{\mathbf{a}}|\rho_{\text{in}}|\Psi_{\mathbf{a}} \rangle
    = \frac{1}{4^{n}} \sum_{\mathbf{b}} (-1)^{\langle \mathbf{a}, \mathbf{b}\rangle} \mathcal{E}(\mathbf{b}).
\end{equation}
When the input state is the pure state $\rho_{\text{in}} = |{\Psi_{\text{in}}(\alpha) }\rangle \langle{\Psi_{\text{in}}(\alpha)}| $ where $|{\Psi_{\text{in}}(\alpha)}\rangle = \sum_{\mathbf{a}}c_{\mathbf{a}}(\alpha) |{\Psi_{\mathbf{a}}}\rangle$ as given in Eq.~\eqref{input state}, 
\begin{equation}
    \label{E and c transform}
    \mathcal{E}(\mathbf{b})
    = \sum_{\mathbf{a}} (-1)^{\langle \mathbf{a}, \mathbf{b}\rangle} |c_{\mathbf{a}}(\alpha)|^{2}, \quad
    |c_{\mathbf{a}}(\alpha)|^{2}
    = \frac{1}{4^{n}} \sum_{\mathbf{b}} (-1)^{\langle \mathbf{a}, \mathbf{b}\rangle} \mathcal{E}(\mathbf{b}).
\end{equation}
Therefore, for our input state in Eq.~\eqref{input state},
\begin{equation}
    \label{function E}
\begin{aligned}
    \mathcal{E}(\mathbf{b})
    &= \sum_{\mathbf{a}} (-1)^{\langle \mathbf{a}, \mathbf{b}\rangle}
    \left[
        \alpha \delta_{\mathbf{a}, \mathbf{0}}
        + (1-\alpha) \left(\frac{1}{2}\right)^{n-|\mathbf{a}|} \left(\frac{1}{6}\right)^{|\mathbf{a}|}
    \right] \\ 
    &= \alpha + (1-\alpha) \sum_{a_{1} \in \mathds{Z}_{2}^{2}} \cdots \sum_{a_{n} \in \mathds{Z}_{2}^{2}}
    (-1)^{\langle \mathbf{a}, \mathbf{b}\rangle}
    \left(\frac{1}{2}\right)^{n-|\mathbf{a}|} \left(\frac{1}{6}\right)^{|\mathbf{a}|} \\
    &= \alpha + (1-\alpha)
    \prod_{j=1}^{n}
    \left[
        \sum_{a \in \mathds{Z}_{2}^{2}} 
        (-1)^{\langle a_{j}, b_{j} \rangle}
        \left(\frac{1}{2}\right)^{1-|a_{j}|} \left(\frac{1}{6}\right)^{|a_{j}|}
    \right] \\
    &= \alpha + (1-\alpha) \prod_{j=1}^{n} \left(\frac{1}{3}\right)^{|b_{j}|} \\ 
    &= \alpha + (1-\alpha) \left(\frac{1}{3}\right)^{|\mathbf{b}|}.
\end{aligned}
\end{equation}

\subsection{Sample complexity}

Since the estimator $\hat{\lambda}(\mathbf{b})$ is bounded within the range $-\frac{1}{\mathcal{E}(\mathbf{b})} \leq \hat{\lambda}(\mathbf{b}) \leq \frac{1}{\mathcal{E}(\mathbf{b})}$, Hoeffding's inequality yields
\begin{equation}
    \label{Hoeffding}
    \Pr(|\hat{\lambda}(\mathbf{b}) - \lambda(\mathbf{b})| \geq \varepsilon)
    \leq 2 \exp( -\frac{1}{2} N(\mathbf{b}) \varepsilon^{2} \mathcal{E}^{2}(\mathbf{b}) ),
\end{equation}
where $N(\mathbf{b})$ denotes the number of measurement outcomes used to estimate $\lambda(\mathbf{b})$. 
Therefore, to achieve an estimation accuracy $\varepsilon$ with success probability at least $1-\delta$, a sufficient number of samples is bounded by $N(\mathbf{b}) \geq \frac{1}{\mathcal{E}^{2}(\mathbf{b})}\times 2 \varepsilon^{-2} \log (2 \delta^{-1})$.
The union bound implies that, to achieve this accuracy and confidence for any of the $4^{n}$ parameters $\lambda(\mathbf{b})$, the required number of samples $N$ needs to satisfy
\begin{equation}
    \label{alpha dependent N}
    N = O( n\alpha^{-2} \times \varepsilon^{-2} \log \delta^{-1}),
\end{equation}
since $\mathcal{E}(\mathbf{b}) \geq \alpha + (1-\alpha) (1/3)^{n} \geq \alpha$ from Eq.~\eqref{function E}.
Finally, by setting $\alpha = \Theta(1/\text{poly}(n))$, we accomplish the learning task with polynomial sample complexity (Eq.~\eqref{alpha dependent N}) by using input states with inverse-polynomially small entanglement (Eq.~\eqref{entanglement entropy}).
This completes the proof of Theorem 1 in the main text. 
Note that when $\alpha=1$, our result reduces to the known bound $N = O(n\times \varepsilon^{-2} \log \delta^{-1})$ obtained by using the $2n$-qubit Bell pair~\cite{chen2022quantum}, which has the maximal entanglement $S_{\text{a}|\text{s}} = n$.

\subsection{Generalization of Theorem 1}

In this section, we consider a more general form of input state than that in Eqs.~(2) and (3) of the main text.
More precisely, we consider the following extended form:
\begin{equation}
    \label{generalized input}
    |{\Psi_{\text{in}}( \alpha,  \varphi_{\text{a}}, \varphi_{\text{s}} )}\rangle
    := \sqrt{\alpha'} |\Psi_{\text{B}}\rangle + \sqrt{1-\alpha} | \varphi_{\text{a}} \rangle \otimes | \varphi_{\text{s}} \rangle,
\end{equation}
where $| \varphi_{\text{a}} \rangle$ and $| \varphi_{\text{s}} \rangle$ are arbitrary states of the ancilla and system, respectively, and the constant $\alpha'$ is determined by $\alpha$ and the overlap $\langle \Psi_{\text{B}}| (| \varphi_{\text{a}} \rangle \otimes | \varphi_{\text{s}} \rangle)$.
Here, we show that any state in Eq.~\eqref{generalized input} satisfying the condition $(| \varphi_{\text{s}} \rangle)^{*} = | \varphi_{\text{a}} \rangle$ (the complex conjugate taken in the computational basis) possesses the same properties as the state in Eq.~(2) of the main text.
Hereafter, we denote the complex-conjugate state of $| \varphi \rangle$ by $|\varphi^{*} \rangle := (| \varphi \rangle)^{*}$.
Using this notation, we present an extended version of Theorem~1 as follows:
\begin{theorem}
    \label{thm: S1}
    \normalfont
    For any $\alpha = \Theta(1/\text{poly}(n))$ and arbitrary $|\varphi\rangle$, the state $|{\Psi_{\text{in}}(\alpha, \varphi, \varphi^{*})}\rangle$ has entanglement $S_{\text{a}|\text{s}} = \Theta(n \alpha)$.
    Furthermore, by using this state with the Bell measurement POVM defined in Eq.~\eqref{Bell POVM}, the ($\varepsilon, \delta, n$)-Pauli channel learning task can be accomplished with $N = O(n\alpha^{-2} \times \varepsilon^{-2}\log \delta^{-1})$.
\end{theorem}
\begin{proof}
For the proof, we use the following identity for the overlap between the Bell state $|\Psi_{\text{B}}\rangle$ and a product state $| \varphi_{\text{a}} \rangle \otimes | \varphi_{\text{s}} \rangle$:
\begin{equation}
    \label{overlap}
    \langle \Psi_{\text{B}}| (| \varphi_{\text{a}} \rangle \otimes | \varphi_{\text{s}} \rangle)
    = \frac{1}{\sqrt{2^{n}}} \langle \varphi_{\text{s}}^{*}|\varphi_{\text{a}} \rangle.
\end{equation}
Hence, when $\varphi_{\text{s}}^{*} = \varphi_{\text{a}} = \varphi$, we have $\langle \Psi_{\text{B}}| (| \varphi_{\text{a}} \rangle \otimes | \varphi_{\text{s}} \rangle) = \frac{1}{\sqrt{2^{n}}}$.
From the normalization $\langle{\Psi_{\text{in}}(\alpha, \varphi, \varphi^{*})}|{\Psi_{\text{in}}(\alpha, \varphi, \varphi^{*})}\rangle = 1$, we have $\alpha' + (1-\alpha) + 2\sqrt{\frac{\alpha'(1-\alpha)}{2^{n}}}=1$,
which leads to $\sqrt{\alpha'} = \sqrt{\alpha + \frac{1-\alpha}{2^{n}}} - \sqrt{\frac{1-\alpha}{2^{n}}}$, identical to Eq.~\eqref{alpha prime in input}.
To compute the entanglement, we use other identities related to the partial trace:
\begin{equation}
    \Tr_{\text{sys}}( |\Psi_{\text{B}}\rangle ( \langle \varphi_{\text{a}}| \otimes \langle \varphi_{\text{s}}|) )
    = \frac{1}{\sqrt{2^{n}}} | \varphi_{\text{s}}^{*}\rangle \langle \varphi_{\text{a}}|, ~~
    \Tr_{\text{sys}}( ( | \varphi_{\text{a}} \rangle \otimes | \varphi_{\text{s}} \rangle) \langle \Psi_{\text{B}} | )
    = \frac{1}{\sqrt{2^{n}}} | \varphi_{\text{a}}\rangle \langle \varphi_{\text{s}}^{*}|.
\end{equation}
Thus, the reduced density matrix is computed as
\begin{equation}
    \Tr_{\text{sys}}
    (
        |{\Psi_{\text{in}}(\alpha, \varphi, \varphi^{*}) }\rangle
        \langle{\Psi_{\text{in}}(\alpha, \varphi, \varphi^{*})}|
    )
    = \frac{\alpha'}{2^{n}} I^{\otimes n}
    + 2 \sqrt{ \frac{ \alpha'(1-\alpha) }{ 2^{n} } } |\varphi \rangle \langle \varphi|
    + (1-\alpha) |\varphi \rangle \langle \varphi|,
\end{equation}
and its eigenvalues are given by
\begin{equation}
\begin{aligned}
    \text{eig}
    \left(
        \Tr_{\text{sys}}
        (
            |{\Psi_{\text{in}}(\alpha, \varphi, \varphi^{*}) }\rangle
            \langle{\Psi_{\text{in}}(\alpha, \varphi, \varphi^{*})}|
        )
    \right)
    =
    \left\lbrace
    \frac{\alpha'}{2^{n}}~(2^{n}-1~\text{degeneracy}), \quad
    \frac{\alpha'}{2^{n}} + 2 \sqrt{ \frac{ \alpha'(1-\alpha) }{ 2^{n} } } + (1-\alpha)
    \right\rbrace.
\end{aligned}
\end{equation}
Note that these eigenvalues are exactly the same as the reduced density matrix of $|\Psi_{\text{in}}(\alpha)$ in Eq.~(2) of the main text (see Eqs.~\eqref{RDM} and~\eqref{RDM eigenvalues}).
Therefore, we confirm that $S_{\text{a}|\text{s}}$ value of $|{\Psi_{\text{in}}(\alpha, \varphi, \varphi^{*})}\rangle$ is also $\Theta(n \alpha)$.

Next, we need to verify that the learning task can be accomplished with $N = O(n\alpha^{-2} \times \varepsilon^{-2}\log \delta^{-1})$.
As follows from Eqs.~\eqref{probability of outcomes},~\eqref{def of E function}, and~\eqref{unbiased estimator}, we need to compute $\mathcal{E}(\mathbf{b}) = \Tr((P_{\mathbf{b}}^{\text{T}}\otimes P_{\mathbf{b}}) \rho_{\text{in}})$, where $\rho_{\text{in}} = |{\Psi_{\text{in}}(\alpha, \varphi, \varphi^{*}) }\rangle \langle{\Psi_{\text{in}}(\alpha, \varphi, \varphi^{*})}|$. 
Since $P_{\mathbf{b}}\otimes P_{\mathbf{b}}^{\mathrm{T}} |\Psi_{\text{B}}\rangle = |\Psi_{\text{B}}\rangle$ for all $P_{\mathbf{b}}$, the quantity $\mathcal{E}(\mathbf{b})$ can be evaluated as
\begin{equation}
    \label{Eb for general input}
\begin{aligned}
    \mathcal{E}(\mathbf{b})
    &= \langle \Psi_{\text{in}}(\alpha, \varphi, \varphi^{*})
    | P_{\mathbf{b}}\otimes P_{\mathbf{b}}^{\mathrm{T}} 
    | \Psi_{\text{in}}(\alpha, \varphi, \varphi^{*}) \rangle \\
    &= \alpha' \langle \Psi_{\text{B}} | \Psi_{\text{B}} \rangle
    + 2\sqrt{\frac{\alpha'(1-\alpha)}{2^{n}}} \langle \varphi|\varphi \rangle
    + (1-\alpha) \langle \varphi| P_{\mathbf{b}} |\varphi \rangle
    \langle \varphi^{*}| P_{\mathbf{b}}^{\mathrm{T}} |\varphi^{*} \rangle \\
    &= \alpha' + 2\sqrt{\frac{\alpha'(1-\alpha)}{2^{n}}} 
    + (1-\alpha) |\langle \varphi| P_{\mathbf{b}} |\varphi \rangle|^{2},
\end{aligned}
\end{equation}
where we have used the property of the Pauli string that $P_{\mathbf{b}}^{\mathrm{T}} = P_{\mathbf{b}}^{*}$. 
Therefore, when $\alpha = \Theta(1/\text{poly}(n))$, we have $\mathcal{E}(\mathbf{b}) = \Omega(\alpha) $ also for the state $|{\Psi_{\text{in}}(\alpha, \varphi, \varphi^{*}) }\rangle$.
In the same manner as in the previous section, we confirm that the required sample complexity is $N = O(n\alpha^{-2} \times \varepsilon^{-2}\log \delta^{-1})$.
\end{proof}
\noindent
Note that Eq.~(2) of the main text is a specific example of $|{\Psi_{\text{in}}(\alpha, \varphi, \varphi^{*}) }\rangle$. 
In addition, the second line of Eq.~\eqref{Eb for general input} clarifies why the constraint $\varphi_{\text{s}}^{*} = \varphi_{\text{a}}$ is necessary to achieve polynomial sample complexity when the Bell measurement POVM is employed.
For an arbitrary $|{\Psi_{\text{in}}( \alpha,  \varphi_{\text{a}}, \varphi_{\text{s}} )}\rangle$ state, the third term in the second line is $(1-\alpha) \langle \varphi_{\text{a}} | P_{\mathbf{b}} |\varphi_{\text{a}} \rangle \langle \varphi_{\text{s}}| P_{\mathbf{b}}^{\mathrm{T}} |\varphi_{\text{s}} \rangle$, which can, in general, be negative.
Thus, it can take the value $(1-\alpha) \langle \varphi_{\text{a}} | P_{\mathbf{b}} |\varphi_{\text{a}} \rangle \langle \varphi_{\text{s}}| P_{\mathbf{b}}^{\mathrm{T}} |\varphi_{\text{s}} \rangle = -\alpha'$, in which case $\mathcal{E}(\mathbf{b})$ reduces to $\mathcal{E}(\mathbf{b}) = 2\sqrt{\frac{\alpha'(1-\alpha)}{2^{n}}}$.
In this situation, the sample complexity scales exponentially with $n$.

\subsection{Mixed state example: Werner state}

In this section, we present an additional example of a state with inverse-polynomially small entanglement, such that the learning task can be completed within polynomial sample complexity when this state is used as an input.
We employ the Werner state $\rho_{\text{W}}(\lambda)$~\cite{werner1989quantum} as the input state, defined as
\begin{equation}
    \label{Werner state}
    \rho_{\text{W}}(\lambda)
    := \frac{1-\lambda}{2^{2n}-1} I_{\text{anc}} \otimes I_{\text{sys}}
    + \left(\lambda - \frac{1-\lambda}{2^{2n}-1} \right) |\Psi_{\text{B}}\rangle \langle\Psi_{\text{B}}|,
\end{equation}
where $0 \leq \lambda \leq 1$ is a constant.
Since $\rho_{\text{W}}(\lambda)$ is a mixed state, the entanglement entropy $S_{\text{a}|\text{s}}$ cannot be used as an entanglement measure.
Instead, we employ the entanglement of formation $\text{EoF}(\rho_{\text{W}}(\lambda))$ between the ancilla and system, which quantifies the number of Bell pairs required to prepare the state~\cite{bennett1996concentrating}.
The EoF value for the Werner state is given by~\cite{terhal2000entanglement}
\begin{equation}
    \text{EoF}(\rho_{\text{W}}(\lambda))
    = \frac{2^{n}\log_{2}(2^{n}-1)}{2^{n}-2}(\lambda -1) + n, \quad
    \lambda\in \left[ \frac{4(2^{n}-1)}{2^{2n}}, 1 \right].
\end{equation}

To obtain the upper bound on the sample complexity, we compute the quantity $\mathcal{E}(\mathbf{b})$ defined in Eq.~\eqref{def of E function} as follows:
\begin{equation}
    \label{Werner E}
    \mathcal{E}(\mathbf{b})
    = \Tr( (P_{\mathbf{b}}^{\text{T}}\otimes P_{\mathbf{b}}) \rho_{\text{W}}(\lambda) )
    = \frac{2^{2n}}{2^{2n}-1}(1-\lambda) \delta_{\mathbf{b}, \mathbf{0}}
    + \left(\lambda - \frac{1-\lambda}{2^{2n}-1} \right).
\end{equation}
Therefore, in the regime $n\gg 1$ with $\lambda = \Theta(1/\text{poly}(n))$, we obtain
\begin{equation}
    \text{EoF}(\rho_{\text{W}}(\lambda)) = \Theta \left( n\lambda \right), \quad
    \mathcal{E}(\mathbf{b}) = \Theta\left( \lambda \right).
\end{equation}
By analogy with Eqs.~\eqref{Hoeffding} and~\eqref{alpha dependent N}, the resulting sample complexity is 
\begin{equation}
    N = O( n\lambda^{-2} \times \varepsilon^{-2} \log \delta^{-1}).
\end{equation}
Thus, we verify that the Werner state also exhibits a property similar to that in Eq.~\eqref{input state}.

\section{Proof of Theorem 2}

To prove the lower bound on the sample complexity stated in Theorem 2 of the main text, we utilize a hypothesis-testing game of channel discrimination, which was recently introduced in~\cite{oh2024entanglementenabled, chen2021exponential, huang2022quantum, liu2025quantum, chen2024tight, chen2025efficient}.
However, the sample complexity for the ($\varepsilon, \delta, w$)-Pauli channel learning task (Definition 1 in the main text), especially the case $w < n$, has not been investigated.
In this section, we present a detailed proof of Theorem 2, focusing on our improvements in the proof technique tailored to the ($\varepsilon, \delta, w$)-Pauli channel learning setting.

\subsection{Hypothesis-testing game}

We develop a hypothesis-testing game, specifically designed to ensure that achieving a high winning probability is a necessary condition for the existence of the ($\varepsilon, \delta, w$)-Pauli channel learning scheme.
By analyzing the sample complexity required to win this game with high probability, we establish the desired lower bound for the learning task.

Before we explain the rules of the hypothesis-testing game, we first introduce two hypotheses used in the game. 
Each hypothesis corresponds to a Pauli channel characterized by Pauli eigenvalues
\begin{equation}
    \label{dep and e channel}
    \mathbf{\Lambda}_{\text{dep}} : \lambda(\mathbf{b}) = \delta_{\mathbf{b}, \mathbf{0}}
    \quad \text{or} \quad
    \mathbf{\Lambda}_{(\mathbf{e}, s)} : \lambda(\mathbf{b})
    = \delta_{\mathbf{b}, \mathbf{0}} + 2s \varepsilon \delta_{\mathbf{b}, \mathbf{e}},
\end{equation}
where the tuple $(\mathbf{e}, s)$ consists of a sign $s \in \{-1, 1\}$ and a $2n$-bit string $\mathbf{e} \in \mathds{Z}_{2}^{2n}$.
Note that the channel $\mathbf{\Lambda}_{\text{dep}}$ is a completely depolarizing channel; for any input state $\rho_{\text{in}}$, the system part of the output is the maximally mixed state, expressed as $\Tr_{\text{sys}}(\rho_{\text{in}})\otimes \frac{1}{2^{n}}I_{\text{sys}}$. 
The second hypothesis $\mathbf{\Lambda}_{(\mathbf{e}, s)}$ resembles the Pauli spike introduced in~\cite{chen2022quantum, chen2025efficient} as it contains a single non-trivial Pauli eigenvalue.
However, the small magnitude $\varepsilon$ makes hypothesis discrimination difficult.

The hypothesis-testing game is described as follows:
\begin{enumerate}
    \item Initially, a referee samples a sign $s \in \{-1, 1\}$ uniformly at random, and the bit string $\mathbf{e}$ is sampled according to a certain probability distribution $\Pr(\mathbf{e})$. 

    \item The referee selects one of the two hypotheses described in Eq.~\eqref{dep and e channel} with equal probability.
    The $N$ copies of the selected channel are sent to the player.

    \item The player performs one measurement on each copy of the channel to collect outcomes.
    All $N$ copies are consumed to collect the measurement outcomes.

    \item Finally, the referee reveals the bit string $\mathbf{e}$ and asks the player to identify whether the selected channel is $\mathbf{\Lambda}_{\text{dep}}$ or not.
\end{enumerate}
If the ($\varepsilon, \delta, w$)-Pauli channel learning task can be accomplished using $N$ channel copies, the player can win this game with high probability, since it implies that the player can estimate any $\lambda(\mathbf{b})$ within the error $\varepsilon$ with probability $1-\delta$.
Specifically, if $|\lambda(\mathbf{e})| > \varepsilon$ for the revealed $\mathbf{e}$, the player can conclude that the answer is not $\mathbf{\Lambda}_{\text{dep}}$; otherwise, the player concludes that it is $\mathbf{\Lambda}_{\text{dep}}$.

\begin{figure}[h]
    \centering
    \begin{tikzpicture}
        \def\width{3.0cm}
        \def\height{1.5cm}
        \def\spacing{4.0cm}
        \def\arrowspacing{0.1cm}
        \node[draw, fill=SkyBlue!20, rounded corners=8pt, minimum width=\width, minimum height=\height, align=center, text width=\width] (A) at (1*\spacing,0)
        {($\varepsilon, \delta, w$)-learning scheme exists};
        \node[draw, fill=SkyBlue!20, rounded corners=8pt, minimum width=\width, minimum height=\height, align=center, text width=\width] (B) at (2*\spacing,0)
        {function of $\delta$ $\leq \Pr(\text{win})$};
        \node[draw, fill=SkyBlue!20, rounded corners=8pt, minimum width=\width, minimum height=\height, align=center, text width=\width] (C) at (3*\spacing,0)
        {$\Pr(\text{win})$ $\leq\text{function of TVD}$};
        \node[draw, fill=SkyBlue!20, rounded corners=8pt, minimum width=\width, minimum height=\height, align=center, text width=\width] (D) at (4*\spacing,0)
        {TVD $\leq\text{function of }N$};
        \draw[->, thick] ([xshift=\arrowspacing]A.east) -- ([xshift=-\arrowspacing]B.west);
        \draw[->, thick] ([xshift=\arrowspacing]B.east) -- ([xshift=-\arrowspacing]C.west);
        \draw[->, thick] ([xshift=\arrowspacing]C.east) -- ([xshift=-\arrowspacing]D.west);
    \end{tikzpicture}
    \caption{Schematics of the proof of the lower bound in Theorem 2 of the main text. Each arrow denotes a necessary condition.}
    \label{lower bound proof flow}
\end{figure}
In Fig.~\ref{lower bound proof flow}, we illustrate the proof technique for establishing the lower bound~\cite{oh2024entanglementenabled, coroi2025exponential}.
From the hypothesis-testing game, the lower bound for the learning task is obtained by the following steps:
(1) The existence of the ($\varepsilon, \delta, w$)-learning scheme guarantees a lower bound on the player's winning probability $\Pr(\text{win})$ as a function of $\delta$.
(2) The winning probability $\Pr(\text{win})$ provides a lower bound on the total variation distance (TVD) between the two distributions of measurement outcomes corresponding to the two hypotheses in Eq.~\eqref{dep and e channel}.
(3) The TVD is upper bounded in terms of the number of channel copies $N$.
In the following sections, we provide detailed explanations for each of these steps.

\subsection{Winning probability and TVD}

For the second box in Fig.~\ref{lower bound proof flow}, the relation between the player's winning probability $\Pr(\text{win})$ and the success probability $1-\delta$ for the ($\varepsilon, \delta, w$)-learning scheme is given by
\begin{equation}
    \label{winning probability and delta}
	\Pr(\text{win}) \geq
    \Pr( |\mathbf{e}| \leq w)\times(1-\delta)
	+ (1-\Pr( |\mathbf{e}| \leq w))\times\frac{1}{2}.
\end{equation}
Here, the first term describes the case in which the referee sends a ``learnable'' channel ($|\mathbf{e}| \leq w$), enabling the player to achieve a high winning probability $1-\delta$.
The second term corresponds to the complementary case, in which an ``unlearnable'' Pauli eigenvalue is sampled, compelling the player to guess randomly with a success probability of $1/2$.
This formulation, which samples the bit string $\mathbf{e}$ according to a certain non-uniform $\Pr(\mathbf{e})$, is a key improvement in our proof technique compared to previous studies~\cite{oh2024entanglementenabled, chen2021exponential, huang2022quantum, liu2025quantum, chen2024tight, chen2025efficient}.
In previous studies, where learning a subset of parameters was not considered, $\Pr(\mathbf{e})$ was taken as a uniform distribution.

To provide the details of the third box in Fig.~\ref{lower bound proof flow}, we introduce the definition of the TVD.
The TVD quantifies the difference between two probability distributions.
In our hypothesis-testing game, these correspond to the probability distributions of the measurement outcomes from the two channels specified in Eq.~\eqref{dep and e channel}, denoted by $\Pr_{\text{dep}}[\mathbf{o}]$ and $\Pr_{(\mathbf{e},s)}[\mathbf{o}]$, respectively.
Here, $\mathbf{o} = \{ o_{l} \}_{l=1}^{N}$ represents the set of measurement outcomes, with each $o_{l}$ obtained from the $l$-th copy of the channel.
As noted in Definition 1 of the main text, the number of measurement outcomes equals the number of channel copies $N$, since concatenated application of the channel is not permitted.
According to the game rule, since the referee reveals only the bit string $\mathbf{e}$ and not the sign $s$, we need to consider the averaged distribution $\underset{s}{\mathds{E}} \Pr_{(\mathbf{e},s)}[\mathbf{o}]$, where $\underset{s}{\mathds{E}}$ denotes averaging over the uniform distribution of the sign $s$.
The definition of the TVD between $\Pr_{\text{dep}}[\mathbf{o}]$ and $\underset{s}{\mathds{E}} \Pr_{(\mathbf{e},s)}[\mathbf{o}]$ is given by
\begin{equation}
    \label{def of TVD}
    \text{TVD}(\Pr_{\text{dep}}[\mathbf{o}], \underset{s}{\mathds{E}} \Pr_{(\mathbf{e},s)}[\mathbf{o}])
	:= \sum_{\mathbf{o}}
	\max \left\{
		0, \Pr_{\text{dep}}[\mathbf{o}]-\underset{s}{\mathds{E}} \Pr_{(\mathbf{e},s)}[\mathbf{o}]
	\right\}.
\end{equation}

It follows from Le Cam's two-point method~\cite{lecam1973convergence} that the player's winning probability (success rate for discriminating between two hypotheses) is bounded by the TVD as
\begin{equation}
    \label{TVD and winning probability}
    \frac{1}{2}(1+\underset{\mathbf{e}}{\mathds{E}}\text{TVD}(\Pr_{\text{dep}}[\mathbf{o}], \underset{s}{\mathds{E}} \Pr_{(\mathbf{e},s)}[\mathbf{o}]))
    \geq \Pr(\text{win}).
\end{equation}
Therefore, by combining Eqs.~\eqref{winning probability and delta} and~\eqref{TVD and winning probability}, we find
\begin{equation}
    \label{TVD and delta}
    \underset{\mathbf{e}}{\mathds{E}}
    \text{TVD}(\Pr_{\text{dep}}[\mathbf{o}], \underset{s}{\mathds{E}}\Pr_{(\mathbf{e},s)}[\mathbf{o}])
    \geq \Pr( |\mathbf{e}| \leq w)(1-2\delta).
\end{equation}
The inequality in Eq.~\eqref{TVD and delta} can be applied to learning tasks involving other subsets of parameters by appropriately modifying the condition $\Pr( |\mathbf{e}| \leq w)$ depending on the specific structure of the task.
Also, altering $\Pr(\mathbf{e})$ changes the averaging procedure $\underset{\mathbf{e}}{\mathds{E}}$.
We emphasize that Eq.~\eqref{TVD and delta} is valid for an arbitrary choice of $\Pr(\mathbf{e})$.
Thus, in the next section, we suitably choose $\Pr(\mathbf{e})$ to find the lower bound for the specific ($\varepsilon, \delta, w$)-Pauli channel learning task.

\subsection{Bound on TVD}

In this section, we derive the relation between the TVD and the number of copies $N$, as denoted in the last box in Fig.~\ref{lower bound proof flow}.
For this step, we follow techniques previously developed in~\cite{oh2024entanglementenabled, chen2021exponential, huang2022quantum, liu2025quantum, chen2024tight, chen2025efficient}.
The probability distributions of the measurement outcomes $\Pr_{\text{dep}}[\mathbf{o}]$ and $\Pr_{(\mathbf{e},s)}[\mathbf{o}]$ can be expressed as
\begin{equation}
    \Pr_{\text{dep}}[\mathbf{o}]
    = \Pr_{\text{dep}}[o_{1}] \Pr_{\text{dep}}[o_{2}|\mathbf{o}_{<2}] \ldots \Pr_{\text{dep}}[o_{N}|\mathbf{o}_{<N}], \quad
    \Pr_{(\mathbf{e},s)}[\mathbf{o}]
    = \Pr_{(\mathbf{e},s)}[o_{1}] \Pr_{(\mathbf{e},s)}[o_{2}|\mathbf{o}_{<2}] \ldots \Pr_{(\mathbf{e},s)}[o_{N}|\mathbf{o}_{<N}].
\end{equation}
Here, $\Pr_{\text{dep}}[o_{l}|\mathbf{o}_{<l}]$ and $\Pr_{(\mathbf{e},s)}[o_{l}|\mathbf{o}_{<l}]$ denote the conditional probability that the $l$-th measurement outcome is $o_{l}$ given the set of previous outcomes $\mathbf{o}_{<l} := \{ o_{1}, o_{2}, \ldots, o_{l-1} \}$, for the channels $\mathbf{\Lambda}_{\text{dep}}$ and $\mathbf{\Lambda}_{(\mathbf{e},s)}$, respectively.
Each conditional probability with the input state $\rho^{\mathbf{o}_{<l}}$ and the POVM elements $E_{o_{l}}^{\mathbf{o}_{<l}}$ are given as
\begin{equation}
    \Pr_{\text{dep}}[o_{l}|\mathbf{o}_{<l}]
    = \Tr( E_{o_{l}}^{\mathbf{o}_{<l}} (\mathds{1}_{\text{anc}}\otimes \mathbf{\Lambda}_{\text{dep}}) (\rho^{\mathbf{o}_{<l}}) )
    , \quad 
    \Pr_{(\mathbf{e},s)}[o_{l}|\mathbf{o}_{<l}]
    = \Tr( E_{o_{l}}^{\mathbf{o}_{<l}} (\mathds{1}_{\text{anc}}\otimes \mathbf{\Lambda}_{\mathbf{e},s}) (\rho^{\mathbf{o}_{<l}}) ).
\end{equation}
Here, the superscript $\mathbf{o}_{<l}$ is used to explicitly indicate that the input state and the POVM elements can be adaptively chosen from the previous outcomes $\mathbf{o}_{<l}$.
From the definition in Eq.~\eqref{dep and e channel}, the output states after applying the channel are given as
\begin{equation}
    (\mathds{1}_{\text{anc}}\otimes \mathbf{\Lambda}_{\text{dep}}) (\rho^{\mathbf{o}_{<l}})
    = \frac{1}{2^{n}} \Tr_{\text{sys}} (\rho^{\mathbf{o}_{<l}}) \otimes I_{\text{sys}} , ~~
    (\mathds{1}_{\text{anc}}\otimes \mathbf{\Lambda}_{(\mathbf{e},s)}) (\rho^{\mathbf{o}_{<l}})
    = \frac{1}{2^{n}} \Tr_{\text{sys}} (\rho^{\mathbf{o}_{<l}}) \otimes I_{\text{sys}}
    + \frac{2s \varepsilon}{2^{n}}
    \Tr_{\text{sys}} ((I_{\text{anc}}\otimes P_{\mathbf{e}}) \rho^{\mathbf{o}_{<l}}) \otimes P_{\mathbf{e}}.
\end{equation}
Thus, the conditional probability is written as
\begin{equation}
\begin{aligned}
    \Pr_{\text{dep}}[o_{l}|\mathbf{o}_{<l}]
    &= \frac{1}{2^{n}} \Tr[ \Tr_{\text{sys}} (E_{o_{l}}^{\mathbf{o}_{<l}}) \Tr_{\text{sys}} (\rho^{\mathbf{o}_{<l}}) ], \\
    \Pr_{(\mathbf{e},s)}[o_{l}|\mathbf{o}_{<l}]
    &= \frac{1}{2^{n}} \Tr[ \Tr_{\text{sys}} (E_{o_{l}}^{\mathbf{o}_{<l}}) \Tr_{\text{sys}} (\rho^{\mathbf{o}_{<l}}) ]
	+ \frac{2s \varepsilon}{2^{n}}
	\Tr[
		\Tr_{\text{sys}} ((I_{\text{anc}}\otimes P_{\mathbf{e}}) E_{o_{l}}^{\mathbf{o}_{<l}})
		\Tr_{\text{sys}} ((I_{\text{anc}}\otimes P_{\mathbf{e}}) \rho^{\mathbf{o}_{<l}})
	].
\end{aligned}
\end{equation}
To compute the TVD, we evaluate the difference between the two probabilities $\Pr_{\text{dep}}[\mathbf{o}]-\underset{s}{\mathds{E}} \Pr_{(\mathbf{e},s)}[\mathbf{o}]$ in Eq.~\eqref{def of TVD} as follows:
\begin{equation}
\begin{aligned}
    \Pr_{\text{dep}}[\mathbf{o}]-\underset{s}{\mathds{E}} \Pr_{(\mathbf{e},s)}[\mathbf{o}]
    &= \Pr_{\text{dep}}[\mathbf{o}]
    \left(
        1 - \underset{s}{\mathds{E}} \frac{\Pr_{(\mathbf{e},s)}[\mathbf{o}]}{\Pr_{\text{dep}}[\mathbf{o}]}
    \right)\\
    &= \Pr_{\text{dep}}[\mathbf{o}]
    \left(
        1- \underset{s}{\mathds{E}}\prod_{l=1}^{N}
        \frac{\Pr_{(\mathbf{e},s)}[o_{l}|\mathbf{o}_{l}]}{\Pr_{\text{dep}}[o_{l}|\mathbf{o}_{l}]}
    \right) \\
    &= \Pr_{\text{dep}}[\mathbf{o}]
    \left(
        1- \underset{s}{\mathds{E}}\prod_{l=1}^{N}
        \frac{ \Tr(E_{o_{l}}^{\mathbf{o}_{<l}} (\mathds{1}_{\text{anc}}\otimes \mathbf{\Lambda}_{(\mathbf{e},s)}) (\rho^{\mathbf{o}_{<l}}) )}
        {\Tr(E_{o_{l}}^{\mathbf{o}_{<l}} (\mathds{1}_{\text{anc}}\otimes \mathbf{\Lambda}_{\text{dep}}) (\rho^{\mathbf{o}_{<l}}) )}
    \right) \\
    &= \Pr_{\text{dep}}[\mathbf{o}]
	\left(
		1- \underset{s}{\mathds{E}} \prod_{l=1}^{N}
		\left[
			1 + 2s \varepsilon
			\frac{
				\Tr[
					\Tr_{\text{sys}} ((I_{\text{anc}}\otimes P_{\mathbf{e}}) E_{o_{l}}^{\mathbf{o}_{<l}})
					\Tr_{\text{sys}} ((I_{\text{anc}}\otimes P_{\mathbf{e}}) \rho^{\mathbf{o}_{<l}})
				]
			}
			{\Tr[ \Tr_{\text{sys}} (E_{o_{l}}^{\mathbf{o}_{<l}}) \Tr_{\text{sys}} (\rho^{\mathbf{o}_{<l}}) ]}
		\right]
	\right).
\end{aligned}
\end{equation}
For notational convenience, we define $G_{\mathbf{e}}^{\mathbf{o}_{\leq l}}$ as
\begin{equation}
    \label{def of Ge}
    G_{\mathbf{e}}^{\mathbf{o}_{\leq l}}
    := \frac{
        \Tr[
            \Tr_{\text{sys}} ((I_{\text{anc}}\otimes P_{\mathbf{e}}) E_{o_{l}}^{\mathbf{o}_{<l}})
            \Tr_{\text{sys}} ((I_{\text{anc}}\otimes P_{\mathbf{e}}) \rho^{\mathbf{o}_{<l}})
        ]
    }
    {\Tr[ \Tr_{\text{sys}} (E_{o_{l}}^{\mathbf{o}_{<l}}) \Tr_{\text{sys}} (\rho^{\mathbf{o}_{<l}}) ]}.
\end{equation}
Following the Supplemental Material in~\cite{oh2024entanglementenabled}, the upper bound on $\Pr_{\text{dep}}[\mathbf{o}]-\underset{s}{\mathds{E}} \Pr_{(\mathbf{e},s)}[\mathbf{o}]$ is given as
\begin{equation}
\begin{aligned}
    \Pr_{\text{dep}}[\mathbf{o}]-\underset{s}{\mathds{E}} \Pr_{(\mathbf{e},s)}[\mathbf{o}]
    &= \Pr_{\text{dep}}[\mathbf{o}]
    \left(
        1-\underset{s}{\mathds{E}}\prod_{l=1}^{N}(1+2s \varepsilon G_{\mathbf{e}}^{\mathbf{o}_{\leq l}})
    \right) \\
    &\leq \Pr_{\text{dep}}[\mathbf{o}]
    \left(
        1-\sqrt{\prod_{l=1}^{N} (1+2\varepsilon G_{\mathbf{e}}^{\mathbf{o}_{\leq l}})(1-2\varepsilon G_{\mathbf{e}}^{\mathbf{o}_{\leq l}})}
    \right) \\ 
    &= \Pr_{\text{dep}}[\mathbf{o}]
    \left(
        1-\prod_{l=1}^{N}\sqrt{1-4\varepsilon^2 (G_{\mathbf{e}}^{\mathbf{o}_{\leq l}})^{2}}
    \right) \\
    &\leq \Pr_{\text{dep}}[\mathbf{o}] 
    \left(
        1- \prod_{l=1}^{N}(1-4\varepsilon^2 (G_{\mathbf{e}}^{\mathbf{o}_{\leq l}})^{2})
    \right) \\ 
    & \leq \Pr_{\text{dep}}[\mathbf{o}] \sum_{l=1}^{N} 4\varepsilon^2 (G_{\mathbf{e}}^{\mathbf{o}_{\leq l}})^{2}.
\end{aligned}
\end{equation}
For the second line, we apply the AM-GM inequality; for the fourth line, we use $\sqrt{1-x} \geq 1-x$ for $0 \leq x \leq 1$; and for the last line, we use the inequality $\prod_{l}(1-x_{l}) \geq 1- \sum_{l}x_{l}$ for all $0\leq x_{l} \leq 1$~\cite{oh2024entanglementenabled}.
Therefore, the upper bound on the TVD is
\begin{equation}
    \label{TVD and EG}
    \underset{\mathbf{e}}{\mathds{E}}
    \text{TVD}(\Pr_{\text{dep}}[\mathbf{o}], \underset{s}{\mathds{E}} \Pr_{(\mathbf{e},s)}[\mathbf{o}])
    = \underset{\mathbf{e}}{\mathds{E}} \sum_{\mathbf{o}}
    \max\left\{0, \Pr_{\text{dep}}[\mathbf{o}]-\underset{s}{\mathds{E}} \Pr_{(\mathbf{e},s)}[\mathbf{o}]\right\}
    \leq \sum_{\mathbf{o}} \Pr_{\text{dep}}[\mathbf{o}]
    4\varepsilon^{2} \sum_{l=1}^{N} \underset{\mathbf{e}}{\mathds{E}} (G_{\mathbf{e}}^{\mathbf{o}_{\leq l}})^{2}.
\end{equation}

The next step is to find the upper bound on $\underset{\mathbf{e}}{\mathds{E}} (G_{\mathbf{e}}^{\mathbf{o}_{\leq l}})^{2}$ as noted in Eq.~\eqref{TVD and EG}.
From the definition in Eq.~\eqref{def of Ge}, we exploit convexity to obtain the upper bound~\cite{chen2022quantum, oh2024entanglementenabled}.
Thus, without loss of generality, we can restrict our consideration to a pure input state $\rho^{\mathbf{o}_{<l}} = |{\Psi}^{\mathbf{o}_{<l}}\rangle\langle {\Psi}^{\mathbf{o}_{<l}}|$ and the rank-1 POVM element $E_{o_{l}}^{\mathbf{o}_{<l}} = w_{o_{l}}^{\mathbf{o}_{<l}} |{M_{o_{l}}^{\mathbf{o}_{<l}}}\rangle \langle{M_{o_{l}}^{\mathbf{o}_{<l}}}|$, where $w_{o_{l}}^{\mathbf{o}_{<l}}$ is a weight factor, explicitly given as
\begin{equation}
    \label{def of Phi and M}
    |{\Psi}^{\mathbf{o}_{<l}}\rangle
    = \sum_{j_{\text{anc}} = 0}^{2^{k}-1} \sum_{j_{\text{sys}} = 0}^{2^{n}-1} 
    \Phi_{j_{\text{anc}}, j_{\text{sys}}}
    |{j_{\text{anc}}}\rangle |{j_{\text{sys}}}\rangle ,\quad
    |{M_{o_{l}}^{\mathbf{o}_{<l}}}\rangle
    = \sum_{j_{\text{anc}} = 0}^{2^{k}-1} \sum_{j_{\text{sys}} = 0}^{2^{n}-1} 
    M_{j_{\text{anc}}, j_{\text{sys}}}
    |{j_{\text{anc}}}\rangle |{j_{\text{sys}}}\rangle.
\end{equation}
The normalization condition of the input state is $\langle{\Psi}^{\mathbf{o}_{<l}}|{\Psi}^{\mathbf{o}_{<l}}\rangle = \Tr(\Phi^{\dagger} \Phi) = 1$, and the POVM has to satisfy $\sum_{o_{l}} E_{o_{l}}^{\mathbf{o}_{<l}} = \sum_{o_{l}} w_{o_{l}}^{\mathbf{o}_{<l}} |{M_{o_{l}}^{\mathbf{o}_{<l}}}\rangle \langle{M_{o_{l}}^{\mathbf{o}_{<l}}}| = I_{\text{anc}} \otimes I_{\text{sys}}$.
Using matrices $\Phi$ and $M$ defined in Eq.~\eqref{def of Phi and M}, we have
\begin{equation}
    \Tr_{\text{sys}} ((I_{\text{anc}}\otimes P_{\mathbf{e}}) E_{o_{l}}^{\mathbf{o}_{<l}})
    = w_{o_{l}}^{\mathbf{o}_{<l}} M P_{\mathbf{e}}^{\text{T}} M^{\dagger}, \quad
    \Tr_{\text{sys}} ((I_{\text{anc}}\otimes P_{\mathbf{e}}) \rho^{\mathbf{o}_{<l}})
    = \Phi P_{\mathbf{e}}^{\text{T}} \Phi^{\dagger}.
\end{equation}
Then, from the definition in Eq.~\eqref{def of Ge}, the quantity $(G_{\mathbf{e}}^{\mathbf{o}_{\leq l}})^{2}$ can be expressed as
\begin{equation}
    \label{G expressed as Phi and M}
    (G_{\mathbf{e}}^{\mathbf{o}_{\leq l}})^{2}
    = \frac
    { \Tr^{2} [M P_{\mathbf{e}}^{\text{T}} M^{\dagger} \Phi P_{\mathbf{e}}^{\text{T}} \Phi^{\dagger}] }
    { \Tr^{2} [M M^{\dagger} \Phi \Phi^{\dagger}] }
    =  \frac
    { \Tr^{2} [P_{\mathbf{e}} M^{\dagger} \Phi P_{\mathbf{e}} \Phi^{\dagger} M] }
    { \Tr^{2} [\Phi^{\dagger} M M^{\dagger} \Phi] },
\end{equation}
since $P_{\mathbf{e}}^{\text{T}}$ differs from $P_{\mathbf{e}}$ only by a sign.
For notational simplicity, we define a matrix $C := \Phi^{\dagger} M$, and note that $\text{rank}(C) \leq 2^{k}$. 
By using this notation, we can rewrite the problem of finding the upper bound on $\underset{\mathbf{e}}{\mathds{E}} (G_{\mathbf{e}}^{\mathbf{o}_{\leq l}})^{2}$ as follows:
\begin{equation}
    \label{optimization prob}
    \max \underset{\mathbf{e}}{\mathds{E}} (G_{\mathbf{e}}^{\mathbf{o}_{\leq l}})^{2}
    = \max_{C: \mathrm{rank}(C) \leq 2^{k}}
    \underset{\mathbf{e}}{\mathds{E}}
    \frac{\Tr^{2}(P_\mathbf{e} C P_\mathbf{e} C^{\dagger})}{\Tr^{2}(CC^{\dagger})}.
\end{equation}

To derive the upper bound in Eq.~\eqref{optimization prob}, we note that the rank of $P_{\mathbf{e}} C^{\dagger} P_{\mathbf{e}} C$ in the numerator is at most $2^{k}$.
Thus, we can find a rank-$2^{k}$ projector $\Pi$ that satisfies $\Tr(P_{\mathbf{e}} C^{\dagger} P_{\mathbf{e}} C) = \Tr(P_{\mathbf{e}} C^{\dagger} P_{\mathbf{e}} C \Pi)$.
By using the Cauchy-Schwarz inequality~\cite{chen2022quantum}, the numerator of Eq.~\eqref{optimization prob} is upper bounded by
\begin{equation}
    \label{bound on PeCPeC}
\begin{aligned}
    \Tr^{2}(P_{\mathbf{e}} C^{\dagger} P_{\mathbf{e}} C)
    &= \Tr^{2}(P_{\mathbf{e}} C^{\dagger} P_{\mathbf{e}} C \Pi) \\
    &\leq \Tr[ (P_{\mathbf{e}} C^{\dagger} P_{\mathbf{e}} C)^{\dagger} (P_{\mathbf{e}} C^{\dagger} P_{\mathbf{e}} C) ]
    \Tr(\Pi^{\dagger} \Pi) \\
    &= 2^{k} \times \Tr[C^{\dagger} P_{\mathbf{e}} C C^{\dagger} P_{\mathbf{e}} C] \\
    &= 2^{k} \times \Tr[C C^{\dagger} P_{\mathbf{e}} C C^{\dagger} P_{\mathbf{e}}].
\end{aligned}
\end{equation}
Therefore, we obtain the upper bound as
\begin{equation}
    \label{EG bounded by sum of Pe}
\begin{aligned}
    \underset{\mathbf{e}}{\mathds{E}} (G_{\mathbf{e}}^{\mathbf{o}_{\leq l}})^{2}
    & \leq 2^{k} \times \frac
    { \sum_{\mathbf{e}} \Pr(\mathbf{e}) \Tr( CC^{\dagger} P_{\mathbf{e}} CC^{\dagger}P_{\mathbf{e}} ) }
    { \Tr^{2}(C C^{\dagger})} \\
    & = 2^{k} \times \frac
    { \sum_{\mathbf{e}} \Pr(\mathbf{e}) \Tr[ (CC^{\dagger} P_{\mathbf{e}} \otimes CC^{\dagger}P_{\mathbf{e}}) \mathds{F} ] }
    { \Tr^{2}(C C^{\dagger})} \\
    & = 2^{k} \times \frac
    { \sum_{\mathbf{e}} \Pr(\mathbf{e}) \Tr[ (CC^{\dagger}\otimes CC^{\dagger}) (P_{\mathbf{e}} \otimes P_{\mathbf{e}}) \mathds{F} ] }
    { \Tr^{2}(C C^{\dagger})} \\
    & = 2^{k} \times \frac
    {
        \Tr[(CC^{\dagger}\otimes CC^{\dagger})
        ( \sum_{\mathbf{e}} \Pr(\mathbf{e} ) P_{\mathbf{e}} \otimes P_{\mathbf{e}}) \mathds{F}]
    }
    {\Tr[ CC^{\dagger}\otimes CC^{\dagger} ]},
\end{aligned}
\end{equation}
where $\mathds{F}$ is the swap operator, which satisfies $\Tr(AB) = \Tr[ (A\otimes B) \mathds{F}]$ for arbitrary matrices $A$ and $B$ of size $2^{n} \times 2^{n}$.

Now, we focus on the weighted sum of Pauli strings, $\sum_{\mathbf{e}} \Pr(\mathbf{e}) P_{\mathbf{e}} \otimes P_{\mathbf{e}}$, to determine the maximum value appearing in Eq.~\eqref{EG bounded by sum of Pe}.
The maximum value of $\underset{\mathbf{e}}{\mathds{E}} (G_{\mathbf{e}}^{\mathbf{o}_{\leq l}})^{2}$ is governed by the largest eigenvalue of $(\sum_{\mathbf{e}} \Pr(\mathbf{e}) P_{\mathbf{e}} \otimes P_{\mathbf{e}} ) \mathds{F} $.
Given that all eigenvalues of $\mathds{F}$ are either 1 or -1, it suffices to diagonalize $\sum_{\mathbf{e}} \Pr(\mathbf{e}) P_{\mathbf{e}} \otimes P_{\mathbf{e}}$.

As appropriate for the ($\varepsilon, \delta, w$)-Pauli channel learning task, we consider the probability distribution $\Pr(\mathbf{e})$ as follows:
\begin{equation}
    \label{power law Pr(e)}
    \Pr(\mathbf{e}) = \frac{1}{(1+3x)^{n}}x^{|\mathbf{e}|},
\end{equation}
where $0< x \leq 1$ is a constant.
Note that the probability distribution is properly normalized as follows:
\begin{equation}
    \sum_{ \mathbf{e} \in \mathds{Z}_{2}^{2n} }\Pr(\mathbf{e})
    = \sum_{u=0}^{n} \binom{n}{u} 3^{u} \times \frac{1}{(1+3x)^{n}}x^{u}
    = \sum_{u=0}^{n} \binom{n}{u}
    \left( \frac{1}{1+3x} \right)^{n-u}
    \left( \frac{3x}{1+3x} \right)^{u}
    = 1,
\end{equation}
since the number of weight $u$ Pauli strings is $\binom{n}{u}$.
The probability that the referee sends a ``learnable'' channel (i.e., $|\mathbf{e}| \leq w$) is given by
\begin{equation}
    \label{prob for learnable channel}
    \Pr( |\mathbf{e}| \leq w) = \sum_{u=0}^{w} \binom{n}{u} 3^{u} \times \frac{1}{(1+3x)^{n}}x^{u}.
\end{equation}
The probability distribution $\Pr(\mathbf{e})$ in Eq.~\eqref{power law Pr(e)} has the property that it factorizes into a product over individual $2$-bit strings, and is given by
\begin{equation}
    \Pr(\mathbf{e}) = \Pr(e_{1}, e_{2}, \ldots, e_{n}) = \prod_{j=1}^{n} \frac{1}{1+3x}x^{|e_{j}|}.
\end{equation}
This property can reduce the weighted sum of Pauli strings to a factorized form, given by
\begin{equation}
    \label{factorized form of sum of Pauli}
\begin{aligned}
    \sum_{\mathbf{e}} \Pr(\mathbf{e}) P_{\mathbf{e}} \otimes P_{\mathbf{e}}
    & = \frac{1}{(1+3x)^{n}} \sum_{\mathbf{e}} x^{|\mathbf{e}|} P_{\mathbf{e}} \otimes P_{\mathbf{e}} \\
    & = \frac{1}{(1+3x)^{n}} \sum_{e_{1}=I,X,Y,Z}\sum_{e_{2}} \cdots \sum_{e_{n}}
    x^{|e_{1}|}x^{|e_{2}|} \cdots x^{|e_{n}|}
    (P_{e_{1}}\otimes P_{e_{2}}\otimes \cdots \otimes P_{e_{n}})
    \otimes
    (P_{e_{1}}\otimes P_{e_{2}}\otimes \cdots \otimes P_{e_{n}}) \\ 
    & = \frac{1}{(1+3x)^{n}}
    \left[
    \bigotimes_{j=1}^{n}	
    \left[ \sum_{e_{j} = I,X,Y,Z} x^{|e_{j}|} P_{e_{j}} \otimes P_{e_{j}} \right]
    \right] \\ 
    & = \frac{1}{(1+3x)^{n}}
    \left[
    \bigotimes_{j=1}^{n}	
    \left[ I_{j}\otimes I_{j} + x(X_{j}\otimes X_{j} + Y_{j}\otimes Y_{j} + Z_{j}\otimes Z_{j}) \right]
    \right].
\end{aligned}
\end{equation}
Since the eigenvalues of $I\otimes I + x(X\otimes X + Y\otimes Y + Z\otimes Z)$ in Eq.~\eqref{factorized form of sum of Pauli} are $1+x, 1+x, 1+x, 1-3x$, the largest eigenvalue of $\sum_{\mathbf{e}} \Pr(\mathbf{e}) P_{\mathbf{e}} \otimes P_{\mathbf{e}}$ is $\left(\frac{1+x}{1+3x}\right)^{n}$.
It leads to the upper bound
\begin{equation}
    \label{upper bound on EG}
    \underset{\mathbf{e}}{\mathds{E}} (G_{\mathbf{e}}^{\mathbf{o}_{\leq l}})^{2}
    \leq 2^{k} \left(\frac{1+x}{1+3x}\right)^{n},
\end{equation}
and from Eqs.~\eqref{TVD and delta},~\eqref{TVD and EG},~and~\eqref{upper bound on EG}, 
\begin{equation}
\begin{aligned}
    \Pr( |\mathbf{e}| \leq w)(1-2\delta)
    & \leq \underset{\mathbf{e}}{\mathds{E}}
    \text{TVD}(\Pr_{\text{dep}}[\mathbf{o}], \underset{s}{\mathds{E}} \Pr_{(\mathbf{e},s)}[\mathbf{o}]) \\
    & \leq \sum_{\mathbf{o}} \Pr_{\text{dep}}[\mathbf{o}]
    4\varepsilon^{2} \sum_{l=1}^{N} \underset{\mathbf{e}}{\mathds{E}} (G_{\mathbf{e}}^{\mathbf{o}_{\leq l}})^{2} \\
    & \leq \sum_{\mathbf{o}} \Pr_{\text{dep}}[\mathbf{o}]
    4\varepsilon^{2} \sum_{l=1}^{N} 2^{k} \left(\frac{1+x}{1+3x}\right)^{n} \\
    & = N \times 4\varepsilon^{2} \times 2^{k} \left(\frac{1+x}{1+3x}\right)^{n}.
\end{aligned}
\end{equation}
Thus, we find the lower bound on the sample complexity $N$ as
\begin{equation}
    \label{lower bound of arbitrary x}
    N \geq \frac{1-2\delta}{4 \varepsilon^{2}} 2^{-k}
    \left(\frac{1+3x}{1+x}\right)^{n} \Pr( |\mathbf{e}| \leq w)
    = \frac{1-2\delta}{4 \varepsilon^{2}} 2^{-k}
    \frac{1}{(1+x)^{n}} \sum_{u=0}^{w} \binom{n}{u}(3x)^{u}.
\end{equation}
Since Eq.~\eqref{lower bound of arbitrary x} is valid for any $0< x \leq 1$, the lower bound can be further optimized by properly choosing the value of $x$.
We denote the central quantity in Eq.~\eqref{lower bound of arbitrary x} as
\begin{equation}
    \label{def of F}
    \mathcal{F}_{w}(x) := \frac{1}{(1+x)^{n}} \sum_{u=0}^{w} \binom{n}{u}(3x)^{u},
\end{equation}
and in the following, we determine the optimal $x$ that maximizes $\mathcal{F}_{w}(x)$.

\subsection{Bound on sample complexity}

In this section, by finding an optimal $x$ that maximizes $\mathcal{F}_{w}(x)$ in Eq.~\eqref{def of F}, we complete the proof of Theorem 2 in the main text.
To analyze the sum $\sum_{u=0}^{w} \binom{n}{u}(3x)^{u}$, we first determine the value of $u=\tilde{u}$ that maximizes $\binom{n}{u}(3x)^{u}$.
Since the quantity $\binom{n}{u}(3x)^{u}$ is related to the binomial distribution as
\begin{equation}
    \frac{1}{(1+3x)^{n}} \binom{n}{u} (3x)^{u}
    = \binom{n}{u} \left(\frac{3x}{1+3x}\right)^{u} \left(\frac{1}{1+3x}\right)^{n-u},
\end{equation}
we find that $\tilde{u}=\lfloor n\frac{3x}{1+3x} \rfloor $ or $\tilde{u}=\lceil n\frac{3x}{1+3x} \rceil $, which is the median of the binomial distribution.

Then, we observe that $\mathcal{F}_{w}(x)$ is given by a sum over $0 \leq u \leq w$ according to Eq.~\eqref{def of F}.
When $w \geq \tilde{u}$, the term corresponding to $u = \tilde{u}$ is included in the summation.
However, if $w < \tilde{u}$, the term $\binom{n}{\tilde{u}} (3x)^{\tilde{u}}$ is absent from the summation $\sum_{u=0}^{w} \binom{n}{u}(3x)^{u}$. 
In this case, the largest term in $\sum_{u=0}^{w} \binom{n}{u}(3x)^{u}$ is $\binom{n}{w}(3x)^{w}$, occurring at $u=w$, as $\binom{n}{u}(3x)^{u}$ is a monotonically increasing function in the range $0 \leq u \leq \tilde{u}$.
Thus, depending on whether $w \geq n\frac{3x}{1+3x} $ or $w < n\frac{3x}{1+3x}$, the sum $\sum_{u=0}^{w} \binom{n}{u}(3x)^{u}$ can be expressed differently as
\begin{equation}
\label{bound on binomial summation}
\begin{aligned}
    \binom{n}{\tilde{u}} (3x)^{\tilde{u}}
    \leq \sum_{u=0}^{w} \binom{n}{u}(3x)^{u}
    \leq w \binom{n}{\tilde{u}} (3x)^{\tilde{u}}
    & ~~ \text{for}~~ w \geq n\frac{3x}{1+3x}, \\
    \binom{n}{w}(3x)^{w}
    \leq \sum_{u=0}^{w} \binom{n}{u}(3x)^{u}
    \leq w\binom{n}{w}(3x)^{w}
    & ~~ \text{for}~~ w < n\frac{3x}{1+3x}. 
\end{aligned}
\end{equation}

Additionally, to handle the binomial coefficients, we employ Stirling's formula:
\begin{equation}
    \label{Stirling approx}
    \sqrt{2\pi n}\left(\frac{n}{e}\right)^{n} e^{\frac{1}{12n+1}}
    \leq n!
    \leq \sqrt{2\pi n}\left(\frac{n}{e}\right)^{n} e^{\frac{1}{12n}}, \quad
    \frac{\exp[n H(\frac{u}{n})]}
    { \sqrt{8u \left(1-\frac{u}{n}\right)} }
    \leq \binom{n}{u}
    \leq \frac{\exp[n H(\frac{u}{n})]}
    {\sqrt{2\pi u \left(1-\frac{u}{n}\right)} }.
\end{equation}
Since an inequality $ 1-\frac{1}{n} \leq u\left(1-\frac{u}{n}\right) \leq \frac{n}{4}$ holds for any $1\leq u\leq n-1$, for $n\geq 2$, we have $ \frac{1}{2} \leq u\left(1-\frac{u}{n}\right) \leq \frac{n}{4} $.
Thus, for $n\geq 2$, the binomial coefficient is bounded as follows:
\begin{equation}
    \label{lower bound on binom}
    \frac{1}{\sqrt{2n}} \exp \left[n H \left( \frac{u}{n} \right) \right]
    \leq \binom{n}{u}
    \leq \frac{1}{\sqrt{\pi}} \exp \left[n H \left( \frac{u}{n} \right) \right].
\end{equation}
Applying Eq.~\eqref{lower bound on binom}, we find
\begin{equation}
    \label{bound on u tilde}
    \frac{1}{\sqrt{2n}} (1+3x)^{n}
    \leq \binom{n}{\tilde{u}} (3x)^{\tilde{u}}
    \leq \frac{1}{\sqrt{\pi}} (1+3x)^{n}.
\end{equation}
Therefore, by applying Eq.~\eqref{bound on u tilde} to Eq.~\eqref{bound on binomial summation}, we obtain the following inequalities for $\mathcal{F}_{w}(x)$:
\begin{equation}
\begin{aligned}
    \label{sum of binomial distribution}
    \frac{1}{\sqrt{2n}} \left(\frac{1+3x}{1+x}\right)^{n}
    \leq \mathcal{F}_{w}(x)
    \leq \frac{w}{\sqrt{\pi}} \left(\frac{1+3x}{1+x}\right)^{n}
    &  ~~ \text{for}~~  w \geq n\frac{3x}{1+3x}, \\ 
    \binom{n}{w} \frac{(3x)^{w}}{(1+x)^{n}}
    \leq \mathcal{F}_{w}(x)
    \leq w \binom{n}{w} \frac{(3x)^{w}}{(1+x)^{n}}
    &  ~~ \text{for}~~  w < n\frac{3x}{1+3x}. 
\end{aligned}
\end{equation}

According to Eq.~\eqref{sum of binomial distribution}, to find the value of $x$ that maximizes $\mathcal{F}_{w}(x)$, we separately consider the two cases:
\begin{enumerate}
    \item $w \geq n\frac{3x}{1+3x}$ ($x \leq \frac{w/n}{3(1-w/n)}$) :
    The quantity $\left(\frac{1+3x}{1+x}\right)^{n}$ is a monotonically increasing function of $x$. Thus, the maximum occurs at the boundary value $x = \tilde{x}_{1} = \frac{w/n}{3(1-w/n)}$ ($w = n\frac{3\tilde{x}_{1}}{1+3\tilde{x}_{1}}$). At this optimal value, we have 
    \begin{equation}
        \frac{1}{\sqrt{2n}} \left(1-\frac{2}{3}\frac{w}{n}\right)^{-n}
        \leq \mathcal{F}_{w}(\tilde{x}_{1})
        \leq \frac{w}{\sqrt{\pi}} \left(1-\frac{2}{3}\frac{w}{n}\right)^{-n}.
    \end{equation}
    \item $w < n\frac{3x}{1+3x}$ ($x > \frac{w/n}{3(1-w/n)}$) :
    Using Eq.~\eqref{sum of binomial distribution}, the quantity $\mathcal{F}_{w}(x)$ satisfies
    \begin{equation}
        \frac{1}{\sqrt{2n}} \exp \left[n H \left( \frac{w}{n} \right) \right]
        \times 3^{w} \times \left(\frac{x^{w/n}}{1+x}\right)^{n}
        \leq \mathcal{F}_{w}(x)
        \leq w \times \frac{1}{\sqrt{\pi}} \exp \left[n H \left( \frac{w}{n} \right) \right]
        \times 3^{w} \times \left(\frac{x^{w/n}}{1+x}\right)^{n}.
    \end{equation}
    Maximizing this expression with respect to $x$ yields 
    \begin{equation}
        \max_{x} \frac{x^{w/n}}{1+x} = \exp \left[ - H \left( \frac{w}{n} \right) \right]
        \text{ when } x = \tilde{x}_{2} = \frac{w/n}{1-w/n}.
    \end{equation}
    Thus, we obtain the following bound:
    \begin{equation}
        \frac{1}{\sqrt{2n}} 3^{w}
        \leq \mathcal{F}_{w}(\tilde{x}_{2})
        \leq \frac{1}{\sqrt{\pi}} w \times 3^{w}.
    \end{equation}
\end{enumerate}
Since $\left(1-\frac{2}{3}\frac{w}{n}\right)^{-1} \leq 3^{w/n}$ holds for $0<w/n\leq 1$, we have $\mathcal{F}_{w}(\tilde{x}_{1}) \leq \mathcal{F}_{w}(\tilde{x}_{2})$.
Thus, it suffices to consider the second case, as taking $x=\tilde{x}_{2} = \frac{w/n}{1-w/n}$ gives the maximum value of $\mathcal{F}_{w}(x)$.
At $w/n = 1/2 $, the optimal $\tilde{x}_{2}$ reaches its maximal value $\tilde{x}_{2} = 1$.
Consequently, for all  $w > n/2$, the optimal choice is fixed to be $x = \tilde{x}_{2} = 1$.
Thus, we find the bound on $\mathcal{F}_{w}(x)$ to be
\begin{equation}
    \mathcal{F}_{w}(x) =
    \begin{cases}
        \Omega(\frac{1}{\sqrt{n}}3^{w}) & w \leq n/2~\text{at } x=\frac{w/n}{1-w/n} \\
        \Omega\left( \frac{\sum_{u=0}^{w} \binom{n}{u}3^{u}}{2^{n}} \right) & w > n/2~\text{at } x=1 \\
    \end{cases}
\end{equation}
Finally, from Eqs.~\eqref{lower bound of arbitrary x} and~\eqref{def of F}, we derive that the lower bound on the sample complexity is given by
\begin{equation}
    \label{final lower bound on N}
    N =
    \begin{cases}
        \Omega \left( \frac{1}{\sqrt{n}}2^{-k} 3^{w} \times \varepsilon^{-2}(1-2\delta) \right) & w \leq n/2 \\ 
        \Omega \left(
            2^{-k} \frac{\sum_{u=0}^{w} \binom{n}{u}3^{u}}{2^{n}}
            \times \varepsilon^{-2}(1-2\delta)  
        \right) & w > n/2
    \end{cases}.
\end{equation}
which is Theorem 2 in the main text.
Note that in the regime $w \geq \frac{3}{4}n$, the quantity $\sum_{u=0}^{w} \binom{n}{u}3^{u}$ is $\Omega(4^{n})$.
Thus, our results improve the previously established lower bound $N = \Omega(2^{(n-k)/3})$ to $N = \Omega(2^{n-k})$ for the case $w=n$~\cite{chen2022quantum}.

\subsection{Proposal for a sharper lower bound on the sample complexity}
\label{Sec: S3E}

Here, we present an expectation for a tighter lower bound on the sample complexity.
As indicated in Eq.~\eqref{TVD and EG}, the crucial step in deriving a lower bound on $N$ is obtaining a upper bound on $\underset{\mathbf{e}}{\mathds{E}} (G_{\mathbf{e}}^{\mathbf{o}_{\leq l}})^{2}$, whose current form is given in Eq.~\eqref{upper bound on EG}. 
Nevertheless, this upper bound remains a loose approximation.
In the limit $x \rightarrow 0$, the probability distribution in Eq.~\eqref{power law Pr(e)} converges to $\Pr(\mathbf{e}) = \delta_{\mathbf{e}, \mathbf{0}}$, at which point $\underset{\mathbf{e}}{\mathds{E}} (G_{\mathbf{e}}^{\mathbf{o}_{\leq l}})^{2}$ saturates to $1$, whereas the right-hand side of Eq.~\eqref{upper bound on EG} converges to $2^k$.
The reason for this gap is as follows: according to Eq.~\eqref{EG bounded by sum of Pe}, the upper bound $2^{k} \left(\frac{1+x}{1+3x}\right)^{n}$ is attained when $CC^{\dagger} = \bigotimes_{j=1}^{n} |\gamma_{j}\rangle \langle \gamma_{j}|$, where $|\gamma_{j}\rangle \langle \gamma_{j}|$ represents the state of the $j$-th qubit, and for each $j$, $|\gamma_{j} \rangle$ is an eigenstate of one of the Pauli matrices $X$, $Y$, or $Z$.
However, for such a choice of $C$, the equality condition in Eq.~\eqref{bound on PeCPeC} cannot be satisfied. 
Therefore, although we cannot prove it, we suggest a sharper upper bound on $\underset{\mathbf{e}}{\mathds{E}} (G_{\mathbf{e}}^{\mathbf{o}_{\leq l}})^{2}$ as follows:
\begin{equation}
    \label{conjecture for EG}
    \text{Expectation: }
    \max \underset{\mathbf{e}}{\mathds{E}} (G_{\mathbf{e}}^{\mathbf{o}_{\leq l}})^{2}
    = \max_{C: \mathrm{rank}(C) \leq 2^{k}}
    \underset{\mathbf{e}}{\mathds{E}}
    \frac{\Tr^{2}(P_\mathbf{e} C P_\mathbf{e} C^{\dagger})}{\Tr^{2}(CC^{\dagger})}
    = \left(\frac{1+x}{1+3x}\right)^{n-k},
\end{equation}
and the equality holds when $C = \left( \bigotimes_{j=1}^{n-k} |\gamma_{j}\rangle \langle \gamma_{j}| \right) \otimes I_{k}$, where $I_{k}$ denotes the identity operator acting on the $k$ qubits.

If our expectation in Eq.~\eqref{conjecture for EG} holds, a tighter lower bound on $N$ than the one in Eq.~\eqref{lower bound of arbitrary x} can be derived:
\begin{equation}
    \label{tighter lower bound of arbitrary x}
    N \geq \frac{1-2\delta}{4 \varepsilon^{2}}
    \left(\frac{1+3x}{1+x}\right)^{n-k} \Pr( |\mathbf{e}| \leq w).
\end{equation}
From Eq.~\eqref{tighter lower bound of arbitrary x}, taking $x$ such that $\frac{w}{n} = \frac{3x}{1+3x}$ leads to a lower bound on $N$ as
\begin{equation}
    N \geq \frac{1-2\delta}{8 \varepsilon^{2}} \left(\frac{1}{1-\frac{2}{3}\frac{w}{n}}\right)^{n-k},
\end{equation}
since the probability $\Pr( |\mathbf{e}| \leq w)$ becomes $1/2$ according to Eq.~\eqref{prob for learnable channel}.
It implies that when $k$ and $w$ scale proportionally with $n$, if $k<n$, an exponential number of channel queries is required regardless of $w$ value; i.e., the boundary in Fig.~2 of the main text saturates at $k/n=1$.
Consequently, we can argue that a sufficiently large $k$ is even more crucial than our Theorem 2 in the main text.

We find that even with the optimized $x$ in Eq.~\eqref{tighter lower bound of arbitrary x} in the same spirit as the previous section, the resulting lower bound on $N$ remains below the upper bound in Sec.~\hyperref[Sec: S5]{S5} (see below).
Accordingly, closing the gap likely requires new ideas that sharpen both the upper and lower bounds.


\section{Proof of Theorem 3}
\label{Sec: S4}

In this section, we establish the upper bound on the sample complexity for ($\varepsilon, \delta, w$)-Pauli channel learning in the case $k=0$, i.e., no ancilla qubits are allowed.
More precisely, we derive an upper bound on the sample complexity $N$ by employing the concepts of stabilizer covering~\cite{flammia2020efficient, chen2022quantum} and the theory of covering arrays~\cite{johnson1974approximation, lovasz1975ratio, stein1974two, sarkar2017upper}.
Furthermore, we provide an algorithm that achieves the derived upper bound based on the density-based greedy algorithm~\cite{bryce2007density, bryce2009densitybased}.
Finally, we show that the derived upper bound matches the lower bound for the $k=0$ case in Theorem 2 in the main text.

As a preliminary step, we introduce the concept of stabilizer covering, as formulated in~\cite{flammia2020efficient, chen2022quantum}.
\begin{definition}[Stabilizer covering]
    \label{def: SC}
    \normalfont
    Let $\mathsf{C} = \{\mathsf{S}_{i}\}_{i}$ be a set whose elements are stabilizer groups $\mathsf{S}_{i}$, and each $\mathsf{S}_{i}$ is generated by $n$ independent group generators.
    For a given set of Pauli strings $\mathsf{P}$, we define $\mathsf{C}$ as a \textit{stabilizer covering of $\mathsf{P}$} if it satisfies
    \begin{equation}
        \label{def of SC}
        \mathsf{P} \subseteq \bigcup_{\mathsf{S}_{i} \in \mathsf{C}} \mathsf{S}_{i},
    \end{equation}
    and this relationship is denoted by $\mathsf{C} \overset{\text{SC}}{\triangleright} \mathsf{P}$.
    Additionally, if $\mathsf{C} \overset{\text{SC}}{\triangleright} \mathsf{P}$ holds, we say that $\mathsf{C}$ \textit{covers} $\mathsf{P}$; equivalently, $\mathsf{P}$ is \textit{covered by} $\mathsf{C}$.
\end{definition}
\noindent
For any given set $\mathsf{P}$, a stabilizer covering exists, although it is generally not unique.
Thus, we define $\text{CN}(\mathsf{P})$ as the minimum size $|\mathsf{C}|$ among all stabilizer coverings $\mathsf{C} \overset{\text{SC}}{\triangleright} \mathsf{P}$:
\begin{equation}
    \text{CN}(\mathsf{P}) := \min_{\mathsf{C} : \mathsf{C} \overset{\text{SC}}{\triangleright} \mathsf{P}} |\mathsf{C}|.
\end{equation}


It is known that the stabilizer covering defined above provides a direct upper bound on the sample complexity $N$. 
Specifically, to estimate all $\lambda(\mathbf{b})$ associated with the Pauli strings $P_{\mathbf{b}} \in \mathsf{P}$ within an error $\varepsilon$, with success probability at least $1-\delta$ (see Definition 1 in the main text), the required number of samples $N$ is given by~\cite{chen2022quantum}
\begin{equation}
    \label{N for CN}
    N = O\left( n \times \text{CN}(\mathsf{P}) \times \varepsilon^{-2} \log \delta^{-1} \right).
\end{equation}
The proof is as follows:
for the given set $\mathsf{P}$, let $\mathsf{C} = \{\mathsf{S}_{i}\}_{i}$ be a stabilizer covering of $\mathsf{P}$.
The use of stabilizer covering exhibits two key properties.
\begin{enumerate}
    \item Stabilizer groups enable ``simultaneous'' estimation of all $\lambda(\mathbf{b})$ such that $P_{\mathbf{b}} \in \mathsf{S}_{i}$ for each $\mathsf{S}_{i} \in \mathsf{C}$.
    In particular, the estimation of such Pauli eigenvalues $\lambda(\mathbf{b})$ employs the input state $|\Psi^{\mathsf{S}_{i}}\rangle$ and the POVM $\{ |\Psi^{\mathsf{S}_{i}}_{ \mathbf{v}^{(i)} }\rangle \langle\Psi^{\mathsf{S}_{i}}_{ \mathbf{v}^{(i)} }| \}_{ \mathbf{v}^{(i)} }$, defined as
    \begin{equation}
        \label{stabilizer input and POVM}
        |\Psi^{\mathsf{S}_{i}}\rangle \langle\Psi^{\mathsf{S}_{i}}|
        := \frac{1}{2^{n}} \sum_{\mathbf{a} \in {\mathsf{S}_{i}}} P_{\mathbf{a}},~~
        |\Psi^{\mathsf{S}_{i}}_{ \mathbf{v}^{(i)} }\rangle \langle\Psi^{\mathsf{S}_{i}}_{ \mathbf{v}^{(i)} }|
        := \frac{1}{2^{n}} \sum_{\mathbf{a} \in {\mathsf{S}_{i}} }
        (-1)^{\langle \mathbf{v}^{(i)}, \mathbf{a} \rangle} P_{\mathbf{a}},
    \end{equation}
    where $\mathbf{v}^{(i)} \in \mathds{Z}_{2}^{2n}/\mathsf{S}_{i}$ is the error syndrome~\cite{chen2022quantum}.
    Using the input and POVM defined in Eq.~\eqref{stabilizer input and POVM}, estimating all $\lambda(\mathbf{b})$ such that $P_{\mathbf{b}} \in \mathsf{S}_{i}$ can be accomplished within $O\left( n \times \varepsilon^{-2} \log \delta^{-1} \right)$ samples~\cite{chen2022quantum}.
    \item Each element in $\mathsf{P}$ is contained in at least one $\mathsf{S}_{i}$, by the definition of the stabilizer covering [Eq.~\eqref{def of SC}].
    Thus, performing the above estimation for each $\mathsf{S}_{i} \in \mathsf{C}$ allows us to estimate all $\lambda(\mathbf{b})$ such that $P_{\mathbf{b}} \in \mathsf{P}$.
\end{enumerate}
Therefore, by using the given stabilizer covering $\mathsf{C}$ of $\mathsf{P}$, the estimation of all the parameters in $\mathsf{P}$ requires $O\left(|\mathsf{C}| \times n \times \varepsilon^{-2} \log \delta^{-1} \right)$ samples.
By selecting $\mathsf{C}$ to be a stabilizer covering of minimum size, i.e., with $|\mathsf{C}| = \text{CN}(\mathsf{P})$, the upper bound in Eq.~\eqref{N for CN} follows.



However, determining the exact value of $\text{CN}(\mathsf{P})$ for arbitrary $\mathsf{P}$ is generally NP-hard~\cite{verteletskyi2020measurement}.
Therefore, our goal is to derive the tightest possible upper bound on $\text{CN}(\mathsf{P})$. 
In the following section, we derive this upper bound using the theory of covering arrays, combined with probabilistic methods
\cite{johnson1974approximation, lovasz1975ratio, stein1974two, sarkar2017upper}.
Additionally, by adapting the density-based greedy algorithm~\cite{bryce2007density, bryce2009densitybased}, we provide an algorithm for constructing a set $\mathsf{C}$ whose size $|\mathsf{C}|$ matches the upper bound we derive.
Finally, we combine the result of Eq.~\eqref{N for CN} with the established upper bound on $\mathrm{CN}(\mathsf{P})$, and apply it to the ($\varepsilon, \delta, w$)-Pauli channel learning task.
In doing so, we complete the proof of Theorem 3 in the main text.

\subsection{Upper bound on $\text{CN}(\mathsf{P})$ from the theory of covering array for the case $k=0$}
\label{Sec: S4A}

We present a systematic method for deriving an upper bound on $\text{CN}(\mathsf{P})$.
Our analysis begins by defining a special type of stabilizer covering.
\begin{definition}[Uniform stabilizer covering]
    \label{def: uniform SC}
    \normalfont
    For a given set of Pauli strings $\mathsf{P}$, a stabilizer covering $\mathsf{U}$ is called \textit{uniform} if it satisfies the following two conditions:
    \begin{enumerate}
        \item
        Each stabilizer group $\mathsf{S}_{i} \in \mathsf{U}$ contains exactly $\Sigma$ distinct Pauli strings from $\mathsf{P}$.
        More formally, for all $\mathsf{S}_{i} \in \mathsf{U}$, $|\mathsf{S}_{i} \cap \mathsf{P}| = \Sigma$.
        We refer to $\Sigma$ as the \textit{covering power}.
        
        
        \item
        Every Pauli string in $\mathsf{P}$ appears in exactly $R$ distinct stabilizer groups in $\mathsf{U}$, i.e., for all $P_{\mathbf{a}} \in \mathsf{P}$, $|\{ \mathsf{S}_{i} \in \mathsf{U} : P_{\mathbf{a}} \in \mathsf{S}_{i} \}| = R$.   
        By condition 1, the relation $|\mathsf{U}|\times \Sigma = |\mathsf{P}| \times R$ holds.
        Note that this condition ensures that $\mathsf{U} \overset{\text{SC}}{\triangleright} \mathsf{P}$.
    \end{enumerate}
\end{definition}

Given a uniform stabilizer covering $\mathsf{U}$ for a set $\mathsf{P}$, a smaller stabilizer covering can be obtained by selecting an appropriate subset of $\mathsf{U}$.
Specifically, by extending the theory of covering arrays
\cite{johnson1974approximation, lovasz1975ratio, stein1974two, sarkar2017upper},
we derive the following upper bound on $\text{CN}(\mathsf{P})$:
\begin{lemma}[Upper bound on the minimum size of stabilizer covering]
    \label{lemma: upper bound on CN}
    \normalfont
    For a given set of Pauli strings $\mathsf{P}$, if there exists a uniform stabilizer covering $\mathsf{U}$ with covering power $\Sigma$, $\text{CN}(\mathsf{P})$ is upper bounded by
    \begin{equation}
        \label{CN and Sigma}
        \text{CN}(\mathsf{P})
        \leq \left\lceil \frac{|\mathsf{P}| \log |\mathsf{P}|}{\Sigma} \right\rceil.
    \end{equation}
\end{lemma}

\begin{proof}

For a Pauli string $P_{\mathbf{a}}$ and a set of Pauli strings $\mathsf{S}$, we define an indicator function $NI(P_{\mathbf{a}}; \mathsf{S})$ as
\begin{equation}
    \label{indicator function}
    NI(P_{\mathbf{a}}; \mathsf{S}) = 
    \begin{cases}
        1 & \text{if } P_{\mathbf{a}} \notin \mathsf{S} \\
        0 & \text{if } P_{\mathbf{a}} \in \mathsf{S}.
    \end{cases}
\end{equation}
Using the indicator function, we denote the number of $\mathsf{S}_{i} \in \mathsf{U}$ that do not include $P_{\mathbf{a}} \in \mathsf{P}$ as
\begin{equation}
    \left| \left\{ \mathsf{S}_{i} \in \mathsf{U} : P_{\mathbf{a}} \notin \mathsf{S}_{i} \right\} \right|
    = \sum_{ \mathsf{S}_{i} \in \mathsf{U} } NI( P_{\mathbf{a}}; \mathsf{S}_{i} )
    = |\mathsf{U}|-R,
\end{equation}
since each $P_{\mathbf{a}} \in \mathsf{P}$ is contained in $R$ distinct stabilizer groups in $\mathsf{U}$ (by condition 2 in Definition~\ref{def: uniform SC}).
As a generalization, for a given $P_{\mathbf{a}} \in \mathsf{P}$, if it is not covered by a subset $\mathsf{C} \subseteq \mathsf{U}$ of size $\mathcal{N}$, all $\mathsf{S}_{i} \in \mathsf{C}$ must belong to the set $\{ \mathsf{S}_{i} \in \mathsf{U} : P_{\mathbf{a}} \notin \mathsf{S}_{i} \}$.
Consequently, the number of such subsets is given by
\begin{equation}
    \left|
    \left\lbrace
        \mathsf{C} \subseteq \mathsf{U} ~:~
        |\mathsf{C}| = \mathcal{N},~
        P_{\mathbf{a}} \notin \bigcup_{ \mathsf{S}_{i} \in \mathsf{C} } \mathsf{S}_{i}
    \right\rbrace
    \right|
    = \sum_{ \substack{ \mathsf{C} \subseteq \mathsf{U} \\ |\mathsf{C}| = \mathcal{N} } }
    NI ( P_{\mathbf{a}}; \bigcup_{\mathsf{S}_{i} \in \mathsf{C}} \mathsf{S}_{i} )
    = \binom{ |\mathsf{U}|-R }{ \mathcal{N} }
    \leq ( |\mathsf{U}|-R )^{\mathcal{N}}.
\end{equation}
We now extend this argument to account for the total number of uncovered Pauli strings in $\mathsf{P}$.
For a given subset $\mathsf{C} \subseteq \mathsf{U}$, the number of Pauli strings in $\mathsf{P}$ that are not covered by $\mathsf{C}$ is
\begin{equation}
    \left| 
        \mathsf{P} \setminus \bigcup_{\mathsf{S}_{i} \in \mathsf{C}} \mathsf{S}_{i}
    \right|
    = 
    \sum_{P_{\mathbf{a}} \in \mathsf{P}} 
    NI ( P_{\mathbf{a}}; \bigcup_{\mathsf{S}_{i} \in \mathsf{C}} \mathsf{S}_{i} ).
\end{equation}
Therefore, the expected number of Pauli strings in $\mathsf{P}$ that are not covered by a subset $\mathsf{C} \subseteq \mathsf{U}$ of size $\mathcal{N}$ is evaluated as
\begin{equation}
\label{expectation value of uncovered string}
\begin{aligned}
    \underset{ \substack{ \mathsf{C} \subseteq \mathsf{U} \\ |\mathsf{C}| = \mathcal{N} } }{\mathds{E}}
    \left[
        \left| 
            \mathsf{P} \setminus \bigcup_{\mathsf{S}_{i} \in \mathsf{C}} \mathsf{S}_{i}
        \right|
    \right]
    &= 
    \frac{
        \sum_{ \substack{ \mathsf{C} \subseteq \mathsf{U} \\ |\mathsf{C}| = \mathcal{N} } }
        \left| 
            \mathsf{P} \setminus \bigcup_{\mathsf{S}_{i} \in \mathsf{C}} \mathsf{S}_{i}
        \right|
        }
        {
            \sum_{ \substack{ \mathsf{C} \subseteq \mathsf{U} \\ |\mathsf{C}| = \mathcal{N} } } 1
        }\\
    &= 
    \frac{
        \sum_{ \substack{ \mathsf{C} \subseteq \mathsf{U} \\ |\mathsf{C}| = \mathcal{N} } }
        \sum_{P_{\mathbf{a}} \in \mathsf{P}} 
        NI ( P_{\mathbf{a}}; \bigcup_{\mathsf{S}_{i} \in \mathsf{C}} \mathsf{S}_{i} ) 
        }
        { |\mathsf{U}|^{\mathcal{N}} }\\
    &= 
    \frac{
        \sum_{P_{\mathbf{a}} \in \mathsf{P}} 
        \sum_{ \substack{ \mathsf{C} \subseteq \mathsf{U} \\ |\mathsf{C}| = \mathcal{N} } }        
        NI ( P_{\mathbf{a}}; \bigcup_{\mathsf{S}_{i} \in \mathsf{C}} \mathsf{S}_{i} ) 
        }
        { |\mathsf{U}|^{\mathcal{N}} }\\
    &\leq \frac{ |\mathsf{P}| (|\mathsf{U}|-R)^{\mathcal{N}} }{ |\mathsf{U}|^{\mathcal{N}} } \\
    &= |\mathsf{P}|\left(1- \frac{\Sigma}{|\mathsf{P}|}\right)^{\mathcal{N}}
    ~(\because~|\mathsf{U}|\times \Sigma = |\mathsf{P}| \times R).
\end{aligned}
\end{equation}
The number of uncovered Pauli strings is a nonnegative integer.
Therefore, if $\mathcal{N}$ is sufficiently large that the expected number in Eq.~\eqref{expectation value of uncovered string} is less than 1, then there exists at least one size-$\mathcal{N}$ subset of $\mathsf{U}$ with zero uncovered Pauli strings.
More explicitly, if $\mathcal{N}$ satisfies $|\mathsf{P}|\left(1- \frac{\Sigma}{|\mathsf{P}|}\right)^{\mathcal{N}} < 1$, there exists a stabilizer covering of size $\mathcal{N}$ constructed from $\mathsf{U}$.
From the inequality $|\mathsf{P}|\left(1- \frac{\Sigma}{|\mathsf{P}|}\right)^{\mathcal{N}} \leq |\mathsf{P}| \exp(-\frac{\Sigma}{|\mathsf{P}|} \mathcal{N})$, we find that an integer $\mathcal{N} = \left\lceil \frac{|\mathsf{P}| \log |\mathsf{P}|}{\Sigma} \right\rceil$ is sufficient to make the expected number of uncovered Pauli strings less than 1.
Since a stabilizer covering of size $\mathcal{N} = \left\lceil \frac{|\mathsf{P}| \log |\mathsf{P}|}{\Sigma} \right\rceil$ exists, we obtain the upper bound on $\mathrm{CN}(\mathsf{P})$ stated in Eq.~\eqref{CN and Sigma}.

\end{proof}

We can find a stabilizer covering with the size given by Eq.~\eqref{CN and Sigma} via the density-based greedy algorithm~\cite{bryce2007density, bryce2009densitybased}.
The procedure begins with the empty set $\mathsf{C}_{\text{greedy}} = \varnothing$, and we define the initial set of Pauli strings uncovered by $\mathsf{C}_{\text{greedy}}$ as $\mathsf{P}_{0} := \mathsf{P}$. 
We then iteratively select stabilizer groups from $\mathsf{U}$ and add to $\mathsf{C}_{\text{greedy}}$ until all Pauli strings in $\mathsf{P}$ are covered by $\mathsf{C}_{\text{greedy}}$.
We denote by $\mathsf{P}_{j}$ the set of uncovered Pauli strings remaining after choosing $j$ stabilizer groups.
The key idea is that, at each iteration $j$, we select a stabilizer group that covers at least $|\mathsf{P}_{j}| \times \frac{R}{|\mathsf{U}|}$ of the currently uncovered Pauli strings~\cite{bryce2007density, bryce2009densitybased}.
The existence of such a stabilizer group is justified as follows:
(1) At iteration $j$, each of the $|\mathsf{P}_{j}|$ uncovered Pauli strings is contained in $R$ distinct groups (by condition 2 in Definition~\ref{def: uniform SC}).
(2) Hence, the total number of uncovered Pauli strings across all stabilizer groups in $\mathsf{U}$ is $|\mathsf{P}_{j}| \times R$, so the average number of uncovered Pauli strings per group is $|\mathsf{P}_{j}| \times \frac{R}{|\mathsf{U}|}$.
(3) Consequently, there exists a stabilizer group that contains at least $|\mathsf{P}_{j}| \times \frac{R}{|\mathsf{U}|}$ uncovered Pauli strings.
By choosing such a stabilizer group in every iteration, the number of uncovered Pauli strings is reduced as
\begin{equation}
    \label{greedy algorithm}
    |\mathsf{P}_{j+1}|
    \leq |\mathsf{P}_{j}| - |\mathsf{P}_{j}| \times \frac{R}{|\mathsf{U}|}
    = |\mathsf{P}_{j}|\left( 1- \frac{\Sigma}{|\mathsf{P}|}\right)
    ~~(\because~|\mathsf{U}|\times \Sigma = |\mathsf{P}| \times R).
\end{equation}
According to Eq.~\eqref{greedy algorithm}, after $\mathcal{N}$ iterations, $|\mathsf{P}_{\mathcal{N}}|$ satisfies
\begin{equation}
    |\mathsf{P}_{\mathcal{N}}|
    \leq |\mathsf{P}_{0}| \left( 1- \frac{\Sigma}{|\mathsf{P}|}\right)^{\mathcal{N}}
    = |\mathsf{P}| \left( 1- \frac{\Sigma}{|\mathsf{P}|}\right)^{\mathcal{N}}.
\end{equation}
Since $|\mathsf{P}_{\mathcal{N}}|$ is an integer, the condition $|\mathsf{P}_{\mathcal{N}}| < 1$ implies that $|\mathsf{P}_{\mathcal{N}}|=0$, and the iteration halts.
Therefore, this algorithm offers a more practical method for finding a stabilizer covering than checking all possible combinations of stabilizer groups.

To apply the upper bound given in Eq.~\eqref{N for CN} for the ($\varepsilon, \delta, w$)-Pauli channel learning task, we focus on the set $\mathsf{P}(w)$ consisting of weight-$w$ Pauli strings.
More precisely, the set $\mathsf{P}(w)$ is defined as
\begin{equation}
    \label{def of Pw}
    \mathsf{P}(w) := \{ P_{\mathbf{a}} ~:~ |P_{\mathbf{a}}| = w \},
\end{equation}
and its size is $|\mathsf{P}(w)| = \binom{n}{w}3^{w}$.
In the following section, we construct a uniform stabilizer covering for the set $\mathsf{P}(w)$.
Based on this construction, we determine the covering power as a function of $n$ and $w$, and we find an upper bound on $\mathrm{CN}(\mathsf{P}(w))$ via Eq.~\eqref{CN and Sigma}.
Since the behavior of the covering power changes at the threshold $w=n/2$ (see the following sections), we analyze the two regimes $w \leq n/2$ and $w>n/2$ separately.

\subsubsection{Upper bound on $\mathrm{CN}(\mathsf{P}(w))$ for $k=0$, $w \leq n/2$ case}


First, we consider the case $k=0$ with $w \leq n/2$, and we find a uniform stabilizer covering $\mathsf{U}^{ (w\leq n/2) }$. 
To construct $\mathsf{U}^{ (w\leq n/2) }$, we consider a tuple $\mathbb{G} \in \{X, Y, Z\}^{n}$, where $\{X, Y, Z\}^{n}$ denotes the set of all tuples consisting of $n$ non-identity Pauli operators.
The elements of the tuple are denoted by $\mathbb{G} = (G_{1}, G_{2}, \ldots, G_{n})$, where each $G_{j}$ is a non-trivial Pauli operator.
Then, for a given $\mathbb{G}$, we define a set $\mathsf{G}^{(n)}(\mathbb{G})$ as 
\begin{equation}
    \label{k=0 generator for Gn}
    \mathsf{G}^{(n)}(\mathbb{G}) := \{ \mathbf{g}^{(1)}, \mathbf{g}^{(2)}, \ldots, \mathbf{g}^{(n)} \},
    ~~ g_{i}^{(j)} =
    \begin{cases}
        G_{j} & i=j \\
        I & \text{otherwise}
    \end{cases}
\end{equation}
As shown in Eq.~\eqref{k=0 generator for Gn}, every element $\mathbf{g}^{(j)} \in \mathsf{G}^{(n)}(\mathbb{G})$ is a weight-$1$ Pauli string such that it acts as operator $G_{j}$ on the $j$-th qubit and as the identity operator $I$ on all other qubits.
In Fig.~\ref{illust of stabilizer group k=0 w<n/2}, we illustrate the set $\mathsf{G}^{(n)}(\mathbb{G})$.

\begin{figure}[h]
\begin{tikzpicture}
    \def\w{1.6}
    \def\h{0.5}
    \def\hmargin{0.2}
    \def\wmargin{0.0}
    \foreach \i in {0,...,3}
    {
        \pgfmathsetmacro\x{\i*\w + 0.8}
        \draw[fill=green!15] (\x,0) rectangle ++(\w,0.8);
    }
    \node at (1*\w+\wmargin,0.4) {$1$};
    \node at (2*\w+\wmargin,0.4) {$2$};
    \node at (3*\w+\wmargin,0.4) {$\cdots$};
    \node at (4*\w+\wmargin,0.4) {$n$};
    \node at (0.5, -1*\h+\hmargin) {$\mathbf{g}^{(1)}$:};
    \node at (0.5, -2*\h+\hmargin) {$\mathbf{g}^{(2)}$:};
    \node at (0.5, -3*\h+\hmargin+0.1) {$\vdots$};
    \node at (0.5, -4*\h+\hmargin) {$\mathbf{g}^{(n)}$:};
    \node at (1*\w+\wmargin, -1*\h+\hmargin) {$G_{1}$};
    \node at (2*\w+\wmargin, -1*\h+\hmargin) {$I$};
    \node at (3*\w+\wmargin, -1*\h+\hmargin) {$\cdots$};
    \node at (4*\w+\wmargin, -1*\h+\hmargin) {$I$};
    \node at (1*\w+\wmargin, -2*\h+\hmargin) {$I$};
    \node at (2*\w+\wmargin, -2*\h+\hmargin) {$G_{2}$};
    \node at (3*\w+\wmargin, -2*\h+\hmargin) {$\cdots$};
    \node at (4*\w+\wmargin, -2*\h+\hmargin) {$I$};
    \node at (1*\w+\wmargin, -4*\h+\hmargin) {$I$};
    \node at (2*\w+\wmargin, -4*\h+\hmargin) {$I$};
    \node at (3*\w+\wmargin, -4*\h+\hmargin) {$\cdots$};
    \node at (4*\w+\wmargin, -4*\h+\hmargin) {$G_{n}$};
\end{tikzpicture}
\caption{
Illustration of the set $\mathsf{G}^{(n)}(\mathbb{G})$.
Each number in the box labels a qubit. 
}
\label{illust of stabilizer group k=0 w<n/2}
\end{figure}

Since all Pauli strings in the constructed set $\mathsf{G}^{(n)}(\mathbb{G})$ mutually commute, we define a stabilizer group $\mathsf{S}^{(n)}(\mathbb{G})$ as
\begin{equation}
    \label{k=0 w<n/2 stabilizer group Sn}
    \mathsf{S}^{(n)}(\mathbb{G})
    := \left\langle \mathsf{G}^{(n)}(\mathbb{G}) \right\rangle.
\end{equation}
Then, the definition of $\mathsf{U}^{ (w\leq n/2) }$ is given by the collection of $\mathsf{S}^{(n)}(\mathbb{G})$ for all $\mathbb{G} \in \{X, Y, Z\}^{n}$:
\begin{equation}
    \mathsf{U}^{ (w\leq n/2) }
    := \{ \mathsf{S}^{(n)}(\mathbb{G}) ~:~ \mathbb{G} \in \{X, Y, Z\}^{n} \}.
\end{equation}
By definition, the set $\mathsf{U}^{ (w\leq n/2) }$ has the same size as $\{X, Y, Z\}^{n}$, i.e., $|\mathsf{U}^{ (w\leq n/2) }| = 3^{n}$.


It is straightforward to verify that the set $\mathsf{U}^{ (w\leq n/2) }$ is a uniform stabilizer covering of $\mathsf{P}(w)$.
Each stabilizer group $\mathsf{S}^{(n)}(\mathbb{G})$ contains exactly $\binom{n}{w}$ distinct weight-$w$ Pauli strings, since every weight $w$-Pauli string in $\mathsf{S}^{(n)}(\mathbb{G})$ is generated by multiplying $w$ distinct $\mathbf{g}^{(j)} \in \mathsf{G}^{(n)}(\mathbb{G})$. 
Namely, the covering power $\Sigma(w)$ is given by
\begin{equation}
    \label{Sigma for k=0 w<n/2}
    \Sigma(w) = \binom{n}{w},
    ~~\text{for}~\mathsf{U}^{ (w\leq n/2) }.
\end{equation}
As $\mathsf{U}^{ (w\leq n/2) }$ accounts for all possible tuples $\mathbb{G}$, it covers the entire set of weight-$w$ Pauli strings.
Furthermore, because there is no bias in selecting any particular $\mathbb{G}$ for $\mathsf{S}^{(n)}(\mathbb{G})$, each weight-$w$ Pauli string appears in the same number of stabilizer groups.
Hence, the second condition is satisfied.

Therefore, by using the obtained covering power in Eq.~\eqref{Sigma for k=0 w<n/2} along with $|\mathsf{P}(w)| = \binom{n}{w}3^{w}$ and Eq.~\eqref{CN and Sigma}, we derive
\begin{equation}
    \text{CN}(\mathsf{P}(w)) = O( n 3^{w} ), \quad
    \text{CN}(\bigcup_{u=0}^{w} \mathsf{P}(u)) \leq 
    \sum_{u=0}^{w} \text{CN}(\mathsf{P}(u)) = O( n 3^{w} ),
\end{equation}
where we have $\log(|\mathsf{P}(w)|) = O(n)$ by Stirling's formula, Eq.~\eqref{Stirling approx}.
From Eq.~\eqref{N for CN}, we obtain the upper bound on the sample complexity for the case $k=0$, $w \leq n/2$ as
\begin{equation}
    \label{upper bound for small w}
    N = O(n^{2} 3^{w} \times \varepsilon^{-2}\log\delta^{-1} ),
\end{equation}
and it completes the proof of the first line of Eq.~(10) in Theorem 3 of the main text.
Since the upper bound Eq.~\eqref{upper bound for small w} coincides with the lower bound given in Eq.~\eqref{final lower bound on N} up to a polynomial factor in $n$, the upper bound we obtain on $\mathrm{CN}(\mathsf{P}(w))$ is also tight.

\subsubsection{Upper bound on $\mathrm{CN}(\mathsf{P}(w))$ for $k=0$, $w > n/2$ case}

For the case $k=0$ with $w > n/2$, we establish a uniform stabilizer covering $\mathsf{U}^{ (w>n/2) }$. 
To describe this construction, we introduce a partition $(\mathcal{A}, \mathcal{B})$ of the set of $n$ qubits $\{1, 2, \ldots, n\}$ such that $|\mathcal{A}| = 2(n-w)$ and $|\mathcal{B}| = 2w-n$.
We denote their elements as $\mathcal{A} = \{ \mathcal{A}_{1}, \mathcal{A}_{2}, \ldots, \mathcal{A}_{2(n-w)} \}$ and $\mathcal{B} = \{ \mathcal{B}_{1}, \mathcal{B}_{2}, \ldots, \mathcal{B}_{2w-n} \}$.
Then, for a given partition $(\mathcal{A}, \mathcal{B})$ with a weight-$n$ Pauli string $\mathbf{g} = g_{1}g_{2}\cdots g_{n}$ where each $g_{j}$ is a non-identity Pauli operator, we define a set $\mathsf{G}^{ (\mathrm{A}) }(\mathbf{g}, \mathcal{A})$ as
\begin{equation}
    \label{k=0 generators for A}
    \mathsf{G}^{ (\mathrm{A}) }(\mathbf{g}, \mathcal{A})
    := \{ \mathbf{a}^{ (1) }, \mathbf{a}^{ (2) }, \ldots, \mathbf{a}^{ (2(n-w)) } \},~~
    a_{i}^{ (j) } =
    \begin{cases}
         g_{ \mathcal{A}_{j} } & i = \mathcal{A}_{j} \\
         I & \text{otherwise}
    \end{cases}
\end{equation}
According to Eq.~\eqref{k=0 generators for A}, the set $\mathsf{G}^{ (\mathrm{A}) }(\mathbf{g}, \mathcal{A})$ consists of $|\mathcal{A}|$ weight-$1$ Pauli strings $\mathbf{a}^{ (j) }$, where each $\mathbf{a}^{ (j) }$ contains exactly one non-identity Pauli operator $g_{\mathcal{A}_{j}}$ acting on the $\mathcal{A}_{j}$-th qubit.
We define another set $\mathsf{G}^{ (\mathrm{B}) }(\mathbf{g}, \mathcal{B})$ as
\begin{equation}
    \label{k=0 generators for B}
    \mathsf{G}^{ (\mathrm{B}) }(\mathbf{g}, \mathcal{B})
    := \{ \mathbf{b}^{ (1) }, \mathbf{b}^{ (2) }, \ldots, \mathbf{b}^{ (2w-n-1) } \},~~
    b_{i}^{ (j) } =
    \begin{cases}
        \mathcal{P}(g_{ \mathcal{B}_{j} }) & i = \mathcal{B}_{j} \\
        \mathcal{P}(g_{ \mathcal{B}_{j+1} }) & i=\mathcal{B}_{j+1} \\
        I & \text{otherwise}
    \end{cases}
\end{equation}
where $\mathcal{P}(X) = Y$, $\mathcal{P}(Y) = Z$, and $\mathcal{P}(Z) = X$. 
As shown in Eq.~\eqref{k=0 generators for B}, the set $\mathsf{G}^{ (\mathrm{B}) }(\mathbf{g}, \mathcal{B})$ consists of $|\mathcal{B}|-1$ weight-$2$ Pauli strings $\mathbf{b}^{ (j) }$ for $j=1, 2, \ldots, 2w-n-1$, where each $\mathbf{b}^{ (j) }$ acts non-trivially only on the $\mathcal{B}_{j}$-th and $\mathcal{B}_{j+1}$-th qubits.
Figure~\ref{illust of stabilizer group k=0 w>n/2} visualizes the sets $\mathsf{G}^{ (\mathrm{A}) }$ and $\mathsf{G}^{ (\mathrm{B}) }$ with the partition $(\mathcal{A}, \mathcal{B})$.

\begin{figure}[H]
\begin{center}
\begin{tikzpicture}
    \def\w{1.6}
    \def\h{0.5}
    \def\hmargin{0.2}
    \def\wmargin{0.0}
    \def\boxmargin{0.8}
    \foreach \i in {0,...,7}
    {
        \pgfmathsetmacro\x{\i*\w + \boxmargin}
        \ifnum\i<4
            \draw[fill=blue!15] (\x,0) rectangle ++(\w,\boxmargin);
        \else
            \draw[fill=red!15] (\x,0) rectangle ++(\w,\boxmargin);
        \fi
    }
    \draw[decorate,decoration={brace,amplitude=5pt},yshift=0.4cm]
    (0*\w+\boxmargin,0.7) -- (4*\w+\boxmargin,0.7) node[midway,yshift=0.4cm] {$|\mathcal{A}| = 2(n-w)$};
    \draw[decorate,decoration={brace,amplitude=5pt},yshift=0.4cm]
    (4*\w+\boxmargin,0.7) -- (8*\w+\boxmargin,0.7) node[midway,yshift=0.4cm] {$|\mathcal{B}| = 2w-n$};
    \node at (1*\w+\wmargin,0.4) {$\mathcal{A}_{1}$};
    \node at (2*\w+\wmargin,0.4) {$\mathcal{A}_{2}$};
    \node at (3*\w+\wmargin,0.4) {$\cdots$};
    \node at (4*\w+\wmargin,0.4) {$\mathcal{A}_{2(n-w)}$};
    \node at (5*\w+\wmargin,0.4) {$\mathcal{B}_{1}$};
    \node at (6*\w+\wmargin,0.4) {$\mathcal{B}_{2}$};
    \node at (7*\w+\wmargin,0.4) {$\cdots$};
    \node at (8*\w+\wmargin,0.4) {$\mathcal{B}_{2w-n}$};
    \node at (0.0, -1*\h+\hmargin) {$\mathbf{g}$:};
    \node at (0.0, -2*\h+\hmargin) {$\mathbf{a}^{ (1) }$:};
    \node at (0.0, -3*\h+\hmargin) {$\mathbf{a}^{ (2) }$:};
    \node at (0.0, -4*\h+\hmargin+0.1) {$\vdots$};
    \node at (0.0, -5*\h+\hmargin) {$\mathbf{a}^{ (2(n-w)) }$:};
    \node at (0.0, -6*\h+\hmargin) {$\mathbf{b}^{ (1) }$:};
    \node at (0.0, -7*\h+\hmargin) {$\mathbf{b}^{ (2) }$:};
    \node at (0.0, -8*\h+\hmargin+0.1) {$\vdots$};
    \node at (0.0, -9*\h+\hmargin) {$\mathbf{b}^{ (2w-n-1) }$:};
    \node at (1*\w+\wmargin, -1*\h+\hmargin) {$g_{ \mathcal{A}_{1} }$};
    \node at (2*\w+\wmargin, -1*\h+\hmargin) {$g_{ \mathcal{A}_{2} }$};
    \node at (3*\w+\wmargin, -1*\h+\hmargin) {$g_{ \mathcal{A}_{\cdots} }$};
    \node at (4*\w+\wmargin, -1*\h+\hmargin) {$g_{ \mathcal{A}_{2(n-w)} }$};
    \node at (5*\w+\wmargin, -1*\h+\hmargin) {$g_{ \mathcal{B}_{1} }$};
    \node at (6*\w+\wmargin, -1*\h+\hmargin) {$g_{ \mathcal{B}_{2} }$};
    \node at (7*\w+\wmargin, -1*\h+\hmargin) {$g_{\mathcal{B}_{\cdots}}$};
    \node at (8*\w+\wmargin, -1*\h+\hmargin) {$g_{ \mathcal{B}_{2w-n} }$};
    \draw (-1.0, -1.5*\h+\hmargin) --  (8.5*\w, -1.5*\h+\hmargin);
    \node at (1*\w+\wmargin, -2*\h+\hmargin) {$g_{ \mathcal{A}_{1} }$};
    \node at (2*\w+\wmargin, -2*\h+\hmargin) {$I$};
    \node at (3*\w+\wmargin, -2*\h+\hmargin) {$\cdots$};
    \node at (4*\w+\wmargin, -2*\h+\hmargin) {$I$};
    \node at (5*\w+\wmargin, -2*\h+\hmargin) {$I$};
    \node at (6*\w+\wmargin, -2*\h+\hmargin) {$I$};
    \node at (7*\w+\wmargin, -2*\h+\hmargin) {$\cdots$};
    \node at (8*\w+\wmargin, -2*\h+\hmargin) {$I$};
    \node at (1*\w+\wmargin, -3*\h+\hmargin) {$I$};
    \node at (2*\w+\wmargin, -3*\h+\hmargin) {$g_{ \mathcal{A}_{2} }$};
    \node at (3*\w+\wmargin, -3*\h+\hmargin) {$\cdots$};
    \node at (4*\w+\wmargin, -3*\h+\hmargin) {$I$};
    \node at (5*\w+\wmargin, -3*\h+\hmargin) {$I$};
    \node at (6*\w+\wmargin, -3*\h+\hmargin) {$I$};
    \node at (7*\w+\wmargin, -3*\h+\hmargin) {$\cdots$};
    \node at (8*\w+\wmargin, -3*\h+\hmargin) {$I$};
    \node at (1*\w+\wmargin, -5*\h+\hmargin) {$I$};
    \node at (2*\w+\wmargin, -5*\h+\hmargin) {$I$};
    \node at (3*\w+\wmargin, -5*\h+\hmargin) {$\cdots$};
    \node at (4*\w+\wmargin, -5*\h+\hmargin) {$g_{ \mathcal{A}_{2(n-w)} }$};
    \node at (5*\w+\wmargin, -5*\h+\hmargin) {$I$};
    \node at (6*\w+\wmargin, -5*\h+\hmargin) {$I$};
    \node at (7*\w+\wmargin, -5*\h+\hmargin) {$\cdots$};
    \node at (8*\w+\wmargin, -5*\h+\hmargin) {$I$};
    \draw (-1.0, -5.5*\h+\hmargin) --  (8.5*\w, -5.5*\h+\hmargin);
    \node at (1*\w+\wmargin, -6*\h+\hmargin) {$I$};
    \node at (2*\w+\wmargin, -6*\h+\hmargin) {$I$};
    \node at (3*\w+\wmargin, -6*\h+\hmargin) {$\cdots$};
    \node at (4*\w+\wmargin, -6*\h+\hmargin) {$I$};
    \node at (5*\w+\wmargin, -6*\h+\hmargin) {$\mathcal{P}(g_{ \mathcal{B}_{1} })$};
    \node at (6*\w+\wmargin, -6*\h+\hmargin) {$\mathcal{P}(g_{ \mathcal{B}_{2} })$};
    \node at (7*\w+\wmargin, -6*\h+\hmargin) {$\cdots$};
    \node at (8*\w+\wmargin, -6*\h+\hmargin) {$I$};
    \node at (1*\w+\wmargin, -7*\h+\hmargin) {$I$};
    \node at (2*\w+\wmargin, -7*\h+\hmargin) {$I$};
    \node at (3*\w+\wmargin, -7*\h+\hmargin) {$\cdots$};
    \node at (4*\w+\wmargin, -7*\h+\hmargin) {$I$};
    \node at (5*\w+\wmargin, -7*\h+\hmargin) {$I$};
    \node at (6*\w+\wmargin, -7*\h+\hmargin) {$\mathcal{P}(g_{ \mathcal{B}_{2} })$};
    \node at (7*\w+\wmargin, -7*\h+\hmargin) {$\mathcal{P}(g_{ \mathcal{B}_{\cdots} })$};
    \node at (8*\w+\wmargin, -7*\h+\hmargin) {$I$};
    \node at (1*\w+\wmargin, -9*\h+\hmargin) {$I$};
    \node at (2*\w+\wmargin, -9*\h+\hmargin) {$I$};
    \node at (3*\w+\wmargin, -9*\h+\hmargin) {$\cdots$};
    \node at (4*\w+\wmargin, -9*\h+\hmargin) {$I$};
    \node at (5*\w+\wmargin, -9*\h+\hmargin) {$I$};
    \node at (6*\w+\wmargin, -9*\h+\hmargin) {$I$};
    \node at (7*\w+\wmargin, -9*\h+\hmargin) {$\mathcal{P}(g_{ \mathcal{B}_{\cdots} })$};
    \node at (8*\w+\wmargin, -9*\h+\hmargin) {$\mathcal{P}(g_{ \mathcal{B}_{2w-n} })$};
\end{tikzpicture}
\caption{
Illustration of $\mathbf{g}$, $\mathsf{G}^{ (\mathrm{A}) }$, and $\mathsf{G}^{ (\mathrm{B}) }$.
Each boxed number indicates the corresponding qubit index.
For simplicity, although we draw $\mathcal{A}$ and $\mathcal{B}$ as contiguous subsets of qubits, any choice of the two subsets is allowed. 
}
\label{illust of stabilizer group k=0 w>n/2}
\end{center}
\end{figure}

By the above construction, for any weight-$n$ Pauli string $\mathbf{g}$ and any partition $(\mathcal{A}, \mathcal{B})$, all Pauli strings in the union $\{ \mathbf{g} \} \cup \mathsf{G}^{ (\mathrm{A}) }(\mathbf{g}, \mathcal{A}) \cup \mathsf{G}^{ (\mathrm{B}) }(\mathbf{g}, \mathcal{B})$ mutually commute.
Thus, we define a stabilizer group $\mathsf{S}^{ (\mathrm{A,B}) }( \mathbf{g}, (\mathcal{A}, \mathcal{B}) )$ as
\begin{equation}
    \mathsf{S}^{ (\mathrm{A,B}) }( \mathbf{g}, (\mathcal{A}, \mathcal{B}) )
    := 
    \left\langle
    \{ \mathbf{g} \}
    \cup \mathsf{G}^{ (\mathrm{A}) }(\mathbf{g}, \mathcal{A})
    \cup \mathsf{G}^{ (\mathrm{B}) }(\mathbf{g}, \mathcal{B})
    \right\rangle.
\end{equation}
Note that $\mathsf{S}^{ (\mathrm{A,B}) }( \mathbf{g}, (\mathcal{A}, \mathcal{B}) )$ is generated by a total of $n$ generators, since $1 + |\mathsf{G}^{ (\mathrm{A}) }| + |\mathsf{G}^{ (\mathrm{B}) }| = 1 + 2(n-w) + 2w-n-1 = n$.
Consequently, the set $\mathsf{U}^{ (w>n/2) }$ is defined as the collection of all possible choices of $\mathsf{S}^{ (\mathrm{A,B}) }( \mathbf{g}, (\mathcal{A}, \mathcal{B}) )$, given by
\begin{equation}
    \label{U for k=0 w>n/2}
    \mathsf{U}^{ (w>n/2) } :=
    \{
        \mathsf{S}^{ (\mathrm{A,B}) }( \mathbf{g}, (\mathcal{A}, \mathcal{B}) )
        ~:~ |\mathbf{g}|=n,
       ~(\mathcal{A}, \mathcal{B}) \text{ partition such that } |\mathcal{A}|=2(n-w),~ |\mathcal{B}|=2w-n
    \}.
\end{equation}
Since there are $3^{n}$ possible choices of $\mathbf{g}$ and $\binom{n}{2w-n}$ ways to choose $(\mathcal{A}, \mathcal{B})$, the size of the set is $|\mathsf{U}^{ (w>n/2) }| = 3^{n}\times \binom{n}{2w-n}$.

It remains to verify that the set $\mathsf{U}^{ (w>n/2) }$ constitutes a uniform stabilizer covering, which involves computing the covering power $\Sigma(w)$.
The verification follows from the observation that, for each $\mathsf{S}^{ (\mathrm{A,B}) }( \mathbf{g}, (\mathcal{A}, \mathcal{B}) )$, $\binom{2(n-w)}{n-w} 2^{2w-n-1}$ weight-$w$ Pauli strings are generated by multiplying generators according to the following rules:
\begin{enumerate}
    \item Selecting $\mathbf{g}$:
    First, we select the weight-$n$ Pauli string $ \mathbf{g} $ in $\mathsf{S}^{ (\mathrm{A,B}) }( \mathbf{g}, (\mathcal{A}, \mathcal{B}) )$.
    
    \item Selecting $n-w$ distinct $\mathbf{a}^{ (j) } \in \mathsf{G}^{ (\mathrm{A}) }$:
    According to Eq.~\eqref{k=0 generators for A}, multiplying $\mathbf{g}$ by $n-w$ distinct $\mathbf{a}^{ (j) }$ yields a weight-$w$ Pauli string, since each multiplication by an $\mathbf{a}^{ (j) }$ removes the non-identity Pauli operator at the $\mathcal{A}_{j}$-th qubit of $\mathbf{g}$.
    The number of such choices is $\binom{|\mathsf{G}^{ (\mathrm{A}) }|}{n-w} = \binom{2(n-w)}{n-w}$.
    
    \item Arbitrarily selecting multiple $\mathbf{b}^{ (j) } \in \mathsf{G}^{ (\mathrm{B}) }$:
    Since $\mathbf{g}$ is already chosen, according to Eq.~\eqref{k=0 generators for B}, multiplying by any combination of $\mathbf{b}^{ (j) }$ does not affect the weight. 
    Therefore, the number of possible choices is $2^{ |\mathsf{G}^{ (\mathrm{B}) }| } = 2^{2w-n-1}$.
\end{enumerate}
Besides this construction, there exist weight-$w$ Pauli strings generated without selecting $\mathbf{g}$, by using only combinations of $\mathbf{a}^{ (j) }$ and $\mathbf{b}^{ (j) }$.
We do not specify the exact number of this type of weight-$w$ Pauli strings, but this number is determined by the sizes of the subsets $|\mathcal{A}|$ and $|\mathcal{B}|$ according to a certain combinatorial rule.
In other words, it does not depend on the specific choice of $(\mathcal{A}, \mathcal{B})$; hence, the total number of weight-$w$ Pauli strings in each $\mathsf{S}^{ (\mathrm{A,B}) }( \mathbf{g}, (\mathcal{A}, \mathcal{B}) )$ is the same.
As a result, the first condition is satisfied with the covering power
\begin{equation}
    \label{Sigma for k=0 w>n/2}
    \Sigma(w) \geq \binom{2(n-w)}{n-w} 2^{2w-n-1},
    ~~\text{for}~\mathsf{U}^{ (w > n/2) }.
\end{equation}
Furthermore, the set $\mathsf{U}^{ (w> n/2) }$ includes all stabilizer groups corresponding to every possible choice of $\mathbf{g}$ and $(\mathcal{A}, \mathcal{B})$, and the construction does not favor any particular choice.
Therefore, each weight-$w$ Pauli string is contained in the same number of stabilizer groups, i.e., the second condition is fulfilled.


From the obtained bound on $\Sigma(w)$ as given in Eq.~\eqref{Sigma for k=0 w>n/2}, we find the upper bound on the sample complexity. 
By employing Stirling's formula Eq.~\eqref{Stirling approx}, we have $\Sigma(w) = \Omega(\frac{1}{\sqrt{n}}2^{n})$.
Therefore, by using Eq.~\eqref{CN and Sigma}, the upper bound on $\mathrm{CN}(\mathsf{P}(w))$ is derived as
\begin{equation}
    \label{upper bound on CN for k=0 w>n/2}
    \text{CN}(\mathsf{P}(w)) = O\left( n^{3/2} \frac{\binom{n}{w} 3^{w}}{2^{n}} \right), \quad
    \text{CN}(\bigcup_{u=0}^{w} \mathsf{P}(u)) \leq 
    \sum_{u=0}^{w} \text{CN}(\mathsf{P}(u)) = O\left( n^{3/2} \frac{ \sum_{u=0}^{w} \binom{n}{u} 3^{u}}{2^{n}} \right).
\end{equation}
Finally, from Eq.~\eqref{N for CN}, we obtain the upper bound on the sample complexity for the case $k=0$, $w > n/2$ as
\begin{equation}
    \label{upper bound for large w}
    N = O\left(
            n^{5/2} \frac{ \sum_{u=0}^{w} \binom{n}{u} 3^{u} }{ 2^{n} }
            \times \varepsilon^{-2}\log\delta^{-1}
        \right),
\end{equation}
which completes the proof of the second line of Eq.~(10) in Theorem 3 of the main text.
The upper bound in Eq.~\eqref{upper bound for large w} also aligns with the lower bound given in Eq.~\eqref{final lower bound on N}, differing only by a polynomial factor of $n$.
Thus, this establishes the tightness of our upper bound on $\mathrm{CN}(\mathsf{P}(w))$.

Note that when $n = w$, the sample complexity in Eq.~\eqref{upper bound for large w} becomes $N = O(n^{5/2} 2^{n})$ since the numerator can be evaluated as $\sum_{u=0}^{w} \binom{n}{u} 3^{u} = 4^{n}$.
This result is looser than the known result $N = O(n 2^{n})$ in~\cite{chen2022quantum} by a polynomial factor of $n$.
The discrepancy arises from the fact that, while the exact value of $\mathrm{CN}(\bigcup_{u=0}^{n} \mathsf{P}(u)) = 2^{n}+1$~\cite{wootters1989optimal, lawrence2002mutually}, our bound based on the covering array theory involves the $\log|\mathsf{P}| = O(n)$ factor as shown in Eq.~\eqref{CN and Sigma}.
An additional $n^{1/2}$ factor arises from Eq.~\eqref{Sigma for k=0 w>n/2}, which follows from Stirling's formula Eq.~\eqref{Stirling approx}.
Although Eq.~\eqref{upper bound on CN for k=0 w>n/2} provides a loose upper bound on $\mathrm{CN}(\bigcup_{u=0}^{n} \mathsf{P}(u))$, it differs only by a factor of $n$, and it can be applied for the case $w<n$ where the exact value of $\mathrm{CN}(\mathsf{P}(w))$ remains unknown.

\section{Upper bound on sample complexity with non-zero ancilla qubits}
\label{Sec: S5}

In this section, we establish an upper bound on the sample complexity $N$ in the presence of ancilla qubits, i.e., $k>0$.
The upper bound is obtained by generalizing the result of Sec.~\hyperref[Sec: S4]{S4}---originally formulated for $k=0$---to the case $k>0$.
Specifically, we extend the concept of the uniform stabilizer covering introduced in Sec.~\hyperref[Sec: S4A]{S4 A} to account for the use of ancilla qubits.
Finally, by constructing a uniform stabilizer covering tailored for the case $k>0$, we derive an upper bound on the sample complexity.

To consider the case $k>0$, we define some notations.
Since we have additional $k$ ancilla qubits, the total number of qubits is $k+n$.
Accordingly, we denote a Pauli string $P_{\mathbf{a}}$ on the full $k+n$ qubits as $P_{\mathbf{a}} = P_{\mathbf{a}^{(\text{anc})}} \otimes P_{\mathbf{a}^{(\text{sys})}}$, where $P_{\mathbf{a}^{(\text{anc})}}$ and $P_{\mathbf{a}^{(\text{sys})}}$ denote Pauli strings acting only on the $k$-qubit ancilla and the $n$-qubit system, respectively.
In the same manner as discussed in Sec.~\hyperref[Sec: S1 A]{S1 A}, we denote $\mathbf{a}^{(\text{anc})}$ and $\mathbf{a}^{(\text{sys})}$ as the $2k$-bit string and the $2n$-bit string associated with $P_{\mathbf{a}^{(\text{anc})}}$ and $P_{\mathbf{a}^{(\text{sys})}}$, respectively.
Consequently, the concatenated string $\mathbf{a} = \mathbf{a}^{(\text{anc})}\mathbf{a}^{(\text{sys})}$ is the $2(k+n)$-bit string associated with $P_{\mathbf{a}} = P_{\mathbf{a}^{(\text{anc})}} \otimes P_{\mathbf{a}^{(\text{sys})}}$.
If we want to specifically refer to only the system part of $P_{\mathbf{a}} = P_{\mathbf{a}^{(\text{anc})}} \otimes P_{\mathbf{a}^{(\text{sys})}}$, we write $\mathrm{Sys}(P_{\mathbf{a}}) := P_{\mathbf{a}^{(\text{sys})}}$, and similarly, we define $\mathrm{Sys}(\mathbf{a}) := \mathbf{a}^{(\text{sys})}$.
Likewise, we define $\mathrm{Anc}(P_{\mathbf{a}}) := P_{\mathbf{a}^{(\text{anc})}}$, and $\mathrm{Anc}(\mathbf{a}) := \mathbf{a}^{(\text{anc})}$.

\subsection{Extended definition of stabilizer covering for $k>0$}

To generalize the results for the case $k=0$, we first extend the concept of the stabilizer covering.
With the assistance of the $k$-qubit ancilla, we consider stabilizer groups generated by $k+n$ generators, where each generator is a Pauli string acting on $k+n$ qubits.
Since the channel acts only on the $n$-qubit system, we introduce an additional notation: for a given stabilizer group $\mathsf{S}$, we define $\mathrm{Sys}(\mathsf{S})$ as
\begin{equation}
    \label{system part of stabilizer group}
    \mathrm{Sys}(\mathsf{S}) := \bigcup_{P_{\mathbf{a}} \in \mathsf{S}} \{ \mathrm{Sys}(P_{\mathbf{a}}) \},
\end{equation}
which denotes the union of the system components of the Pauli strings in $\mathsf{S}$.
By extending the definition of the stabilizer group for $k>0$ as in Eq.~\eqref{system part of stabilizer group}, the same property of stabilizer groups as in the case $k=0$ holds: even when $k>0$, for a given stabilizer group $\mathsf{S}$, all Pauli eigenvalues $\lambda(\mathbf{b})$ such that $P_{\mathbf{b}} \in \mathrm{Sys}(\mathsf{S})$ can be estimated by using $O\left( n \times \varepsilon^{-2} \log \delta^{-1} \right)$ samples~\cite{chen2022quantum}.
To incorporate this property, we extend the definition of the stabilizer covering. 
\begin{definition}[Stabilizer covering with ancilla qubits]
    \label{def: SC with k}
    \normalfont
    We consider a $k$-qubit ancilla and an $n$-qubit system.
    Let $\mathsf{C} = \{\mathsf{S}_{i}\}_{i}$ be a set of stabilizer groups $\mathsf{S}_{i}$ on the full $k+n$ qubits.
    Each $\mathsf{S}_{i}$ is generated by $k+n$ independent group generators.
    For a given set $\mathsf{P}$ of Pauli strings on the $n$-qubit system, we also refer to $\mathsf{C}$ as a \textit{stabilizer covering} of $\mathsf{P}$ if it satisfies
    \begin{equation}
        \label{def of SC with k}
        \mathsf{P} \subseteq \bigcup_{\mathsf{S}_{i} \in \mathsf{C}} \mathrm{Sys}(\mathsf{S}_{i}).
    \end{equation}
    We denote this relationship as $\mathsf{C} \overset{\text{SC}}{\triangleright} \mathsf{P}$.
    In addition, we also use $\text{CN}(\mathsf{P})$ to denote the minimum size of any stabilizer covering of $\mathsf{P}$.
\end{definition}
\noindent

With the extended stabilizer covering condition given in Definition~\ref{def: SC with k}, the upper bound on the required sample complexity $N$ in Eq.~\eqref{N for CN} can be generalized to the case $k>0$.
It is known that even for $k>0$, the upper bound on $N$ to estimate all Pauli eigenvalues associated with a given set $\mathsf{P}$ is given by~\cite{chen2022quantum}
\begin{equation}
    \label{N for CN with k}
    N = O\left( n \times \text{CN}(\mathsf{P}) \times \varepsilon^{-2} \log \delta^{-1} \right)
    ~\text{even for } k > 0.
\end{equation}
This follows from the same analysis as in the case $k=0$.
Therefore, in the following section, we derive an upper bound on $\text{CN}(\mathsf{P})$ to obtain an upper bound on $N$ via Eq.~\eqref{N for CN with k}.

\subsection{Upper bound on $\text{CN}(\mathsf{P})$ for the case $k>0$}

To derive an upper bound on $\text{CN}(\mathsf{P})$, we consider the concept of uniform stabilizer covering introduced in the case $k=0$. 
Building on the preceding definitions, we naturally generalize this concept to the case $k>0$.
\begin{definition}[Uniform stabilizer covering with ancilla qubits]
    \label{def: uniform SC with k}
    \normalfont
    We consider a $k$-qubit ancilla and an $n$-qubit system.
    For a given set $\mathsf{P}$ of Pauli strings on the $n$-qubit system, a stabilizer covering $\mathsf{U}$ is said to be \textit{uniform} when it fulfills the following two conditions:
    \begin{enumerate}
        \item
        For all $\mathsf{S}_{i} \in \mathsf{U}$, $|\mathrm{Sys}(\mathsf{S}_{i}) \cap \mathsf{P}| = \Sigma$.
        We also refer to the value $\Sigma$ as a covering power.
        
        
        \item    
        For all $P_{\mathbf{a}} \in \mathsf{P}$, $|\{ \mathsf{S}_{i} \in \mathsf{U} : P_{\mathbf{a}} \in \mathrm{Sys}(\mathsf{S}_{i}) \}| = R$.   
        The relation $|\mathsf{U}|\times \Sigma = |\mathsf{P}| \times R$ also holds.
        Furthermore, this condition ensures that $\mathsf{U} \overset{\text{SC}}{\triangleright} \mathsf{P}$.
    \end{enumerate}
\end{definition}
    
    


By extending the definitions of uniform stabilizer covering and covering power, we derive the upper bound on $\mathrm{CN}(\mathsf{P})$ for $k>0$ using the same method as in Lemma~\ref{lemma: upper bound on CN}.
Since our proof that uses the probabilistic method remains valid even for $k>0$, we obtain the same upper bound as given in Eq.~\eqref{CN and Sigma}.
\begin{lemma}[Upper bound on the minimum size of stabilizer covering with ancilla qubits]
    \label{lemma: upper bound on CN with k}
    \normalfont
    We consider a $k$-qubit ancilla and an $n$-qubit system.
    For a given set $\mathsf{P}$ of Pauli strings on the $n$-qubit system, if there exists a uniform stabilizer covering $\mathsf{U}$ with covering power $\Sigma$, then $\text{CN}(\mathsf{P})$ is upper bounded by
    \begin{equation}
        \label{CN and Sigma with k}
        \text{CN}(\mathsf{P})
        \leq \left\lceil \frac{|\mathsf{P}| \log |\mathsf{P}|}{\Sigma} \right\rceil.
    \end{equation}
\end{lemma}
\noindent
Moreover, a stabilizer covering of this size can also be constructed using the same density-based greedy algorithm.

Based on the generalized results in Eqs.~\eqref{N for CN with k} and~\eqref{CN and Sigma with k}, the upper bound on the sample complexity for the case $k>0$ can also be obtained by constructing a uniform stabilizer covering and computing its covering power.
A key difference from the case $k=0$ is that the availability of ancilla qubits enables a distinctive stabilizer group construction scheme.
In particular, we employ Bell pairs to construct stabilizer groups~\cite{chen2022quantum}.
More precisely, we define a $2k$-qubit Bell pair generator set $\mathsf{G}^{ (\text{Bell}) }$ as follows:
given a size-$k$ subset $\mathcal{S} = \{ \mathcal{S}_{1}, \mathcal{S}_{2}, \ldots, \mathcal{S}_{k} \}$, which is a subset of system qubits, the \textit{$2k$-qubit Bell pair generator set} $\mathsf{G}^{ (\text{Bell}) }(\mathcal{S})$ associated with $\mathcal{S}$ is defined as
\begin{equation}
    \label{Bell pair generator}
\begin{gathered}
    \mathsf{G}^{ (\text{Bell}) }(\mathcal{S})
    := \{ \mathbf{x}^{ (j) }, \mathbf{z}^{ (j) } 
    ~:~ j\in \{1, 2, \ldots, k\}
    \}, ~\text{where}\\
    \mathrm{Anc}(\mathbf{x}^{ (j) })_{i}
    =\begin{cases}
        X & i = j \\
        I & i \neq j
    \end{cases}~~
    \mathrm{Sys}(\mathbf{x}^{ (j) })_{i}
    =\begin{cases}
        X & i = \mathcal{S}_{j} \\
        I & i \neq \mathcal{S}_{j}
    \end{cases} \\
    \mathrm{Anc}(\mathbf{z}^{ (j) })_{i}
    =\begin{cases}
        Z & i = j \\
        I & i \neq j
    \end{cases} ~~
    \mathrm{Sys}(\mathbf{z}^{ (j) })_{i}
    =\begin{cases}
        Z & i = \mathcal{S}_{j} \\
        I & i \neq \mathcal{S}_{j}
    \end{cases} 
\end{gathered}
\end{equation}
In Fig.~\ref{illust of Bell pair generator}, we illustrate the $2k$-qubit Bell pair generator set.
The key advantage of $\mathsf{G}^{ (\text{Bell}) }(\mathcal{S})$ is its ability to generate every Pauli string whose non-identity components in the system part are supported on $\mathcal{S}$.

\begin{figure}[h]
\begin{tikzpicture}
    \def\w{1.7}
    \def\h{0.5}
    \def\hmargin{0.2}
    \def\wmargin{-0.05}
    \def\boxmargin{0.8}
    \foreach \i in {0,...,8}
    {
        \pgfmathsetmacro\x{\i*\w + \boxmargin}
        \ifnum\i<8
            \draw[fill=yellow!30] (\x,0) rectangle ++(\w,\boxmargin);
        \else
            \draw (\x,0) rectangle ++(\w,\boxmargin);
        \fi
    }
    \draw[decorate,decoration={brace,amplitude=5pt},yshift=0.4cm]
    (4*\w+\boxmargin,1.3) -- (9*\w+\boxmargin,1.3) node[midway,yshift=0.4cm] {$n$-qubit system};
    \draw[decorate,decoration={brace,amplitude=5pt},yshift=0.4cm]
    (0*\w+\boxmargin,1.3) -- (4*\w+\boxmargin,1.3) node[midway,yshift=0.4cm] {$k$-qubit ancilla};
    \draw[decorate,decoration={brace,amplitude=5pt},yshift=0.4cm]
    (4*\w+\boxmargin,0.7) -- (8*\w+\boxmargin,0.7) node[midway,yshift=0.4cm] {$|\mathcal{S}| = k$};
    \draw[decorate,decoration={brace,amplitude=5pt},yshift=0.4cm]
    (8*\w+\boxmargin,0.7) -- (9*\w+\boxmargin,0.7) node[midway,yshift=0.4cm] {$ [n] \setminus \mathcal{S}$};
    \node at (1*\w+\wmargin,0.4) {$1$};
    \node at (2*\w+\wmargin,0.4) {$2$};
    \node at (3*\w+\wmargin,0.4) {$\cdots$};
    \node at (4*\w+\wmargin,0.4) {$k$};
    \node at (5*\w+\wmargin,0.4) {$\mathcal{S}_{1}$};
    \node at (6*\w+\wmargin,0.4) {$\mathcal{S}_{2}$};
    \node at (7*\w+\wmargin,0.4) {$\cdots$};
    \node at (8*\w+\wmargin,0.4) {$\mathcal{S}_{k}$};
    \node at (9*\w+\wmargin,0.4) {$\cdots$};
    \node at (0.0, -1*\h+\hmargin) {$\mathbf{x}^{ (1) }~ (\mathbf{z}^{ (1) })$ :};
    \node at (0.0, -2*\h+\hmargin) {$\mathbf{x}^{ (2) }~ (\mathbf{z}^{ (2) })$ :};
    \node at (0.0, -3*\h+\hmargin+0.1) {$\vdots$}; 
    \node at (0.0, -4*\h+\hmargin) {$\mathbf{x}^{ (k) }~ (\mathbf{z}^{ (k) })$ :};
    \node at (1*\w+\wmargin, -1*\h+\hmargin) {$X(Z)$};
    \node at (2*\w+\wmargin, -1*\h+\hmargin) {$I$};
    \node at (3*\w+\wmargin, -1*\h+\hmargin) {$\cdots$};
    \node at (4*\w+\wmargin, -1*\h+\hmargin) {$I$};
    \node at (5*\w+\wmargin, -1*\h+\hmargin) {$X(Z)$};
    \node at (6*\w+\wmargin, -1*\h+\hmargin) {$I$};
    \node at (7*\w+\wmargin, -1*\h+\hmargin) {$\cdots$};
    \node at (8*\w+\wmargin, -1*\h+\hmargin) {$I$};
    \node at (9*\w+\wmargin, -1*\h+\hmargin) {$\cdots$};
    \node at (1*\w+\wmargin, -2*\h+\hmargin) {$I$};
    \node at (2*\w+\wmargin, -2*\h+\hmargin) {$X(Z)$};
    \node at (3*\w+\wmargin, -2*\h+\hmargin) {$\cdots$};
    \node at (4*\w+\wmargin, -2*\h+\hmargin) {$I$};
    \node at (5*\w+\wmargin, -2*\h+\hmargin) {$I$};
    \node at (6*\w+\wmargin, -2*\h+\hmargin) {$X(Z)$};
    \node at (7*\w+\wmargin, -2*\h+\hmargin) {$\cdots$};
    \node at (8*\w+\wmargin, -2*\h+\hmargin) {$I$};
    \node at (9*\w+\wmargin, -2*\h+\hmargin) {$\cdots$};
    \node at (1*\w+\wmargin, -4*\h+\hmargin) {$I$};
    \node at (2*\w+\wmargin, -4*\h+\hmargin) {$I$};
    \node at (3*\w+\wmargin, -4*\h+\hmargin) {$\cdots$};
    \node at (4*\w+\wmargin, -4*\h+\hmargin) {$X(Z)$};
    \node at (5*\w+\wmargin, -4*\h+\hmargin) {$I$};
    \node at (6*\w+\wmargin, -4*\h+\hmargin) {$I$};
    \node at (7*\w+\wmargin, -4*\h+\hmargin) {$\cdots$};
    \node at (8*\w+\wmargin, -4*\h+\hmargin) {$X(Z)$};
    \node at (9*\w+\wmargin, -4*\h+\hmargin) {$\cdots$};
\end{tikzpicture}
\caption{
Schematic diagram of the $2k$-qubit Bell pair generator set $\mathsf{G}^{ (\text{Bell}) }(\mathcal{S})$.
We denote the set $\{1,2,\ldots, n\}$ by $[n]$, and each number in the box represents a qubit index.
For simplicity, we illustrate $\mathcal{S}$ as a contiguous segment, although any choice is allowed. 
}
\label{illust of Bell pair generator}
\end{figure}

In the following section, for the case $k>0$, we construct a uniform stabilizer covering of $\mathsf{P}(w)$.
By leveraging the $k$-qubit ancilla, we employ the $2k$-qubit Bell pair generator set defined in Eq.~\eqref{Bell pair generator}.
Analogous to the case $k=0$, the behavior of the covering power exhibits a transition at $2w = k+n$; thus, we consider two regimes: $2w \leq k+n$ and $2w > k+n$.



\subsubsection{Upper bound on $\mathrm{CN}(\mathsf{P}(w))$ for $2w \leq k+n$ case}

For the case $2w \leq k+n$, we find a uniform stabilizer covering $\mathsf{U}^{ (2w \leq k+n) }$, adopting a similar strategy to that used in the case $k=0$, $w\leq n/2$.
In order to construct $\mathsf{U}^{ (2w \leq k+n) }$, we consider a partition $(\mathcal{S}, \mathcal{T})$ of $n$ system qubits $\{1, 2, \ldots, n\}$ such that $|\mathcal{S}| = k$ and $|\mathcal{T}| = n-k$.
The elements of each subset are denoted by $\mathcal{S} = \{ {\mathcal{S}}_{1}, {\mathcal{S}}_{2}, \ldots, {\mathcal{S}}_{k} \}$ and $\mathcal{T} = \{ {\mathcal{T}}_{1}, {\mathcal{T}}_{2}, \ldots, {\mathcal{T}}_{n-k} \}$.
In addition, we consider a tuple $\mathbb{G} = (G_{1}, G_{2}, \ldots, G_{n-k}) \in \{X, Y, Z\}^{n-k}$, where $\{X, Y, Z\}^{n-k}$ now denotes the set of all length-$(n-k)$ tuples consisting of non-identity Pauli operators.
Given $\mathbb{G}$ and $(\mathcal{S}, \mathcal{T})$, we define $\mathsf{G}^{ (\text{Bell}) }(\mathcal{S})$ according to Eq.~\eqref{Bell pair generator}, and analogously to Eq.~\eqref{k=0 generator for Gn}, we define $\mathsf{G}^{ (\mathrm{T}) }(\mathbb{G}, \mathcal{T})$ as follows: 
\begin{equation}
    \label{k>0 generator for GT}
\begin{gathered}
    \mathsf{G}^{ (\mathrm{T}) }(\mathbb{G}, \mathcal{T})
    := \{ \mathbf{t}^{ (1) }, \mathbf{t}^{ (2) }, \ldots, \mathbf{t}^{ (n-k) } \},~\text{where} \\
    \mathrm{Anc}(\mathbf{t}^{ (j) })_{i} = I~ \forall i \in \{1, 2, \ldots, k\},~~
    \mathrm{Sys}(\mathbf{t}^{ (j) })_{i}
    =
    \begin{cases}
        G_{j} & i=\mathcal{T}_{j}\\
        I & i \neq \mathcal{T}_{j}
    \end{cases} 
\end{gathered}
\end{equation}
In Fig.~\ref{illust of stabilizer group 2w<k+n}, we depict the partition $(\mathcal{S}, \mathcal{T})$ and the sets $\mathsf{G}^{ (\text{Bell}) }(\mathcal{S})$ and $\mathsf{G}^{ (\mathrm{T}) }(\mathbb{G}, \mathcal{T})$.

\begin{figure}[h]
\begin{tikzpicture}
    \def\w{1.7}
    \def\h{0.5}
    \def\hmargin{0.2}
    \def\wmargin{-0.05}
    \def\boxmargin{0.8}
    \foreach \i in {0,...,4}
    {
        \pgfmathsetmacro\x{\i*\w + \boxmargin}
        \ifnum\i<1
            \draw[fill=yellow!30] (\x,0) rectangle ++(\w,\boxmargin);
        \else
            \draw[fill=green!15] (\x,0) rectangle ++(\w,\boxmargin);
        \fi
    }
    \draw[decorate,decoration={brace,amplitude=5pt},yshift=0.4cm]
    (0*\w+\boxmargin,0.7) -- (1*\w+\boxmargin,0.7) node[midway,yshift=0.4cm] {$|\mathcal{S}| = k$};
    \draw[decorate,decoration={brace,amplitude=5pt},yshift=0.4cm]
    (1*\w+\boxmargin,0.7) -- (5*\w+\boxmargin,0.7) node[midway,yshift=0.4cm] {$|\mathcal{T}| = n-k$};
    \node at (1*\w+\wmargin,0.4) {$\mathcal{S}$};
    \node at (2*\w+\wmargin,0.4) {$\mathcal{T}_{1}$};
    \node at (3*\w+\wmargin,0.4) {$\mathcal{T}_{2}$};
    \node at (4*\w+\wmargin,0.4) {$\cdots$};
    \node at (5*\w+\wmargin,0.4) {$\mathcal{T}_{n-k}$};
    \node at (0.0, -1*\h+\hmargin) {$\mathbf{x}^{ (j) }, \mathbf{z}^{ (j) }$ :};
    \node at (0.0, -2*\h+\hmargin) {$\mathbf{t}^{ (1) }$ :};
    \node at (0.0, -3*\h+\hmargin) {$\mathbf{t}^{ (2) }$ :};
    \node at (0.0, -4*\h+\hmargin+0.1) {$\vdots$}; 
    \node at (0.0, -5*\h+\hmargin) {$\mathbf{t}^{ (n-k) }$ :};
    \node at (1*\w+\wmargin, -1*\h+\hmargin) {$ \mathsf{G}^{ (\text{Bell}) }(\mathcal{S}) $};
    \node at (2*\w+\wmargin, -1*\h+\hmargin) {$I$};
    \node at (3*\w+\wmargin, -1*\h+\hmargin) {$I$};
    \node at (4*\w+\wmargin, -1*\h+\hmargin) {$\cdots$};
    \node at (5*\w+\wmargin, -1*\h+\hmargin) {$I$};    
    \draw (-1.0, -1.5*\h+\hmargin) --  (5.5*\w, -1.5*\h+\hmargin);
    \node at (1*\w+\wmargin, -2*\h+\hmargin) {$I$};
    \node at (2*\w+\wmargin, -2*\h+\hmargin) {$G_{1}$};
    \node at (3*\w+\wmargin, -2*\h+\hmargin) {$I$};
    \node at (4*\w+\wmargin, -2*\h+\hmargin) {$\cdots$};
    \node at (5*\w+\wmargin, -2*\h+\hmargin) {$I$};
    \node at (1*\w+\wmargin, -3*\h+\hmargin) {$I$};
    \node at (2*\w+\wmargin, -3*\h+\hmargin) {$I$};
    \node at (3*\w+\wmargin, -3*\h+\hmargin) {$G_{2}$};
    \node at (4*\w+\wmargin, -3*\h+\hmargin) {$\cdots$};
    \node at (5*\w+\wmargin, -3*\h+\hmargin) {$I$};
    \node at (1*\w+\wmargin, -5*\h+\hmargin) {$I$};
    \node at (2*\w+\wmargin, -5*\h+\hmargin) {$I$};
    \node at (3*\w+\wmargin, -5*\h+\hmargin) {$I$};
    \node at (4*\w+\wmargin, -5*\h+\hmargin) {$\cdots$};
    \node at (5*\w+\wmargin, -5*\h+\hmargin) {$G_{n-k}$};
\end{tikzpicture}
\caption{
Schematic diagram of $\mathsf{G}^{ \mathrm{Bell} }$ and $\mathsf{G}^{ (\mathrm{T}) }$ for the case $2w \leq k+n$.
Only the $n$-qubit system part is shown in this figure.
The full structure of $\mathsf{G}^{ \mathrm{Bell} }$ is illustrated in Fig.~\ref{illust of Bell pair generator}, and $\mathsf{G}^{ (\mathrm{T}) }$ acts as the identity on the $k$-qubit ancilla part.
Although $\mathcal{S}$ and $\mathcal{T}$ are drawn as contiguous sets, they can be chosen arbitrarily.
}
\label{illust of stabilizer group 2w<k+n}
\end{figure}

As illustrated in Fig.~\ref{illust of stabilizer group 2w<k+n}, all Pauli strings in $\mathsf{G}^{ (\text{Bell}) }(\mathcal{S}) \cup \mathsf{G}^{ (\mathrm{T}) }(\mathbb{G}, \mathcal{T})$ mutually commute.
Accordingly, we define a stabilizer group $\mathsf{S}^{ (\mathrm{S}, \mathrm{T}) }(\mathbb{G}, (\mathcal{S}, \mathcal{T}))$ as 
\begin{equation}
    \mathsf{S}^{ (\mathrm{S}, \mathrm{T}) }(\mathbb{G}, (\mathcal{S}, \mathcal{T}))
    := \left\langle
    \mathsf{G}^{ (\text{Bell}) }(\mathcal{S})
    \cup \mathsf{G}^{ (\mathrm{T}) }(\mathbb{G}, \mathcal{T})
    \right\rangle.
\end{equation}
Then, the set $\mathsf{U}^{ (2w \leq k+n) }$ is defined as
\begin{equation}
    \mathsf{U}^{ (2w \leq k+n) }
    := \{ \mathsf{S}^{ (\mathrm{S}, \mathrm{T}) }(\mathbb{G}, (\mathcal{S}, \mathcal{T}))
    ~:~ \mathbb{G} \in \{X, Y, Z\}^{n-k},
    ~(\mathcal{S}, \mathcal{T}) \text{ partition such that } |\mathcal{S}|=k,~ |\mathcal{T}|=n-k \}.
\end{equation}
From the definition of $\mathsf{U}^{ (2w \leq k+n) }$, its size is $|\mathsf{U}^{ (2w \leq k+n) }| = \binom{n}{k}3^{n-k}$, where $\binom{n}{k}$ is the number of possible partitions $(\mathcal{S}, \mathcal{T})$, and $3^{n-k}$ is the number of tuples in $\{X, Y, Z\}^{n-k}$.

\begin{figure}[h]
\centering
\begin{tikzpicture}
\def\S{2}
\def\A{3.5}

\fill[yellow!30] (0,0) rectangle (\S,1);
\fill[green!15] (\S,0) rectangle ({\S+\A},1);

\draw (0,0) rectangle ({\S+\A},1); 
\draw (\S,0) -- (\S,1); 

\node at (\S/2,0.5) {$m$};
\node at (\S+\A/2,0.5) {$w-m$};

\draw[decorate,decoration={brace,amplitude=5pt},yshift=0.4cm]
(0,1) -- (\S,1) node[midway,yshift=0.4cm] {$|\mathcal{S}| = k$};
\draw[decorate,decoration={brace,amplitude=5pt},yshift=0.4cm]
(\S,1) -- ({\S+\A},1) node[midway,yshift=0.4cm] {$|\mathcal{T}| = n-k$};

\node at (\S/2,-0.5) {$\binom{k}{m}3^{m}$};
\node at (\S+\A/2,-0.5) {$\binom{n-k}{w-m}$};

\node at (-1.8,1.5) {\textbf{partition}};
\node[anchor=base] at (-1.8,0.4) {\textbf{weight}};
\node at (-1.8,-0.45) {\textbf{\# of choices}};

\end{tikzpicture}
\caption{
Illustration of constructing weight-$w$ Pauli strings in $\mathsf{S}^{ (\mathrm{S}, \mathrm{T}) }(\mathbb{G}, (\mathcal{S}, \mathcal{T}))$.
Weight-$w$ Pauli strings are obtained by taking weight-$m$ components from $\mathcal{S}$ and weight-$(w-m)$ components from $\mathcal{T}$.
There are $\binom{k}{m}3^{m}$ weight-$m$ Pauli strings in $\mathcal{S}$ and $\binom{n-k}{w-m}$ weight-$(w-m)$ Pauli strings in $\mathcal{T}$.
}
\label{counting sigma for case 2w<k+n}
\end{figure}

We need to verify that the constructed set $\mathsf{U}^{ (2w \leq k+n) }$ constitutes a uniform stabilizer covering and compute the corresponding covering power $\Sigma(w, k)$.
This can be justified by observing that each $\mathsf{S}^{ (\mathrm{S}, \mathrm{T}) }(\mathbb{G}, (\mathcal{S}, \mathcal{T}))$ contains $\Sigma(w, k)$ weight-$w$ Pauli strings, where 
\begin{equation}
    \label{Sigma for 2w<k+n}
     \Sigma(w, k) = \sum_{m=0}^{k}\binom{k}{m} 3^{m} \binom{n-k}{w-m},
     ~~\text{for}~\mathsf{U}^{ (2w \leq k+n) }.
\end{equation}
Figure~\ref{counting sigma for case 2w<k+n} illustrates the combinatorial generation of the $\Sigma(w, k)$ weight-$w$ Pauli strings based on the following rules:
\begin{enumerate}
    \item Selecting a weight-$m$ Pauli string on $\mathcal{S}$ ($0\leq m \leq k$):
    We begin by selecting an arbitrary weight-$m$ Pauli string on $\mathcal{S}$.
    By employing $\mathsf{G}^{\mathrm{Bell}}(\mathcal{S})$, any weight-$m$ Pauli string $P_{\mathbf{a}}$ such that the non-identity components of $\mathrm{Sys}(P_{\mathbf{a}})$ are supported on $\mathcal{S}$ can be generated from $\mathsf{G}^{\mathrm{Bell}}(\mathcal{S})$.
    Therefore, a total of $\binom{k}{m}3^{m}$ weight-$m$ Pauli strings on $\mathcal{S}$ can be chosen, as illustrated in Fig.~\ref{counting sigma for case 2w<k+n}.
    
    \item Selecting $w-m$ distinct $\mathbf{t}^{ (j) } \in \mathsf{G}^{ (\mathrm{T}) }$:
    Having constructed the weight-$m$ contribution from $\mathcal{S}$, we now choose $w-m$ distinct weight-$1$ generators in $\mathsf{G}^{ (\mathrm{T}) }$, to complete the construction of a weight-$w$ Pauli string.
    The number of such choices is given by $\binom{|\mathsf{G}^{ (\mathrm{T}) }|}{w-m} = \binom{n-k}{w-m}$.
\end{enumerate}
By summing over all cases $0 \leq m \leq k$, we derive Eq.~\eqref{Sigma for 2w<k+n}.
Hence, the above construction ensures the first condition with the covering power in Eq.~\eqref{Sigma for 2w<k+n}.
Since all possible choices of $\mathbb{G}$ and $(\mathcal{S}, \mathcal{T})$ are accounted for in the construction of $\mathsf{U}^{ (2w \leq k+n) }$, it follows that $\mathsf{P}(w)$ is entirely covered.
Moreover, since the construction is symmetric over all such choices, it ensures that each weight-$w$ Pauli string is included in the same number of stabilizer groups.
Thus, the second condition is satisfied.

From the covering power in Eq.~\eqref{Sigma for 2w<k+n}, an upper bound on $N$ can be derived by using Eqs.~\eqref{N for CN with k} and~\eqref{CN and Sigma with k}.
This upper bound does not match the lower bound in Theorem 2 of the main text, although it is derived using a similar strategy to the case $k=0$, $w \leq n/2$, where the upper bound is tight.


\subsubsection{Upper bound on $\mathrm{CN}(\mathsf{P}(w))$ for $2w > k+n$ case}

For the case $2w > k+n$, we construct a uniform stabilizer covering $\mathsf{U}^{ (2w > k+n) }$, by extending the approach used for the case $k=0$, $w > n/2$.
To construct stabilizer groups in $\mathsf{U}^{ (2w > k+n) }$, we consider a partition $(\mathcal{S}, \mathcal{A}, \mathcal{B})$ of the set of $n$ system qubits such that $|\mathcal{S}| = k$, $|\mathcal{A}| = 2(n-w)$, and $|\mathcal{B}| = 2w-n-k$.
Their elements are denoted as $\mathcal{S} = \{ {\mathcal{S}}_{1}, {\mathcal{S}}_{2}, \ldots, {\mathcal{S}}_{k} \}$, $\mathcal{A} = \{ {\mathcal{A}}_{1}, {\mathcal{A}}_{2}, \ldots, {\mathcal{A}}_{2(n-w)} \}$ and $\mathcal{B} = \{ {\mathcal{B}}_{1}, {\mathcal{B}}_{2}, \ldots, {\mathcal{B}}_{2w-n-k} \}$.
Additionally, as in the case $k=0$, $w > n/2$, we consider a weight-$(n-k)$ Pauli string $\mathbf{g}$, such that $\mathrm{Anc}(\mathbf{g})$ is an identity on the ancilla qubits, and $\mathrm{Sys}(\mathbf{g}) = g_{1}g_{2}\cdots g_{n-k}$ where each $g_{j}$ is a non-identity Pauli operator.
Then, given $(\mathcal{S}, \mathcal{A}, \mathcal{B})$ and $\mathbf{g}$, we define $\mathsf{G}^{ (\text{Bell}) }(\mathcal{S})$ as in Eq.~\eqref{Bell pair generator}, and in analogy with the case $k=0$, $w > n/2$, we define the sets $\mathsf{G}^{ (\mathrm{A}) }(\mathbf{g}, \mathcal{A})$ and $\mathsf{G}^{ (\mathrm{B}) }(\mathbf{g}, \mathcal{B})$ as follows: 
\begin{equation}
    \label{k>0 generators for A}
\begin{gathered}
    \mathsf{G}^{ (\mathrm{A}) }(\mathbf{g}, \mathcal{A})
    := \{ \mathbf{a}^{ (1) }, \mathbf{a}^{ (2) }, \ldots, \mathbf{a}^{ (2(n-w)) } \},  ~\text{where} \\
    \mathrm{Anc}(\mathbf{a}^{ (j) })_{i} =I~ \text{for all}~ i \in \{1, 2, \cdots, k\}, ~~
    \mathrm{Sys}(\mathbf{a}^{ (j) })_{i}
    =\begin{cases}
        g_{\mathcal{A}_{j}} & i = \mathcal{A}_{j}\\
        I & \text{otherwise}
    \end{cases}
\end{gathered}
\end{equation}
\begin{equation}
\begin{gathered}
    \label{k>0 generators for B}
    \mathsf{G}^{ (\mathrm{B}) }(\mathbf{g}, \mathcal{B})
    := \{ \mathbf{b}^{ (1) }, \mathbf{b}^{ (2) }, \ldots, \mathbf{b}^{ (2w-n-k-1) } \}, ~\text{where} \\
    \mathrm{Anc}(\mathbf{b}^{ (j) })_{i}=I~ \text{for all}~ i \in \{1, 2, \cdots, k\}, ~~
    \mathrm{Sys}(\mathbf{b}^{ (j) })_{i} =
    \begin{cases}
        \mathcal{P}(g_{ \mathcal{B}_{j} }) & i = \mathcal{B}_{j} \\
        \mathcal{P}(g_{ \mathcal{B}_{j+1} }) & i=\mathcal{B}_{j+1} \\
        I & \text{otherwise}
    \end{cases}
\end{gathered}
\end{equation}
In Fig.~\ref{illust of stabilizer group 2w>k+n}, we illustrate the partition $(\mathcal{S}, \mathcal{A}, \mathcal{B})$ and the sets $\mathsf{G}^{ (\text{Bell}) }$, $\mathsf{G}^{ (\mathrm{A}) }$ and $\mathsf{G}^{ (\mathrm{B}) }$.

\begin{figure}[h]
\begin{tikzpicture}
    \def\w{1.7}
    \def\h{0.5}
    \def\hmargin{0.2}
    \def\wmargin{-0.05}
    \def\boxmargin{0.8}
    \foreach \i in {0,...,8}
    {
        \pgfmathsetmacro\x{\i*\w + \boxmargin}
        \ifnum\i<1
            \draw[fill=yellow!30] (\x,0) rectangle ++(\w,\boxmargin);
        \else
            \ifnum\i<5
                \draw[fill=blue!15] (\x,0) rectangle ++(\w,\boxmargin);
            \else
                \draw[fill=red!15] (\x,0) rectangle ++(\w,\boxmargin);
            \fi
        \fi
    }
    \draw[decorate,decoration={brace,amplitude=5pt},yshift=0.4cm]
    (0*\w+\boxmargin,0.7) -- (1*\w+\boxmargin,0.7) node[midway,yshift=0.4cm] {$|\mathcal{S}| = k$};
    \draw[decorate,decoration={brace,amplitude=5pt},yshift=0.4cm]
    (1*\w+\boxmargin,0.7) -- (5*\w+\boxmargin,0.7) node[midway,yshift=0.4cm] {$|\mathcal{A}| = 2(n-w)$};
    \draw[decorate,decoration={brace,amplitude=5pt},yshift=0.4cm]
    (5*\w+\boxmargin,0.7) -- (9*\w+\boxmargin,0.7) node[midway,yshift=0.4cm] {$|\mathcal{B}| = 2w-n-k$};
    \node at (1*\w+\wmargin,0.4) {$\mathcal{S}$};
    \node at (2*\w+\wmargin,0.4) {$\mathcal{A}_{1}$};
    \node at (3*\w+\wmargin,0.4) {$\mathcal{A}_{2}$};
    \node at (4*\w+\wmargin,0.4) {$\cdots$};
    \node at (5*\w+\wmargin,0.4) {$\mathcal{A}_{2(n-w)}$};
    \node at (6*\w+\wmargin,0.4) {$\mathcal{B}_{1}$};
    \node at (7*\w+\wmargin,0.4) {$\mathcal{B}_{2}$};
    \node at (8*\w+\wmargin,0.4) {$\cdots$};
    \node at (9*\w+\wmargin,0.4) {$\mathcal{B}_{2w-n-k}$};
    \node at (0.0, -1*\h+\hmargin) {$\mathbf{x}^{ (j) }, \mathbf{z}^{ (j) }$ :};
    \node at (0.0, -2*\h+\hmargin) {$\mathbf{g}$:};
    \node at (0.0, -3*\h+\hmargin) {$\mathbf{a}^{ (1) }$:};
    \node at (0.0, -4*\h+\hmargin) {$\mathbf{a}^{ (2) }$:};
    \node at (0.0, -5*\h+\hmargin+0.1) {$\vdots$};
    \node at (0.0, -6*\h+\hmargin) {$\mathbf{a}^{ (2(n-w)) }$:};
    \node at (0.0, -7*\h+\hmargin) {$\mathbf{b}^{ (1) }$:};
    \node at (0.0, -8*\h+\hmargin) {$\mathbf{b}^{ (2) }$:};
    \node at (0.0, -9*\h+\hmargin+0.1) {$\vdots$};
    \node at (0.0, -10*\h+\hmargin) {$\mathbf{b}^{ (2w-n-k-1) }$:};
    \node at (1*\w+\wmargin, -1*\h+\hmargin) {$ \mathsf{G}^{(\text{Bell})}(\mathcal{S}) $};
    \node at (2*\w+\wmargin, -1*\h+\hmargin) {$I$};
    \node at (3*\w+\wmargin, -1*\h+\hmargin) {$I$};
    \node at (4*\w+\wmargin, -1*\h+\hmargin) {$\cdots$};
    \node at (5*\w+\wmargin, -1*\h+\hmargin) {$I$};
    \node at (6*\w+\wmargin, -1*\h+\hmargin) {$I$};
    \node at (7*\w+\wmargin, -1*\h+\hmargin) {$I$};
    \node at (8*\w+\wmargin, -1*\h+\hmargin) {$\cdots$};
    \node at (9*\w+\wmargin, -1*\h+\hmargin) {$I$};
    \draw (-1.0, -1.5*\h+\hmargin) --  (9.5*\w, -1.5*\h+\hmargin);
    \node at (1*\w+\wmargin, -2*\h+\hmargin) {$I$};
    \node at (2*\w+\wmargin, -2*\h+\hmargin) {$g_{ \mathcal{A}_{1} }$};
    \node at (3*\w+\wmargin, -2*\h+\hmargin) {$g_{ \mathcal{A}_{2} }$};
    \node at (4*\w+\wmargin, -2*\h+\hmargin) {$g_{ \mathcal{A}_{\cdots} }$};
    \node at (5*\w+\wmargin, -2*\h+\hmargin) {$g_{ \mathcal{A}_{2(n-w)} }$};
    \node at (6*\w+\wmargin, -2*\h+\hmargin) {$g_{ \mathcal{B}_{1} }$};
    \node at (7*\w+\wmargin, -2*\h+\hmargin) {$g_{ \mathcal{B}_{2} }$};
    \node at (8*\w+\wmargin, -2*\h+\hmargin) {$g_{\mathcal{B}_{\cdots}}$};
    \node at (9*\w+\wmargin, -2*\h+\hmargin) {$g_{ \mathcal{B}_{2w-n-k} }$};
    \draw (-1.0, -2.5*\h+\hmargin) --  (9.5*\w, -2.5*\h+\hmargin);
    \node at (1*\w+\wmargin, -3*\h+\hmargin) {$I$};
    \node at (2*\w+\wmargin, -3*\h+\hmargin) {$g_{ \mathcal{A}_{1} }$};
    \node at (3*\w+\wmargin, -3*\h+\hmargin) {$I$};
    \node at (4*\w+\wmargin, -3*\h+\hmargin) {$\cdots$};
    \node at (5*\w+\wmargin, -3*\h+\hmargin) {$I$};
    \node at (6*\w+\wmargin, -3*\h+\hmargin) {$I$};
    \node at (7*\w+\wmargin, -3*\h+\hmargin) {$I$};
    \node at (8*\w+\wmargin, -3*\h+\hmargin) {$\cdots$};
    \node at (9*\w+\wmargin, -3*\h+\hmargin) {$I$};
    \node at (1*\w+\wmargin, -4*\h+\hmargin) {$I$};
    \node at (2*\w+\wmargin, -4*\h+\hmargin) {$I$};
    \node at (3*\w+\wmargin, -4*\h+\hmargin) {$g_{ \mathcal{A}_{2} }$};
    \node at (4*\w+\wmargin, -4*\h+\hmargin) {$\cdots$};
    \node at (5*\w+\wmargin, -4*\h+\hmargin) {$I$};
    \node at (6*\w+\wmargin, -4*\h+\hmargin) {$I$};
    \node at (7*\w+\wmargin, -4*\h+\hmargin) {$I$};
    \node at (8*\w+\wmargin, -4*\h+\hmargin) {$\cdots$};
    \node at (9*\w+\wmargin, -4*\h+\hmargin) {$I$};
    \node at (1*\w+\wmargin, -6*\h+\hmargin) {$I$};
    \node at (2*\w+\wmargin, -6*\h+\hmargin) {$I$};
    \node at (3*\w+\wmargin, -6*\h+\hmargin) {$I$};
    \node at (4*\w+\wmargin, -6*\h+\hmargin) {$\cdots$};
    \node at (5*\w+\wmargin, -6*\h+\hmargin) {$g_{ \mathcal{A}_{2(n-w)} }$};
    \node at (6*\w+\wmargin, -6*\h+\hmargin) {$I$};
    \node at (7*\w+\wmargin, -6*\h+\hmargin) {$I$};
    \node at (8*\w+\wmargin, -6*\h+\hmargin) {$\cdots$};
    \node at (9*\w+\wmargin, -6*\h+\hmargin) {$I$};
    \draw (-1.0, -6.5*\h+\hmargin) --  (9.5*\w, -6.5*\h+\hmargin);
    \node at (1*\w+\wmargin, -7*\h+\hmargin) {$I$};
    \node at (2*\w+\wmargin, -7*\h+\hmargin) {$I$};
    \node at (3*\w+\wmargin, -7*\h+\hmargin) {$I$};
    \node at (4*\w+\wmargin, -7*\h+\hmargin) {$\cdots$};
    \node at (5*\w+\wmargin, -7*\h+\hmargin) {$I$};
    \node at (6*\w+\wmargin, -7*\h+\hmargin) {$\mathcal{P}(g_{ \mathcal{B}_{1} })$};
    \node at (7*\w+\wmargin, -7*\h+\hmargin) {$\mathcal{P}(g_{ \mathcal{B}_{2} })$};
    \node at (8*\w+\wmargin, -7*\h+\hmargin) {$\cdots$};
    \node at (9*\w+\wmargin, -7*\h+\hmargin) {$I$};
    \node at (1*\w+\wmargin, -8*\h+\hmargin) {$I$};
    \node at (2*\w+\wmargin, -8*\h+\hmargin) {$I$};
    \node at (3*\w+\wmargin, -8*\h+\hmargin) {$I$};
    \node at (4*\w+\wmargin, -8*\h+\hmargin) {$\cdots$};
    \node at (5*\w+\wmargin, -8*\h+\hmargin) {$I$};
    \node at (6*\w+\wmargin, -8*\h+\hmargin) {$I$};
    \node at (7*\w+\wmargin, -8*\h+\hmargin) {$\mathcal{P}(g_{ \mathcal{B}_{2} })$};
    \node at (8*\w+\wmargin, -8*\h+\hmargin) {$\mathcal{P}(g_{ \mathcal{B}_{\cdots} })$};
    \node at (9*\w+\wmargin, -8*\h+\hmargin) {$I$};
    \node at (1*\w+\wmargin, -10*\h+\hmargin) {$I$};
    \node at (2*\w+\wmargin, -10*\h+\hmargin) {$I$};
    \node at (3*\w+\wmargin, -10*\h+\hmargin) {$I$};
    \node at (4*\w+\wmargin, -10*\h+\hmargin) {$\cdots$};
    \node at (5*\w+\wmargin, -10*\h+\hmargin) {$I$};
    \node at (6*\w+\wmargin, -10*\h+\hmargin) {$I$};
    \node at (7*\w+\wmargin, -10*\h+\hmargin) {$I$};
    \node at (8*\w+\wmargin, -10*\h+\hmargin) {$\mathcal{P}(g_{ \mathcal{B}_{\cdots} })$};
    \node at (9*\w+\wmargin, -10*\h+\hmargin) {$\mathcal{P}(g_{ \mathcal{B}_{2w-n-k} })$};
\end{tikzpicture}
\caption{
Illustration of $\mathsf{G}^{ \mathrm{Bell} }$, $\mathsf{G}^{ (\mathrm{A}) }$, and $\mathsf{G}^{ (\mathrm{B}) }$.
As in Fig.~\ref{illust of stabilizer group 2w<k+n}, we illustrate only the $n$-qubit system part.
Although $\mathcal{S}$, $\mathcal{A}$, and $\mathcal{B}$ are visualized as contiguous sets, arbitrary choices are allowed. 
}
\label{illust of stabilizer group 2w>k+n}
\end{figure}

From the above construction, $\mathsf{G}^{ \mathrm{Bell} }(\mathcal{S}) \cup \{ \mathbf{g} \} \cup \mathsf{G}^{ (\mathrm{A}) }(\mathbf{g}, \mathcal{A}) \cup \mathsf{G}^{ (\mathrm{B}) }(\mathbf{g}, \mathcal{B})$ is also a set of $k+n$ mutually commuting Pauli strings. 
Therefore, we define a stabilizer group $\mathsf{S}^{ (\mathrm{S}, \mathrm{A}, \mathrm{B}) }(\mathbf{g}, (\mathcal{S}, \mathcal{A}, \mathcal{B}))$ as
\begin{equation}
    \mathsf{S}^{ (\mathrm{S}, \mathrm{A}, \mathrm{B}) }(\mathbf{g}, (\mathcal{S}, \mathcal{A}, \mathcal{B}))
    := \left\langle
    \mathsf{G}^{ \mathrm{Bell} }(\mathcal{S})
    \cup \{ \mathbf{g} \}
    \cup \mathsf{G}^{ (\mathrm{A}) }(\mathbf{g}, \mathcal{A})
    \cup \mathsf{G}^{ (\mathrm{B}) }(\mathbf{g}, \mathcal{B})
    \right\rangle.
\end{equation}
Consequently, the set $\mathsf{U}^{ (2w > k+n) }$ is defined as
\begin{equation}
    \mathsf{U}^{ (2w > k+n) }
    := \{
    \mathsf{S}^{ (\mathrm{S}, \mathrm{A}, \mathrm{B}) }(\mathbf{g}, (\mathcal{S}, \mathcal{A}, \mathcal{B}))
    ~:~ |\mathbf{g}| = n-k,
        ~(\mathcal{S}, \mathcal{A}, \mathcal{B}) \text{ partition such that }
        |\mathcal{S}|=k,~ |\mathcal{A}|=2(n-w),~ |\mathcal{B}|=2w-n-k
    \}.
\end{equation}
Based on the construction, the size of $\mathsf{U}^{ (2w > k+n) }$ is $|\mathsf{U}^{ (2w > k+n) }| = \binom{n}{k} \times \binom{n-k}{2(n-w)} \times 3^{n-k}$, where $\binom{n}{k}$ corresponds to the choice of the size-$k$ subset $\mathcal{S}$, $\binom{n-k}{2(n-w)}$ corresponds to the choice of $\mathcal{A}$, and $3^{n-k}$ is the number of the weight-$(n-k)$ Pauli strings $\mathbf{g}$.

\begin{figure}[h]
\centering
\begin{tikzpicture}
\def\S{2}
\def\A{4.5}
\def\B{3}

\fill[yellow!30] (0,0) rectangle (\S,1);
\fill[blue!15] (\S,0) rectangle ({\S+\A},1);
\fill[red!15] ({\S+\A},0) rectangle ({\S+\A+\B},1);

\draw (0,0) rectangle ({\S+\A+\B},1); 
\draw (\S,0) -- (\S,1); 
\draw ({\S+\A},0) -- ({\S+\A},1); 

\node at (\S/2,0.5) {$m$};
\node at (\S+\A/2,0.5) {$n - w + k - m$};
\node at (\S+\A+\B/2,0.5) {$2w - n - k$};

\draw[decorate,decoration={brace,amplitude=5pt},yshift=0.4cm]
(0,1) -- (\S,1) node[midway,yshift=0.4cm] {$|\mathcal{S}| = k$};
\draw[decorate,decoration={brace,amplitude=5pt},yshift=0.4cm]
(\S,1) -- ({\S+\A},1) node[midway,yshift=0.4cm] {$|\mathcal{A}| = 2(n-w)$};
\draw[decorate,decoration={brace,amplitude=5pt},yshift=0.4cm]
({\S+\A},1) -- ({\S+\A+\B},1) node[midway,yshift=0.4cm] {$|\mathcal{B}| = 2w - n - k$};

\node at (\S/2,-0.5) {$\binom{k}{m}3^{m}$};
\node at (\S+\A/2,-0.5) {$\binom{2(n-w)}{(n-k) - (w-m)}$};
\node at (\S+\A+\B/2,-0.5) {$2^{2w - n - k-1}$};

\node at (-1.8,1.5) {\textbf{partition}};
\node[anchor=base] at (-1.8,0.4) {\textbf{weight}};
\node at (-1.8,-0.45) {\textbf{\# of choices}};

\end{tikzpicture}
\caption{
Illustration of constructing weight-$w$ Pauli strings in $\mathsf{S}^{ (\mathrm{S}, \mathrm{A}, \mathrm{B}) }(\mathbf{g}, (\mathcal{S}, \mathcal{A}, \mathcal{B}))$.
As in the case $2w \leq k+n$, a weight-$w$ Pauli string is formed by selecting $m$ components from $\mathcal{S}$ and the remaining $(w-m)$ components from $\mathcal{A}\cup \mathcal{B}$.
However, in this case, we begin with a fixed weight-$(n-k)$ Pauli string $\mathbf{g}$ and reduce its weight on $\mathcal{A}\cup \mathcal{B}$ to $w-m$ by multiplying $(n-k) - (w-m)$ weight-1 generators from $\mathsf{G}^{ (\mathrm{A}) }$.
Therefore, the number of choices for each partition is given by $\binom{k}{m}3^{m}$ for $\mathcal{S}$, $\binom{|\mathsf{G}^{ (\mathrm{A}) }|}{ (n-k)-(w-m) }$ for $\mathcal{A}$, and $2^{ |\mathsf{G}^{ (\mathrm{B}) }| }$ for $\mathcal{B}$.
}
\label{counting sigma for case 2w>k+n}
\end{figure}

We proceed to verify that $\mathsf{U}^{ (2w > k+n) }$ is a uniform stabilizer covering and evaluate the covering power $\Sigma(w, k)$.
This is established by showing that each stabilizer group in $\mathsf{U}^{ (2w > k+n) }$ contains $\Sigma(w, k)$ distinct weight-$w$ Pauli strings, where $\Sigma(w, k)$ satisfies the bound
\begin{equation}
    \label{Sigma for 2w>k+n}
     \Sigma(w, k)
     \geq \sum_{m=0}^{k}\binom{k}{m} 3^{m} \binom{2(n-w)}{ (n-k)-(w-m) } 2^{2w-n-k-1},
     ~~\text{for}~\mathsf{U}^{ (2w > k+n) }.
\end{equation}
In Fig.~\ref{counting sigma for case 2w>k+n}, we illustrate how the weight-$w$ Pauli strings---whose number is denoted in Eq.~\eqref{Sigma for 2w>k+n}---are obtained using the following rules:
\begin{enumerate}
    \item Selecting a weight-$m$ Pauli string on $\mathcal{S}$ ($0\leq m \leq k$):
    It is the same as the case $2w \leq k+n$.
    A total of $\binom{k}{m}3^{m}$ weight-$m$ Pauli strings supported on $\mathcal{S}$ can be chosen, as shown in Fig.~\ref{counting sigma for case 2w>k+n}.
    
    \item Selecting $\mathbf{g}$:
    Similar to the case $k=0$, $w > n/2$, we select the weight-$(n-k)$ Pauli string $ \mathbf{g} \in \mathsf{S}^{ (\mathrm{S}, \mathrm{A}, \mathrm{B}) }(\mathbf{g}, (\mathcal{S}, \mathcal{A}, \mathcal{B})) $.
    As $\mathbf{g}$ acts non-trivially on every system qubit in $\mathcal{A}\cup \mathcal{B}$, at this step, the Pauli weight on $\mathcal{A} \cup \mathcal{B}$ is $n-k$.
    Especially, the Pauli weight on $\mathcal{A}$ is $2(n-w)$, and that on $\mathcal{B}$ is $2w-n-k$.
    
    \item Selecting $(n-k) - (w-m)$ distinct $\mathbf{a}^{ (j) } \in \mathsf{G}^{ (\mathrm{A}) }$:
    As in the case $k=0$, $w > n/2$, each multiplication of $\mathbf{g}$ by an $\mathbf{a}^{ (j) }$ reduces one Pauli weight.
    Thus, by choosing $ (n-k) - (w-m) $ distinct $\mathbf{a}^{ (j) }$, the total weight on $\mathcal{A} \cup \mathcal{B}$ is reduced to $w-m$ from $n-k$.
    In particular, the Pauli weight on $\mathcal{A}$ is reduced to $2(n-w) - [(n-k) - (w-m)] = n-w + k-m$ from $2(n-w)$, as illustrated in Fig.~\ref{counting sigma for case 2w>k+n}.
    The number of choices is $\binom{|\mathsf{G}^{ (\mathrm{A}) }|}{ (n-k)-(w-m) } = \binom{2(n-w)}{ (n-k)-(w-m) }$.
    
    \item Arbitrarily selecting multiple $\mathbf{b}^{ (j) } \in \mathsf{G}^{ (\mathrm{B}) }$:
    As in the case $k=0$, $w > n/2$, multiplying by any combination of $\mathbf{b}^{ (j) }$ does not change the weight of the constructed Pauli string.
    Therefore, the number of choices is $2^{ |\mathsf{G}^{ (\mathrm{B}) }| } = 2^{2w-n-k-1}$.
\end{enumerate}
By considering all cases $0 \leq m \leq k$, we obtain the value given in Eq.~\eqref{Sigma for 2w>k+n}.
Similar to the case $k=0$, $w>n/2$, there exist weight-$w$ Pauli strings generated without selecting $\mathbf{g}$, and the number of such Pauli strings is the same across all $\mathsf{S}^{ (\mathrm{S}, \mathrm{A}, \mathrm{B}) }(\mathbf{g}, (\mathcal{S}, \mathcal{A}, \mathcal{B}))$.
Therefore, the set $\mathsf{U}^{ (2w > k+n) }$ satisfies the first condition.
By the definition of $\mathsf{U}^{ (2w > k+n) }$, all possible choices of the partitions and $\mathbf{g}$ are included, so $\mathsf{P}(w)$ is fully covered.
Moreover, since the construction treats all choices equivalently, every weight-$w$ Pauli string appears in an equal number of stabilizer groups. 
This ensures that the second condition is satisfied.

Finally, using the covering power in Eq.~\eqref{Sigma for 2w>k+n}, we obtain an upper bound on $N$ via Eqs.~\eqref{N for CN with k} and~\eqref{CN and Sigma with k}.
As in the case $2w \leq k+n$, this upper bound does not coincide with the lower bound in Theorem 2 of the main text.
We leave as an open question how to close the gap between the lower and upper bounds.
As noted in Sec.~\hyperref[Sec: S3E]{S3 E}, the lower bound in Theorem 2 is not tight, and it could be improved by showing Eq.~\eqref{conjecture for EG}.
However, even if the expectation is confirmed, it still falls short of reaching the upper bound established in this section.
Thus, it remains possible that a scheme more efficient than our stabilizer-covering method could be found.

\subsection{Upper bound on $N$ for the case $k>0$}


In this section, we analyze the proposed upper bound on $N$ for the case $k>0$. 
More precisely, we study the quantity $ \left\lceil \frac{|\mathsf{P}| \log |\mathsf{P}|}{\Sigma} \right\rceil $ (see Eqs.~\eqref{N for CN with k} and~\eqref{CN and Sigma with k}), which represents the sample complexity obtained from our extended stabilizer-covering scheme. 
In particular, we show that when both $k$ and $w$ scale proportionally with $n$ (see Fig.~2 of the main text), whenever $k/n < 1$, the sample complexity grows exponentially under our scheme; i.e., $ \left\lceil \frac{|\mathsf{P}(w)| \log |\mathsf{P}(w)|}{\Sigma(w,k)} \right\rceil = \exp(\Omega (n))$. 
In other words, in Fig.~2 of the main text, only the line corresponding to $k=n$ represents the region where the learning task can be achieved with a polynomial sample complexity.

For fixed $n$ and $w$ values, the case with enough $k$ (i.e., $2w \leq k+n$) is easier than the case $2w > k+n$ to accomplish the learning task.
Therefore, to verify the above argument, it is enough to consider the case $2w \leq k+n$.
At first, we find the upper bound on the covering power $\Sigma(w, k)$ for the case $2w \leq k+n$ given in Eq.~\eqref{Sigma for 2w<k+n} as follows:
\begin{equation}
    \Sigma(w, k) \leq \binom{k}{\tilde{m}} 3^{\tilde{m}} \binom{n-k}{w-\tilde{m}},
\end{equation}
where $\tilde{m} = \lceil \frac{kw}{n} \rceil$.
Note that the value $\frac{kw}{n}$ corresponds to the mean of the hypergeometric distribution whose probability mass function is given by $\Pr(X=x) = \frac{ \binom{k}{x}\binom{n-k}{w-x} }{ \binom{n}{w} }$.
Using Stirling's formula in Eq.~\eqref{Stirling approx}, we obtain the upper bound on $\Sigma(w, k)$ as
\begin{equation}
    \Sigma(w, k)
    \leq \binom{k}{\tilde{m}} 3^{\tilde{m}} \binom{n-k}{w-\tilde{m}}
    = O\left( 3^{\frac{kw}{n}}\times \frac{1}{\sqrt{k(n-k)}} \exp[k H(w/n) + (n-k)H(w/n)] \right)
    = O\left( \frac{1}{\sqrt{n}} 3^{\frac{kw}{n}} \times \binom{n}{w} \right).
\end{equation}
Using this upper bound, we derive
\begin{equation}
    \left\lceil \frac{|\mathsf{P}(w)| \log |\mathsf{P}(w)|}{\Sigma(w,k)} \right\rceil
    = \Omega(n^{3/2} 3^{w\left(1-\frac{k}{n}\right)}),
\end{equation}
since $|\mathsf{P}(w)| = 3^{w}\binom{n}{w}$.
Therefore, unless $k=n$, our stabilizer-covering scheme requires an exponential sample complexity when $w$ scales proportionally with $n$.



\bibliography{On_the_Fundamental_Resource_for_Exponential_Advantage_in_Quantum_Channel_Learning}